\documentclass[11pt,english,aps,tightenlinesletterpaper,superscriptaddress,secnumarabic,nofootinbib]{revtex4}
\usepackage[T1]{fontenc}
\usepackage[latin9]{inputenc}
\usepackage{babel}
\usepackage{amsmath}
\usepackage{amssymb}
\usepackage{graphicx}
\usepackage{esint}
\usepackage[unicode=true,pdfusetitle,
 bookmarks=true,bookmarksnumbered=false,bookmarksopen=false,
 breaklinks=false,pdfborder={0 0 1},backref=section,colorlinks=false]
 {hyperref}
\hypersetup{
 citecolor=blue}

\makeatletter
\@ifundefined{textcolor}{}
{%
 \definecolor{BLACK}{gray}{0}
 \definecolor{WHITE}{gray}{1}
 \definecolor{RED}{rgb}{1,0,0}
 \definecolor{GREEN}{rgb}{0,1,0}
 \definecolor{BLUE}{rgb}{0,0,1}
 \definecolor{CYAN}{cmyk}{1,0,0,0}
 \definecolor{MAGENTA}{cmyk}{0,1,0,0}
 \definecolor{YELLOW}{cmyk}{0,0,1,0}
}

\usepackage{amsfonts}\usepackage{bbm}\usepackage{euscript}
\usepackage{dcolumn}
\usepackage{bm}
\usepackage{epsfig}\epsfclipon
\usepackage{slashed}

\makeatother

\begin{document}

\title{Fermionic Isocurvature Perturbations}

\author{Daniel J. H. Chung}

\email{danielchung@wisc.edu}

\author{Hojin Yoo}

\email{hyoo6@wisc.edu}

\author{Peng Zhou}

\email{pzhou@wisc.edu}

\affiliation{Department of Physics, University of Wisconsin-Madison, Madison,
WI 53706, USA}

\date{\today}
\begin{abstract}
Isocurvature perturbations in the inflationary literature typically
involve quantum fluctuations of bosonic field degrees of freedom.
In this work, we consider isocurvature perturbations from fermionic
quantum fluctuations during inflation. When a stable massive fermion
is coupled to a non-conformal sector different from the scalar metric
perturbations, observably large amplitude scale invariant isocurvature
perturbations can be generated. In addition to the computation of
the isocurvature two-point function, an estimate of the local non-Gaussianities
is also given and found to be promising for observations in a corner
of the parameter space. The results provide a new class of cosmological
probes for theories with stable massive fermions. On the technical
side, we explicitly renormalize the composite operator in curved spacetime
and show that gravitational Ward identities play an important role
in suppressing certain contributions to the fermionic isocurvature
perturbations.
\end{abstract}
\maketitle
\begin{widetext} \tableofcontents{}\vspace{5mm}
 \end{widetext}

\section{Introduction\label{sec:Introduction}}

The Cosmic Microwave Background (CMB) measurements \cite{Ade:2013xsa,Ade:2013lta,Ade:2013rta,Ade:2013sta,Ade:2013ydc,Hinshaw:2012fq,Komatsu:2010fb,Brown:2009uy,Reichardt:2008ay,Fowler:2010cy,Lueker:2009rx}
and the Large Scale Structure (LSS) observations \cite{Percival:2007yw,Eisenstein:2005su}
are consistent with single field inflationary models which can seed
approximately adiabatic, scale-invariant, and Gaussian primordial
density perturbations \cite{Starobinsky:1980te,Sato:1980yn,Linde:1981mu,Mukhanov:1981xt,Albrecht:1982wi,Hawking:1982my,Guth1982,Starobinsky:1982ee,Bardeen:1983qw}.
However, from the multi-field nature of the Standard Model of particle
physics, one may naturally guess that there would be more than one
light degrees of freedom during inflation which may be responsible
for generating isocurvature primordial perturbation initial conditions.
Indeed, in any slow-roll inflationary scenario, non-inflaton degrees
of freedom must eventually turn on in order to reheat successfully.%
\footnote{Even though the reheat degrees of freedom do not need to be dynamically
important during the quasi-dS era, multiple fields are certainly lurking
in the scenario.%
} Hence, isocurvature scenarios are theoretically well motivated.

Isocurvature perturbations have been studied in various scenarios,
such as double inflation \cite{Silk:1986vc,Polarski:1994rz,Langlois:1999dw,Yamaguchi:2001zh},
curvaton scenario \cite{Lyth:2001nq,Enqvist:2001zp,Moroi:2001ct,Lyth:2002my,Langlois:2013dh,Enqvist:2012tc,Harigaya:2012up,Enomoto:2012uy,Enqvist:2012xn,Dimopoulos:2011gb,Alabidi:2010ba,Lin:2010ua,Assadullahi:2007uw,Moroi:2002rd,Bartolo:2002vf},
axions \cite{Seckel:1985tj,Preskill:1982cy,Abbott:1982af,Dine:1982ah,Steinhardt:1983ia,Turner:1985si,Kolb:1990vq,Fox:2004kb,Beltran:2006sq,Hertzberg:2008wr}
and gravitationally produced superheavy dark matter \cite{Linde:1996gt,Chung:1998zb,Chung:2004nh,Chung:2011xd}.
Isocurvature perturbations also can generate rich density perturbation
phenomenology. For example, unlike standard single field inflationary
scenarios, isocurvature perturbations are able to generate large primordial
local non-Gaussianities \cite{Bartolo:2001cw,Chung:2011xd,Kawasaki:2008sn,Langlois:2008vk,Linde:1996gt,Kofman:1989ed,Geyer:2004bx,Ferrer:2004nv,Boubekeur:2005fj,Barbon:2006us,Lyth:2006gd,Koyama:2007if,Lalak:2007vi,Huang:2007hh,Lehners:2008vx,Beltran:2008tc,Kawasaki:2008jy,Langlois:2009jp,Chen:2010xka,Langlois:2010fe,Langlois:2011zz,Mulryne:2011ni,Gong:2011cd,DeSimone:2012gq,Enqvist:2012vx,Kawasaki:2011pd,Langlois:2013dh,Kawasaki:2013ae,Nurmi:2013xv}.
However, most previous studies of isocurvature perturbations focused
on bosonic degrees of freedom such as axions and curvatons. Fermionic
isocurvature degrees of freedom such as gravitinos were only discussed
in the literature associated with the decay products of the inflaton
or other scalars \cite{Kawasaki:2006gs,Kawasaki:2006hm,Pradler:2006hh,Pradler:2006qh,Endo:2007ih,Endo:2007sz,Takahashi:2009dr,Takahashi:2009cx}.
Furthermore, these fermions discussed in the literature were characterized
only by their dependence on the entropy temperature fluctuation $\delta T$
which was assumed to be directly linked to the curvature perturbation
$\zeta$, in a manner consistent with the ``separate universe''
picture of $\delta N$ formalism \cite{Starobinsky:1986fxa,Sasaki:1995aw,Sasaki:1998ug}.
Such previously discussed fermionic isocurvature scenarios lead to
strong correlation or anticorrelation with the curvature perturbation
$\zeta$. One can intuitively characterize these previous fermionic
isocurvature works as having no fermionic quantum fluctuation information
from the inflationary era.

In contrast, we examine in this paper a fermionic isocurvature scenario
that is not (significantly) correlated with $\zeta$ and has fermionic
quantum fluctuation information during inflation encoded in the isocurvature
correlator. In our scenario, the horizon length scale interaction
dynamics of the fermion particles is important, in sharp contrast
with the ``separate universe'' picture of $\delta N$ formalism.
As we will show, although classical gravitational field interactions
alone are sufficient to generate enough fermions during the exit process
of inflation to saturate the phenomenologically required cold dark
matter abundance \cite{Chung:2011ck,Kuzmin:1998kk,Kuzmin:1999zk},
fermion propagators in the classical FRW background is insufficient
to produce any observable isocurvature perturbations because of the
fact that massless fermions enjoy a classical conformal symmetry.%
\footnote{Even with the massive fermions, we will be naturally concerned with
light fermions where $m_{\psi}/H\ll1$.%
} Hence, any large fermion isocurvature correlator must involve couplings
to a conformal symmetry breaking sector. 

For illustrating the existence of such fermionic isocurvature perturbations,
we minimally extend the single field inflation by adding a stable
massive fermion field coupled through a Yukawa coupling to a light
non-inflaton scalar field whose mass is much lighter than the fermion
field (hence, there are no decays of the scalars to the fermions).
The light non-inflaton scalar field (which is minimally coupled to
gravity) serves as a conformal symmetry breaking sector through which
the fermions will attain appreciable correlations. We compute the
isocurvature two-point function of fermions that are gravitationally
produced during inflation and identify the phenomenologically viable
parameter space. We also estimate the local non-Gaussianity and show
that it may be observationally large in a particular parametric regime. 

At the technical level, treating fermionic isocurvature fluctuations
during inflation requires composite operator renormalization in quasi-dS
spacetime because the fermionic energy-momentum tensor is a composite
bilinear operator (i.e.~fermions cannot get VEVs) and the leading
two-point function contribution involves a one loop 1PI diagram. To
our knowledge, this paper is the first paper to apply composite fermion
operator renormalization in inflationary spacetime to treat isocurvature
perturbations. Indeed, an improper treatment of the operator renormalization
can in principle lead to answers that are many orders of magnitude
off as we pointed out with bosonic composite operators \cite{Chung:2013sla}.
We also show that a gravitational Ward identity plays an important
role in suppressing the scalar metric perturbation interaction contribution
to the isocurvature two-point function (thereby justifying our introduction
of another scalar sector).

This paper is presented in the following order. In Sec.~\ref{sec:Fermion-Isocurvature-Model},
we motivate and discuss the fermion isocurvature model. Next, we review
the definition of the gauge-invariant variables and the quantum operator
associated with the cold dark matter (CDM) isocurvature in Sec.~\ref{sec:Operator-for-Isocurvature}.
In subsection \ref{sub:Regularization-and-Renormalizati}, we present
the regulator and the renormalization conditions for our isocurvature
operator. We explain the constraints on the Yukawa coupling coming
from the self-consistency of our simplified scenario in Sec.~\ref{sub:debyemass}.
In Sec. \ref{sec:2-pt}, we compute the isocurvature 2-point function.
The leading order and the next leading order results are given in
subsection \ref{sub:LO} and \ref{sub:NLO}, and the power spectrum
is presented in subsection \ref{sub:Powerspectrum}. In Sec.~\ref{sec:result},
we discuss the numerical implications of our results and non-Gaussianities.
Afterwards in Sec.~\ref{sub:gauge-int}, we discuss the explicit
computation of how a diffeomorphism Ward identity plays a role in
suppressing the scalar metric perturbation contribution to the isocurvature
two-point function. Finally, in Sec.~\ref{sec:conclusion} we summarize
and conclude. Some technical details of the computations are given
in the Appendices.

\section{Fermion Isocurvature Model\label{sec:Fermion-Isocurvature-Model}}

As is well known, if any small mass fermion field degrees of freedom
exist during inflation which is usually assumed to be a Bunch-Davies
vacuum state, fermion particles will be produced gravitationally (see
e.g. \cite{birrell1982ix,DeWitt1975,Kuzmin:1998kk,Kuzmin:1999zk,Chung:2011ck}).
The inhomogeneities of the gravitationally produced fermions will
generically not align with the inhomogeneities of the inflaton, depending
on its interactions. If most of the radiation in the universe comes
from the inflaton decay, then the misalignment of the inhomogeneities
of the fermions and the inflaton will lead to isocurvature perturbations
\cite{peebles1980large,Efstathiou:1986,Liddle:2000cg}.

Now, to motivate our fermion model with Yukawa interactions, it is
important to understand why interactions to conformal symmetry breaking
sector is required. It is also well known that massless fermion classical
action enjoys a conformal symmetry:
\begin{equation}
g_{\mu\nu}\rightarrow e^{2\sigma(x)}g_{\mu\nu}
\end{equation}
\begin{equation}
\psi\rightarrow e^{-3\sigma(x)/2}\psi.
\end{equation}
Since FRW spacetime can be written as a conformal transformation of
the Minkowski space (i.e. $a=\exp(\sigma)$), we would expect for
a tree level fermion propagating on an FRW spacetime without any interactions
with a conformal symmetry breaking sector
\begin{equation}
\langle\bar{\psi}\psi(t,\vec{x})\bar{\psi}\psi(t,\vec{y})\rangle_{conn}=\langle\bar{\psi}_{M}\psi_{M}(t,\vec{x})\bar{\psi}_{M}\psi_{M}(t,\vec{y})\rangle_{conn}a^{-6}
\end{equation}
where $\psi_{M}$ is the Minkowski fermion. At leading order, there
are no other scales in this function except $|\vec{x}-\vec{y}|$.
Hence, we conclude
\begin{equation}
\langle\bar{\psi}\psi(t,\vec{x})\bar{\psi}\psi(t,\vec{y})\rangle_{conn}\sim\frac{1}{a^{6}|\vec{x}-\vec{y}|^{6}}\label{eq:suppressedmasslesscorrelator}
\end{equation}
in the massless limit.%
\footnote{The scaling behavior of the two-point correlator is similar to that
of correlators considered in Ref. \cite{Green:2013rd} in the context
of conformal field theories. %
} We expect this to be the dominant contribution in the limit that
$m_{\psi}/H\ll1$. When $m_{\psi}/H\gg1$, we also expect there can
be factors multiplying this that vanishes exponentially fast as $m_{\psi}/H\rightarrow\infty$
(we show this explicitly in Sec.~\ref{sub:LO}). Hence, we expect
Eq.~(\ref{eq:suppressedmasslesscorrelator}) to be the leading order
of magnitude composite correlator if the theory is approximately conformally
invariant. As we will show below, the comoving gauge isocurvature
perturbations is proportional to 
\begin{equation}
\langle\frac{\delta\rho_{\psi}^{(C)}}{\bar{\rho}_{\psi}}\frac{\delta\rho_{\psi}^{(C)}}{\bar{\rho}_{\psi}}\rangle\sim\frac{\langle\bar{\psi}\psi(t,\vec{x})\bar{\psi}\psi(t,\vec{y})\rangle_{conn}}{\left\langle \bar{\psi}\psi\right\rangle ^{2}}.\label{eq:comovingcorrelator}
\end{equation}
where one sees the appearance of the suppressed correlator computed
in Eq.~(\ref{eq:suppressedmasslesscorrelator}). Because of this
suppression, fermionic isocurvature perturbations require nontrivial
interactions with a conformal symmetry breaking sector.

If the conformal symmetry breaking sector is just the $\zeta$ sector
of the inflaton, then its effective coupling to the fermions is suppressed
because there is an infinitesimal shift symmetry of the $\zeta$ coming
from a residual diffeomorphism symmetry in the comoving gauge. (We
will explain this explicitly in Sec.~\ref{sub:gauge-int} in terms
of a Ward identity.) Hence, to generate an observable fermionic correlator
during the horizon exit, another conformal symmetry breaking sector
must be introduced which does not suffer from derivative coupling
suppression similar to $\zeta$.%
\footnote{Although we have not investigated the suppression for the tensor perturbation
interactions with a full computation, we expect a similar suppression
of the tensor perturbation interactions.%
} We thus introduce a Yukawa coupling to a light non-inflaton scalar
and demonstrate that this interaction can induce observable isocurvature
fluctuations.%
\footnote{ Note that this introduction of a light scalar is not particularly
attractive from a model building perspective since we provide no explicit
mechanism to protect its light mass. We defer the challenge of building
an attractive model to a future work since the purpose of this paper
is to demonstrate the basic physics mechanism.%
}

Given this motivation, let us now specify the model studied in this
paper. We use one real scalar $\phi$ slow-roll inflaton degree of
freedom that dominates the energy density during inflation and then
perturbatively decays to the SM sector to reheat the universe. We
also introduce another minimally coupled light real scalar degree
of freedom $\sigma$ which has no coupling to $\phi$ or the SM sector
(necessary for reheating) stronger than gravity.%
\footnote{For now, we will consider this as a tuning and will not address serious
model building issues in this paper. It is plausible that this kind
of scenario can be realized in the context of SUSY hidden sector.%
} As we explained, the main role of $\sigma$ is to provide a conformal
symmetry breaking sector which can couple to the Dirac fermions $\psi$
through a Yukawa coupling. We assume $\psi$ is charged under a conserved
discrete charge such that the one particle states are stable and can
act as dark matter. Note that since we do not require all of the dark
matter to come from the fermions, this system is consistent with the
existence of the weakly interacting massive particle (WIMP) dark matter.
Because $\psi$ is too weakly interacting with the SM to be produced
directly, gravitationally production of $\psi$ during and after inflation
is significant and gives rise to non-thermal cold dark matter (CDM)
and its isocurvature perturbations. 

Such a model is described by the action%
\footnote{Our metric convention is $\left(-,+,+,+\right).$%
} 
\begin{eqnarray}
S & = & \int(dx)\left\{ \mathcal{L}_{inf}\left[g_{\mu\nu},\phi\right]+\mbox{\ensuremath{\mathcal{L}}}_{SM+CDM}\left[g_{\mu\nu},\left\{ \Psi\right\} \right]+\mbox{\ensuremath{\mathcal{L}}}_{RH}\left[g_{\mu\nu},\phi,\left\{ \Psi\right\} \right]\right.\nonumber \\
 &  & \left.+-\frac{1}{2}g^{\mu\nu}\partial_{\mu}\sigma\partial_{\nu}\sigma-\frac{1}{2}m_{\sigma}^{2}\sigma^{2}-\frac{y}{4!}\sigma^{4}+\bar{\psi}(i\gamma^{a}\nabla_{e_{a}}-m_{\psi})\psi-\lambda\sigma\bar{\psi}\psi\right\} ,\label{eq:full_Lag}
\end{eqnarray}
where $M_{p}^{2}=\frac{1}{8\pi G}=1$, $(dx)\equiv\sqrt{-g}d^{4}x$,
and $\mathcal{L}_{inf}$ and $\mathcal{L}_{SM+CDM}$ are the Lagrangians
for the inflaton and the SM+CDM sectors, and $\mathcal{L}_{RH}$ describes
the sector responsible for reheating. Because an interesting parameter
region exists for our scenario in which the $\psi$ constitute a tiny
fraction of the total dark matter content, the Lagrangian $\mathcal{L}_{SM+CDM}$
describes the CDM sector different from $\psi$ to make the scenario
phenomenologically viable. Note that natural heavy dark matter candidates
for $\psi$ exist in the context of string phenomenology \cite{Ellis:1990iu,Benakli:1998ut}.
Furthermore, many extensions of the Standard Model also possess superheavy
dark matter candidates (see, e.g., \cite{Kusenko:1997si,Han:1998pa,Dvali:1999tq,Hamaguchi:1998nj,Hamaguchi:1999cv,Coriano:2001mg,Cheng:2002iz,Shiu:2003ta,Berezinsky:2008bg,Kephart:2001ix,Kephart:2006zd}).
Since there are many scalar field degrees of freedom in typical BSMs,
the possibility of identifying one of these scalars with $\sigma$
is also plausible. Although the cosmological phenomenology of weakly
interacting dark matter on large scales have been investigated already
in literature (see, e.g., \cite{Chung:2004nh,Chung:2011xd,Linde:1996gt,Lyth:2001nq,Lyth:2002my,Beltran:2006sq,Beltran:2008tc,Fox:2004kb,Hertzberg:2008wr,Bartolo:2004if}),
our work is the first to describe fermionic fluctuation correlations
during inflation. Note that although Eq.~(\ref{eq:full_Lag}) has
a quartic term $\sigma^{4}$, we will focus on the parametric region
in which the quartic coupling $y$ will be small and tuned against
radiative generated quartic couplings from the Yukawa interaction
to keep the effects of the $\sigma$ interactions to a minimum. Hence,
our effective parametric domain will be controlled by $\{\lambda,m_{\sigma},m_{\psi}\}$.

We focus on a particular parametric region of $\{\lambda,m_{\sigma},m_{\psi}\}$
such that $\sigma$ only assists in generating large scale density
perturbations of $\psi$, and the density perturbations and the relic
abundance from the $\sigma$ particles vanish or are suppressed compared
to those from the $\psi$ particles. For example, requiring the correlator
$\langle\sigma\sigma\rangle|_{t_{*}}$ relevant for the isocurvature
perturbations not be suppressed gives the condition $m_{\sigma}/H(t_{*})<1$
where $t_{*}$ is the time at which the fermion production ends. This
implies $m_{\sigma}<m_{\psi}$ is the relevant parameter region. Furthermore,
in order to prevent any large isocurvature perturbations and relic
abundance of $\sigma$, we assume that the $\sigma$ particles decay
before $\sigma$ becomes an important fluid component of the evolution
of the universe (e.g. before matter-radiation equality). Note however
that this restriction is a matter of simplicity. In general, we note
that a weakly interacting and stable $\sigma$ may also be phenomenologically
allowed without problems regarding the relic abundance and the isocurvature
from $\sigma$. Moreover, for simplicity, we restrict $\lambda$ such
that 1) $\sigma\sigma\to\bar{\psi}\psi$ via the Yukawa interactions
is suppressed compared to the gravitational process in producing $\bar{\psi}\psi$
2) any $\sigma+\mbox{gravity}\rightarrow\bar{\psi}\psi$ processes
are estimated to be unimportant. This restriction is approximately
equivalent to being in a parametric region where tree-level propagator
neglecting resumption of the Yukawa interactions is valid.

In addition, in order to detach our model from the details of the
inflationary model of $\phi$, we focus on the light fermion $\psi$,
such that $m_{\psi}<H_{e}$, where $H_{e}$ is the Hubble scale at
the end of inflation. This is because the gravitational particles
production is generally sensitive to how the inflation ends in a such
way that an extra suppression factor $\exp\left(-cm_{\psi}^{2}/H_{e}^{2}\right)$
(where $c$ is a number depending on how the inflation connected with
the post inflationary era) appears in the estimation of the gravitationally
produced particle number density $n_{\psi}$. (Throughout the paper,
we will sometime distinguish $H_{e}$ from $H_{inf}$ which is defined
to be the expansion rate during inflation.) On the other hands, if
$m_{\psi}<H_{e},$ the factor becomes simply an $O(1)$ number, and
particularly, for fermions we can estimate the number density $n_{\psi}(t_{*})$
as $O(0.1)m_{\psi}^{3}$ at $H(t_{*})\sim m_{\psi}$ regardless of
how the inflation ends \cite{Chung:2011ck}. The physics of this universality
is tied to the conformal symmetry of the fermions in the massless
limit. 

At this point, we emphasize that our model is different from other
fermionic (e.g., gravitino) isocurvature models in literature (e.g.~\cite{Takahashi:2009cx,Takahashi:2009dr,Kawasaki:2008pa}).
We explicitly predict the amplitudes of fermion density perturbations
from a joint effect of the gravitational particle production and $\sigma$
modulation on $m_{\psi}$ via the matter loop diagrams. In contrast,
in Refs. \cite{Takahashi:2009cx,Takahashi:2009dr,Kawasaki:2008pa}
the fermions are produced from the on shell inflatons and/or curvatons
(the latter has the closest identification in our model with $\sigma$)
after the end of inflation. A sharp observable contrast of our model
with these other models is that our scenario predicts an uncorrelated
type of isocurvature (i.e. curvature-isocurvature cross correlation
is negligible) while these other models purportedly generate correlated
type of isocurvature. This is a consequence of the fact that these
other models do not describe any fermionic fluctuations during inflation
while in our model, the expansion during inflation imparts work to
virtual fermionic fluctuations to put them on shell.

\section{Operator for Isocurvature Perturbation\label{sec:Operator-for-Isocurvature}}

Recall that the scalar perturbation of the metric is parametrized
as 
\begin{equation}
\delta g_{\mu\nu}^{(S)}=\left(\begin{array}{cc}
-E & aF_{,i}\\
aF_{,i} & a^{2}[A\delta_{ij}+B_{,ij}]
\end{array}\right).
\end{equation}
The gauge-invariant variables are constructed by combining metric
perturbations and other perturbations, such as density perturbations.
For example, the conventional first-order gauge-invariant perturbation
associated with the energy density of a fluid $a$ is defined (see,
e.g., \cite{Weinberg:2008ug,Liddle:2000cg} and references therein)
by 
\begin{equation}
\zeta_{a}\equiv\frac{A}{2}-H\frac{\delta\rho_{a}}{\dot{\bar{\rho}}_{a}}.
\end{equation}
In particular, we define the conventional curvature perturbation as
\begin{equation}
\zeta\equiv\frac{A}{2}-H\frac{\delta\rho_{tot}}{\dot{\bar{\rho}}_{tot}},
\end{equation}
where 
\begin{equation}
\delta\rho_{tot}=\sum_{i}\delta\rho_{i},\quad\bar{\rho}_{tot}=\sum_{i}\bar{\rho}_{i}.
\end{equation}
This quantity $\zeta$ is conserved when modes are stretched out of
the horizon even through the reheating era as long as it is set by
the adiabatic initial condition, i.e., $\zeta=\zeta_{a}$ for any
fluid $a$. Furthermore, if perturbations are generated solely by
inflaton during inflation, such as the single field inflation, superhorizon
perturbations automatically satisfy the adiabatic initial condition
and the perturbations are conserved so that we can match them with
those during the early radiation dominated (RD) era, $\zeta_{\phi}(t_{inf})=\zeta_{\gamma}(t_{RD})=\zeta_{m}(t_{RD})=\cdots$.

On the other hand, an isocurvature perturbation is defined by a relative
density perturbation between two different fluids
\begin{equation}
\delta_{Sij}\equiv3\left(\zeta_{i}-\zeta_{j}\right)=-3H\left(\frac{\delta\rho_{i}}{\dot{\bar{\rho_{i}}}}-\frac{\delta\rho_{j}}{\dot{\bar{\rho_{j}}}}\right).\label{eq:deltaS_ij}
\end{equation}
In general, it may arise during inflation if there are more than one
degree of freedom. Although their mixing with perturbations of different
fluids can lead to the failure of the conservation of the curvature
perturbation $\zeta$, such effects are negligible as for any species
$i$ whose $\bar{\rho}_{i}+\bar{P}_{i}$ is sufficiently smaller than
$\bar{\rho}_{\mbox{tot}}+\bar{P}_{\mbox{tot}}$ until the Universe
reaches radiation domination. Particularly, for gravitationally produced
fermions, we have 
\begin{equation}
\left.\frac{\bar{\rho}_{\psi}+\bar{P}_{\psi}}{\bar{\rho}_{\mbox{tot}}+\bar{P}_{\mbox{tot}}}\right|_{t_{*}}\sim\left.\frac{\bar{\rho}_{\psi}}{\bar{\rho}_{tot}}\right|_{t_{*}}\sim\frac{m_{\psi}^{2}}{M_{p}^{2}}\ll\Delta_{\zeta}^{2},
\end{equation}
where $t_{*}$ is the time that the gravitational fermion production
ends, $H(t_{*})\sim m_{\psi}.$ Hence, we expect the superhorizon
curvature perturbation to be approximately conserved through the reheating,
$\zeta(t_{RD})\approx\zeta_{\phi}(t_{inf}).$

The dominant fraction of the produced fermions are non-relativistic.%
\footnote{This is a valid assumption because gravitationally excited fermion
modes that contributions to the energy density are less than the fermion
mass, i.e., $\left|\beta_{k}\right|^{2}$ for $k/a\lesssim m_{\psi}$,
where $\beta_{k}$ is the Bogoliubov coefficient. See Appendix \ref{sec:prod_review}
for the detail.%
} Then the fermion energy density behaves as %
\footnote{One can find that $\bar{\rho}_{\psi}\propto a^{-3}(t)$ for $t>t_{*}$
if $\bar{\rho}_{\psi}$ is renormalized by the adiabatic subtraction.
See Appendix \ref{sec:prod_review} and Ref. \cite{birrell1982ix}.
Then we can treat $\psi$ as a pressure less matter. %
}
\begin{equation}
\frac{d}{dt}\bar{\rho}_{\psi}(t)\approx-3H\bar{\rho}_{\psi}\quad\mbox{for }t>t_{*},
\end{equation}
and from Eq.~(\ref{eq:deltaS_ij}) a general CDM isocurvature is
written as
\begin{equation}
\delta_{S}=\frac{\delta\rho_{CDM}}{\bar{\rho}_{CDM}}-\frac{3}{4}\frac{\delta\rho_{\gamma}}{\bar{\rho}_{\gamma}}.
\end{equation}
As discussed in Sec. \ref{sec:Fermion-Isocurvature-Model}, the CDM
may include decay products of the inflaton $\phi$. Thus the CDM density
perturbation is generally expressed as
\begin{equation}
\frac{\delta\rho_{CDM}}{\bar{\rho}_{CDM}}=\omega_{\psi}\frac{\delta\rho_{\psi}}{\bar{\rho}_{\psi}}+\left(1-\omega_{\psi}\right)\frac{\delta\rho_{m}}{\bar{\rho}_{m}},
\end{equation}
where the subscript $m$ denotes the CDM component associated with
the inflaton decay products (such as WIMPs of minimal supersymmetric
models), and 
\begin{equation}
\omega_{\psi}\equiv\bar{\rho}_{\psi}/\left(\bar{\rho}_{\psi}+\bar{\rho}_{m}\right).
\end{equation}
 In particular, in the comoving gauge ($\delta\rho_{\phi}/\dot{\bar{\rho}}_{\phi}=\delta\rho_{m}/\dot{\bar{\rho}}_{m}=\delta\rho_{\gamma}/\dot{\bar{\rho}}_{\gamma}=0$),
the CDM isocurvature becomes 
\begin{equation}
\delta_{S}^{(C)}\approx\omega_{\psi}\frac{\delta\rho_{\psi}^{(C)}}{\bar{\rho}_{\psi}},
\end{equation}
where the superscript denotes the gauge choice.

Under the non-relativistic assumption, we also approximate the fermion
mass term $m_{\psi}\bar{\psi}\psi$ as its energy density%
\footnote{Using the adiabatic vacuum prescription, the renormalized energy density
is approximated in the non-relativistic case as
\[
\left\langle \left(\rho_{\psi}\right)_{r}\right\rangle \approx m_{\psi}\left\langle N_{\psi}\right\rangle =2m_{\psi}\int\frac{d^{3}k}{\left(2\pi^{3}\right)}\frac{1}{a^{3}}\left|\beta_{k}\right|^{2},
\]
where $N_{\psi}$ is a fermion number operator, and the subscript
r denotes that the operator is a renormalized composite operator.
This quantity is in accord with
\[
m_{\psi}\left\langle \left(\bar{\psi}\psi\right)_{r}\right\rangle =2m_{\psi}\int\frac{d^{3}k}{\left(2\pi^{3}\right)}\frac{m_{\psi}}{\omega_{p}}\left|\beta_{k}\right|^{2}\approx2m_{\psi}\int\frac{d^{3}k}{\left(2\pi^{3}\right)}\left|\beta_{k}\right|^{2}.
\]
 In particular, $\left(\bar{\psi}\mbox{\ensuremath{\psi}}\right)$
has an advantage in constructing gauge-invariant variables because
it is manifestly 4-scalar, but $N_{\psi}$.%
}
\begin{equation}
\rho_{\psi}\approx m_{\psi}\bar{\psi}\psi,\label{eq:approx_energy_density}
\end{equation}
and then the the fermion isocurvature perturbation becomes
\begin{equation}
\delta_{S}^{(C)}\approx\omega_{\psi}\frac{\rho_{\psi}-\left\langle \rho_{\psi}\right\rangle }{\left\langle \rho_{\psi}\right\rangle }=\omega_{\psi}\frac{\bar{\psi}\psi-\left\langle \bar{\psi}\psi\right\rangle }{\left\langle \bar{\psi}\psi\right\rangle }.\label{eq:def_iso}
\end{equation}
Notice that as it is a quantum composite operator, we renormalize
it with regulators and counter-terms invariant under the underlying
gauge symmetry, diffeomorphism in this case. In the following subsection,
we present the technical detail of the composite operator renormalization.
From now on, we will use the comoving gauge in calculating the correlation
function and drop the superscript $(C)$ for convenience.

\subsection{Regularization and Renormalization for Isocurvature Perturbation\label{sub:Regularization-and-Renormalizati}}

In this subsection, we explain our regularization procedure and renormalization
scheme that determines the counter-terms. The most crucial renormalization
condition that the isocurvature perturbations are sensitive to is
Eq.~(\ref{eq:ren-cond-Z8}).

For the convenience of preserving covariance and incorporating the
adiabatic vacuum boundary condition, we use Pauli-Villars (PV) regularization
\cite{zinn,Weinberg:2010wq}. This involves the replacements 
\begin{equation}
\psi\rightarrow\psi+\sum_{n}\psi_{n},\qquad\sigma\rightarrow\sigma+\sum_{n}\sigma_{n},\label{eq:PV}
\end{equation}
 and the addition of the Pauli-Villars part in the free Lagrangian
\begin{eqnarray}
\mathcal{L}_{PV} & = & \sum_{n=1}C_{n}(-\frac{1}{2}g^{\mu\nu}\partial_{\nu}\sigma_{n}\partial_{\nu}\sigma_{n}-\frac{1}{2}M_{n}^{2}\sigma_{n}^{2})\\
 &  & +\sum_{n=1}D_{n}\bar{\psi}_{n}(i\gamma^{a}\nabla_{a}-m_{n})\psi_{n}.
\end{eqnarray}
For notational simplicity, we let $C_{0}=1,\, M_{0}=m_{\sigma}$ and
$D_{0}=1,\, m_{0}=m_{\psi}$, and let index $N=0,1,\cdots$ and $n=1,2,\cdots$
. We require the following constraints for scalar regulators 
\begin{equation}
\sum_{N}C_{N}^{-1}=0,\quad\sum_{N}C_{N}^{-1}M_{N}^{2}=0,\quad\sum_{N}C_{N}^{-1}M_{N}^{4}=0,\cdots
\end{equation}
 and the following constraints for fermion regulators 
\begin{equation}
\sum_{N}D_{N}^{-1}=0,\quad\sum_{N}D_{N}^{-1}m_{N}=0,\quad\sum_{N}D_{N}^{-1}m_{N}^{2}=0,\cdots
\end{equation}
where we need to introduce sufficient numbers of PV fields and constraints
to cancel all the divergences. Notice that the additional constraints
in the fermions with odd powers of $m_{N}$.

\begin{figure}
\begin{centering}
\includegraphics[scale=0.75]{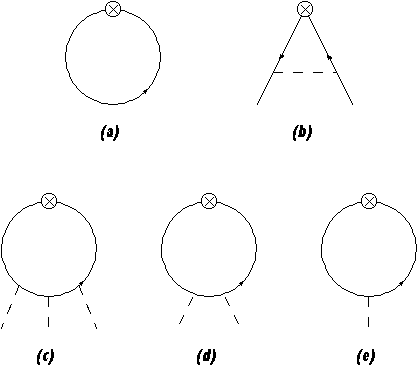}
\par\end{centering}

\caption{\label{fig:Diagrams-determining-counter-terms}Diagrams determining
counter-terms where the solid line corresponds to the fermion line
and the dashed lines corresponds to $\sigma$ lines. It is convenient
to truncate the external $\sigma$ legs on diagrams c), d), and e)
with zero momentum insertion, making these mass insertions.}
\end{figure}

With the operator dimension and the symmetry considered, the renormalized
operator is written as 
\begin{eqnarray}
(\bar{\psi}\psi)_{x,r} & = & (\bar{\psi}_{x})_{r}(\psi_{x})_{r}(1+\delta Z_{1})+\delta Z_{2}(\sigma_{x,r})^{3}+\delta Z_{3}(\sigma_{x,r})^{2}\nonumber \\
 &  & +\delta Z_{4}\sigma_{x,r}+\delta Z_{5}+\delta Z_{6}\square\sigma_{x,r}+\delta Z_{7}R+\delta Z_{8}R\sigma_{x,r}\label{eq:def_psibarpsi}
\end{eqnarray}
where each field operator should be understood as including a sum
of the PV fields as in Eq.~(\ref{eq:PV}). Then we give the renormalization
conditions to determine the counter terms. For $\delta Z_{i}$ which
are not coupled to $R,R_{\mu\nu},R_{\beta\mu\nu}^{\alpha}$ and their
derivatives, we can go to the Minkowski space and impose the renormalization
conditions there. (Of course, we do not need to separate the curved
space contribution and the flat space contribution with two computations,
but we present this here this way here for clarity in the physical
partition.) We define the renormalized operator $\bar{\psi}\psi$
at one-loop order, such that it measures the number density of the
fermion particles. First, we require its expectation value in the
flat space vacuum to vanish:
\begin{eqnarray}
\langle vac|\bar{\psi}\psi(x)|vac\rangle_{flat}+\sum_{n=1}\langle vac|\bar{\psi}_{n}\psi_{n}(x)|vac\rangle_{flat}+\delta Z_{5} & = & 0\label{eq:Z5}\\
\Rightarrow-\int\frac{d^{4}p}{(2\pi)^{4}}\sum_{N}D_{N}^{-1}\mbox{Tr}\left\{ \frac{1}{i}\frac{-\slashed{p}+m_{N}}{p^{2}+m_{N}^{2}-i\epsilon}\right\} +\delta Z_{5} & = & 0.
\end{eqnarray}
This corresponds to the evaluation of diagram (a) in Fig.~\ref{fig:Diagrams-determining-counter-terms}.

Next, we impose the renormalization condition consistent with the
fact that as far as the fermion sector is concerned, a shift of $\sigma$
by a constant in the tree-level action is equivalent to a shift in
the mass of the fermion. More explicitly, we demand that if $\sigma$
is shifted as $\sigma\rightarrow\sigma+c$, the one-point function
satisfies 
\begin{equation}
\langle vac|(\bar{\psi}\psi)_{x,r}|vac\rangle_{flat}=\langle vac|\left[(\bar{\psi}\psi)_{x,r}+\Delta(\bar{\psi}\psi)_{x,r}\right]|vac\rangle_{flat,\mathcal{L}_{I}=-\lambda c\bar{\psi}_{y}\psi_{y}}
\end{equation}
where $\Delta(\bar{\psi}\psi)_{x,r}$ corresponds to a shift in the
$\sigma$ dependent composite operator counter-terms and $\mathcal{L}_{I}$
corresponds to $c$ dependent mass shift Lagrangian term. This leads
to diagrams (c)-(e) in Fig.~\ref{fig:Diagrams-determining-counter-terms}
with the external $\sigma$ propagators truncated and fixes $\delta Z_{2},\delta Z_{3},\delta Z_{4}$:
\begin{eqnarray}
-(-i\lambda)^{3}\int\frac{d^{4}k}{(2\pi)^{4}}\mbox{Tr}\left\{ \left(\sum_{M}D_{M}^{-1}\frac{1}{i}\frac{-\slashed{k}+m_{M}}{k^{2}+m_{M}^{2}-i\epsilon}\right)^{4}\right\} +\delta Z_{2} & = & 0\\
-(-i\lambda)^{2}\int\frac{d^{4}k}{(2\pi)^{4}}\mbox{Tr}\left\{ \left(\sum_{M}D_{M}^{-1}\frac{1}{i}\frac{-\slashed{k}+m_{M}}{k^{2}+m_{M}^{2}-i\epsilon}\right)^{3}\right\} +\delta Z_{3} & = & 0,
\end{eqnarray}
and
\begin{eqnarray}
-i\lambda\int d^{4}y\langle(\bar{\psi}\psi)_{x}(\bar{\psi}\psi)_{y}\rangle+\delta Z_{4} & = & 0\\
\Rightarrow-i\lambda\int\frac{d^{4}k}{(2\pi)^{4}}(-)\mbox{Tr}\left\{ \left(\sum_{M}D_{M}^{-1}\frac{1}{i}\frac{-\slashed{k}+m_{M}}{k^{2}+m_{M}^{2}-i\epsilon}\right)^{2}\right\} +\delta Z_{4} & = & 0.\label{eq:flatspacedelz4}
\end{eqnarray}
Furthermore, we require $\bar{\psi}\psi$ to have no loop corrections
when contracted with on-shell fermion. This leads to the diagram (b)
of Fig.~\ref{fig:Diagrams-determining-counter-terms} (where we have
set the composite operator momentum to be $0$ for convenience) and
fixes $\delta Z_{1}$:

\begin{eqnarray}
\delta Z_{1}+(i\lambda)^{2}\int\frac{d^{4}k}{(2\pi)^{4}}\sum_{L,M,N}C_{L}^{-1}D_{M}^{-1}D_{N}^{-1}\frac{1}{i}\frac{1}{k^{2}+M_{L}^{2}-i\epsilon}\nonumber \\
\times\frac{1}{i}\frac{[-\slashed{k}-\slashed{p}+m_{M}]}{(k+p)^{2}+m_{M}^{2}-i\epsilon}\times\frac{1}{i}\frac{[-\slashed{k}-\slashed{p}+m_{N}]}{(k+p)^{2}+m_{N}^{2}-i\epsilon} & = & 0.\label{eq:numberthis-1}
\end{eqnarray}
Similarly, we demand $\bar{\psi}\psi$ to have no loop corrections
when contracted with on-shell scalar line. Explicitly, the diagram
corresponds to the diagram (e) of Fig.~\ref{fig:Diagrams-determining-counter-terms}
determining $\delta Z_{6}$ :

\begin{eqnarray}
-i\lambda\int d^{4}y\langle(\bar{\psi}\psi)_{x}(\bar{\psi}\psi)_{y}\rangle e^{ip\cdot y}+\delta Z_{4}-p^{2}\delta Z_{6} & = & 0\\
\Rightarrow i\lambda\int\frac{d^{4}k}{(2\pi)^{4}}\mbox{Tr}\left\{ \sum_{M}D_{M}^{-1}\frac{1}{i}\frac{-\slashed{k}+m_{M}}{k^{2}+m_{M}^{2}-i\epsilon}\qquad\right.\nonumber \\
\left.\times\sum_{N}D_{N}^{-1}\frac{1}{i}\frac{-\slashed{k}-\slashed{p}+m_{N}}{(k+p)^{2}+m_{N}^{2}-i\epsilon}\right\} +\delta Z_{4}-p^{2}\delta Z_{6} & = & 0,
\end{eqnarray}
where $p^{2}=-m_{\sigma}^{2}$. 

For $\delta Z_{i}$ that depend on curved spacetime nature, we match
the renormalized result to that from the adiabatic subtraction. In
order to fix $\delta Z_{7}$, we impose the number density $\langle in|(\bar{\psi}\psi)_{r,x}|in\rangle$
to be the density defined by the adiabatic prescription (See, e.g.,
\cite{birrell1982ix,Chung:1998bt,Chung:2011ck,Kuzmin:1998kk,Kuzmin:1999zk,Chung:1998zb,Chung:2001cb}):
\begin{eqnarray}
n_{\psi} & \equiv & \langle in|\bar{\psi}\psi(x)|in\rangle+\sum_{n=1}\langle in|\bar{\psi}_{n}\psi(x)_{n}|in\rangle+\delta Z_{5}+\delta Z_{7}R(x)\\
 & = & \langle in|\bar{\psi}\psi(x)|in\rangle-\langle WKB,vac,t_{x}|\bar{\psi}\psi(x)|WKB,vac,t_{x}\rangle,\label{eq:def_Npsi}
\end{eqnarray}
where $\left|WKB,vac,t_{x}\right\rangle $ is the WKB vacuum defined
at $t_{x}$ by the adiabatic prescription. The diagram of interest
is diagram (a) of Fig.~\ref{fig:Diagrams-determining-counter-terms},
and the divergent part of $\delta Z_{7}$ determined this way is linear
in the fermion mass.

In order to determine $\delta Z_{8}$, we repeat the consideration
analogous to Eq.~(\ref{eq:flatspacedelz4}) on a background field
$\sigma(x)=c$, where $c$ is an infinitesimal constant. Since a constant
$\sigma$ shift is equivalent to a shift of the fermion mass, we want
to choose $\delta Z_{8}$ to get 
\begin{eqnarray}
\lambda\partial_{m}n_{\psi}(x) & = & -i\lambda\int_{CTP}(dy)\sum_{N,M}\langle in|P\{\bar{\psi}_{M}(x)\psi_{N}(x)\bar{\psi}_{N}(y)\psi_{M}(y)\}|in\rangle_{conn}\nonumber \\
 &  & +\delta Z_{4}+\delta Z_{8}R(x),\label{eq:ren-cond-Z8}
\end{eqnarray}
where the subscript CTP denotes closed-time-path, and P is the path-ordering
operator for a ``in-in'' exception value. (For example, see Refs.
\cite{Weinberg:2005ww,Calzetta:2008tw}). Note that diagram of interest
corresponds to (e) of Fig.~\ref{fig:Diagrams-determining-counter-terms}.
As we will see later, this renormalization condition plays a crucial
role in determining the isocurvature correlator. The solution for
all the $\delta Z_{i}$ can be expressed in terms of Feynman parameter
integrals. However, such explicit expressions are not relevant to
determine the isocurvature correlation function. In contrast the left
hand side of Eq.~(\ref{eq:ren-cond-Z8}) is important.

To summarize, we have given a prescription to regularize and renormalize
the composite operator $\bar{\psi}\psi$. The renormalization conditions
ensure that $\langle in|(\bar{\psi}\psi)_{r,x}|in\rangle$ agrees
with that defined by the adiabatic prescription in curved spacetime,
and they also ensure that a constant shift in $\sigma$ is equivalent
to a constant shift in the fermion mass. Note that because the gravitational
production of fermions are still in flux when $m_{\psi}<H$ , we evaluate
the number density $n_{\psi}$ later than $t_{*}$, where $H(t_{*})\sim m_{\psi}$,
as far as the renormalization conditions are concerned.

\section{Scenario Constraints on Scalar Field $\sigma$ }

\label{sub:debyemass} In this section, we explain the constraints
on the Yukawa coupling $\lambda$ that comes from requiring $\sigma$
to behave as an unscreened long range force carrier whose on-shell
particle states do not significantly participate in $\psi$ production. 

We will find that $\langle\sigma\sigma\rangle|_{t_{*}}$ power spectrum
relevant for the isocurvature perturbations is not suppressed if $m_{\sigma}/H(t_{*})<1$
where $t_{*}$ is the time at which $H(t_{*})=m_{\psi}$ (i.e. $t_{*}$
is the time at which the fermion + anti-fermion number freezes \cite{Chung:2011ck}).
This implies $m_{\sigma}<m_{\psi}$ is the relevant parameter region
for the scenario of this paper. Furthermore, in order to prevent any
large isocurvature perturbations and relic abundance of $\sigma$,
we assume that the $\sigma$ particles decay before $\sigma$ becomes
an important fluid component of the evolution of the universe (e.g.
before matter-radiation equality). Note however that this restriction
is a matter of simplicity. There exist parameter regions in $\left(m_{\sigma},\lambda\right)$
such that $\sigma$ survives as a long-lived weakly interacting particle
(i.e. a dark matter). However, in such cases, the constraints from
the relic abundance and the isocurvature of $\sigma$ restrict the
$\sigma$ mass to be very small, e.g., $m_{\sigma}\lesssim10^{-6}\mbox{eV}$
for $H_{inf}\sim10^{13}\mbox{GeV}$. (See, e.g, \cite{Hertzberg:2008wr,Fox:2004kb,Kolb:1990vq,Sikivie:2006ni}
for the parametric bounds for the QCD axion produced by inflation.)
In principle, it is possible to build a model that has such small
$m_{\sigma}$ with help of some underlying symmetry, such as a shift
symmetry.

Although we assume that $m_{\sigma}<m_{\psi}$, $\sigma$ would generally
acquire a plasma mass correction through interactions with an ensemble
of $\psi$ particles. Thus we consider the effect of the produced
$\psi$ on the $\sigma$ correlator and show that the effect is negligible.
We expect the fermions do not affect scalar modes before horizon exit
because the mass correction by the fermion is still small compared
to the Hubble friction during inflation. After the scalar mode exits
the horizon, the fermions exert a tiny computable drag on $\sigma$.
The equation of motion of $\sigma$ from the action (\ref{eq:full_Lag})%
\footnote{The counter-terms appearing in the action includes
\[
S_{c.t.}\ni\int(dx)\left[-\frac{1}{2}\delta Z_{\sigma}\left(\partial\sigma\right)^{2}-\frac{1}{2}\delta m_{\sigma}^{2}\sigma^{2}+\delta Z_{0}\sigma+\delta Z_{R}R\sigma+\delta Z_{\xi}R\sigma^{2}\right].
\]
Note that the the linear $\sigma$ terms exist in the action because
the action does not preserve the $\mathrm{Z}_{2}$ symmetry due to
the Yukawa coupling.%
} is written as
\begin{eqnarray}
0 & = & \left\langle in\left|\left[(\square-m_{\sigma}^{2})\sigma_{x}-\lambda\bar{\psi}\psi_{x}+\delta Z_{0}+\delta Z_{R}R_{x}+\delta Z_{\sigma}\square\sigma_{x}-\delta m_{\sigma}^{2}\sigma_{x}+\delta Z_{\xi}R_{x}\sigma_{x}\right]\left[\cdots\right]\right|in\right\rangle \\
 & = & (\square_{x}-m_{\sigma}^{2})\left\langle \sigma_{x}\left[\cdots\right]\right\rangle +i\lambda^{2}\int^{x}(dz)\langle[\bar{\psi}\psi_{x},\bar{\psi}\psi_{z}]\rangle\left\langle \sigma_{z}\left[\cdots\right]\right\rangle +\left(\delta Z_{\sigma}\square_{x}-\delta m_{\sigma}^{2}+\delta Z_{\xi}R_{x}\right)\left\langle \sigma_{x}\left[\cdots\right]\right\rangle \nonumber \\
 &  & +\left(\delta Z_{0}+\delta Z_{R}R_{x}-\lambda\langle\bar{\psi}\psi_{x}\rangle\right)\left\langle \left[\cdots\right]\right\rangle +O(\lambda^{3},y),\label{eq:EOM_sigma}
\end{eqnarray}
where $\left[\cdots\right]$ denotes any quantum operators in the
correlation function. We choose the counter term $\delta Z_{0}$ and
$\delta Z_{R}$ such that the tadpole $\left\langle \sigma\right\rangle $
vanishes, i.e., $\left(\delta Z_{0}+\delta Z_{R}R-\lambda\left\langle \bar{\psi}\psi\right\rangle \right)=0$,
where the PV regulator is assumed. Moreover, when $\sigma$ varies
very slowly outside the horizon, we factor $\left\langle \sigma_{z}\left[\cdots\right]\right\rangle $
out of the integral in Eq.~(\ref{eq:EOM_sigma}), and we renormalize
the integral using the counter terms $\left(\delta Z_{\sigma}\square_{x}-\delta m_{\sigma}^{2}+\delta Z_{\xi}R_{x}\right)\left\langle \sigma_{x}\left[\cdots\right]\right\rangle $
such that the result is consistent with that obtained by the adiabatic
subtraction%
\footnote{In other words, we identify $-\delta m_{\sigma}^{2}$ and $\delta Z_{\xi}$
with $\delta Z_{4}$ and $\delta Z_{8}$ in Eq.~(\ref{eq:ren-cond-Z8}),
and $\delta Z_{\sigma}\square$ is neglected since $\sigma$ is slowly
varying.%
}:
\begin{equation}
i\lambda^{2}\int^{x}(dz)\langle[\bar{\psi}\psi_{x},\bar{\psi}\psi_{z}]\rangle+\left(-\delta m_{\sigma}^{2}+\delta Z_{\xi}R_{x}\right)=-\lambda^{2}\left(\frac{\partial n_{\psi}}{\partial m_{\psi}}\right),
\end{equation}
where $n_{\psi}$ is the renormalized fermion number density defined
by Eq.~(\ref{eq:def_Npsi}), and we have used Eq.~(\ref{eq:ren-cond-Z8})
in the derivation. Therefore, we find the effective mass of $\sigma$
when it slowly varies (i.e., $k/a\ll H$ and $m_{\sigma}\ll H$)
\begin{equation}
m_{\sigma}^{eff}=m_{\sigma}^{2}+\Delta m_{\sigma}^{2}(t)\approx m_{\sigma}^{2}+\lambda^{2}\frac{\partial n_{\psi}(t)}{\partial m_{\psi}}.\label{eq:masscorr}
\end{equation}
Because we estimate $n_{\psi}\lesssim O(0.1)\left(m_{\psi}H\right)^{3/2}$
when $m_{\psi}\lesssim H$,%
\footnote{Note that the adiabatic prescription to determine the number density
$n_{\psi}$ does not apply for modes $m_{\psi}<k/a<\sqrt{m_{\psi}H}$
when $m_{\psi}<H$ because vacuum varies non-adiabatically, i.e.,
the adiabaticity parameter $\epsilon_{k}\equiv\frac{m_{\psi}k_{p}H}{\omega_{k}}\gtrsim1$,
where $k_{p}=k/a$ and $\omega_{k}=\sqrt{k_{p}^{2}+m_{\psi}^{2}}$.
See Appendix \ref{sec:prod_review} for detail. However, we can estimate
the upper bound of the number density as 
\[
n_{\psi}(t)=\int\frac{d^{3}k_{p}}{(2\pi)^{3}}\left|\beta_{k}\right|^{2}\lesssim\int^{\sqrt{m_{\psi}H}}\frac{d^{3}k_{p}}{(2\pi)^{3}}\frac{1}{2}\sim O(0.1)\left(m_{\psi}H\right)^{3/2}\quad\mbox{for }t<t_{*}.
\]
} based on dimensional analysis, we expect that the mass correction
by the $\psi$ loop is 
\begin{equation}
\Delta m_{\sigma}^{2}(t)\approx\lambda^{2}\frac{\partial n_{\psi}(t)}{\partial m_{\psi}}\sim\begin{cases}
O(0.1\mbox{ or less})\lambda^{2}m_{\psi}^{1/2}H^{3/2} & \mbox{for }m_{\psi}<H(t)\\
O(0.1)\lambda^{2}m_{\psi}^{2} & \mbox{for }m_{\psi}>H(t)
\end{cases}.
\end{equation}
Therefore, in general, before the fermion production ends $m_{\psi}<H$,
this scalar mass correction $\Delta m_{\sigma}^{2}$ does not ruin
the stability of our scenario $m_{\sigma}^{2}+\Delta m_{\sigma}^{2}(t)<m_{\psi}^{2}<H^{2}(t)$
as long as $m_{\sigma}^{2}<m_{\psi}^{2}$. 

Next, we ask the question of which parametric region would be consistent
with the simplifying assumption that $\psi$ particles are primarily
produced gravitationally and not by $\sigma$. To this end, we first
consider the annihilation $\sigma\sigma\to\bar{\psi}\psi$. The annihilation
is the most significant at the end of inflation because $\psi$ particles
produced from $\sigma$ before the end of inflation are diluted, and
$\sigma\sigma\to\bar{\psi}\psi$ after the end of inflation is also
limited because the allowed kinematic phase space is redshifted. Thus
we compare the number density of the produced $\psi$ from $\sigma$
at the end of inflation, $n_{\sigma\to\psi}$ with that of gravitationally
produced $\psi$, $n_{\psi}(t_{*})\sim m_{\psi}^{3}$, and we require
their ratio to be small:
\begin{eqnarray}
\left(\frac{a_{e}}{a(t_{*})}\right)^{3}\frac{n_{\sigma\to\psi}(t_{e})}{n_{\psi}(t_{*})} & \sim & \left(\frac{a_{e}}{a(t_{*})}\right)^{3}\frac{\left.n_{\sigma}\Gamma(\mbox{\ensuremath{\sigma\sigma\to\psi\psi}})\Delta t\right|_{t_{e}}}{n_{\psi}(t_{*})}\label{eq:intermedstep}\\
 & \sim & \left(\frac{H(t_{*})}{H_{e}}\right)^{2}\frac{H_{e}^{3}\cdot\frac{\lambda^{4}}{16\pi^{2}}H_{e}\cdot\frac{1}{H_{e}}}{H^{3}(t_{*})}\sim\frac{\lambda^{4}}{16\pi^{2}}\frac{H_{e}}{m_{\psi}}\lesssim1,\label{eq:condition_number_density}
\end{eqnarray}
where the subscript $e$ means a variable is evaluated at the end
of inflation $t_{e}$.

Even though $m_{\sigma}<m_{\psi},$ the decay production of $\psi$
through $\sigma\rightarrow\bar{\psi}\psi$ may still be possible if
$\sigma$ is sufficiently off shell due to its interactions with finite
density of $\psi$ in the subhorizon region (the subhorizon physics
here is different from the superhorizon physics considered in Eq.~(\ref{eq:masscorr})).
To turn off this channel, we require that the $\sigma$ mass corrections
from the fermion number density at the time of end of inflation be
small. This requires
\begin{equation}
\lambda^{\kappa}\frac{H_{e}/(2\pi)}{m_{\psi}}\lesssim1\label{eq:gencond}
\end{equation}
where $\kappa\gtrsim O(1).$ To see how $\kappa\gtrsim O(1)$ can
come about, consider the following estimate of subhorizon thermal
effect. The maximum effective number density of fermions at the end
of inflation is
\begin{equation}
n_{\psi}(t_{e})\lesssim4m_{\psi}\left(\frac{H_{e}}{2\pi}\right)^{2}.
\end{equation}
The energy density associated with these fermions is
\begin{equation}
\Delta V\sim n_{\psi}(t_{e})\sqrt{\left(\frac{H_{e}}{2\pi}\right)^{2}+\lambda^{2}\sigma^{2}}
\end{equation}
where we neglected $m_{\sigma}\ll H_{e}/(2\pi)$. This leads to an
effective $m_{\sigma}$ correction of 
\begin{equation}
\Delta m_{\sigma}^{2}\sim n_{\psi}(t_{e})\frac{\lambda^{2}}{H_{e}/(2\pi)}\lesssim4\lambda^{2}m_{\psi}\left(\frac{H_{e}}{2\pi}\right).
\end{equation}
Kinematically blocking the $\sigma$ decay into $\psi$, we find
\begin{equation}
4\lambda^{2}\left(\frac{H_{e}}{2\pi}\right)<m_{\psi}\label{eq:scalar_constraint2}
\end{equation}
which corresponds to $\kappa=2$. Note that this condition is more
restrictive than Eq.~(\ref{eq:condition_number_density}).

In sum, requiring $\sigma$ to behave as an unscreened long range
force carrier whose on-shell particle states do not significantly
participate in $\psi$ production gives a constraint on $\lambda$.
The strongest condition is given by Eq.~(\ref{eq:gencond}) with
$\kappa\gtrsim O(1)$.

\section{Isocurvature two-point function}

\label{sec:2-pt} 
\begin{figure}
\begin{centering}
\includegraphics[width=0.5\linewidth]{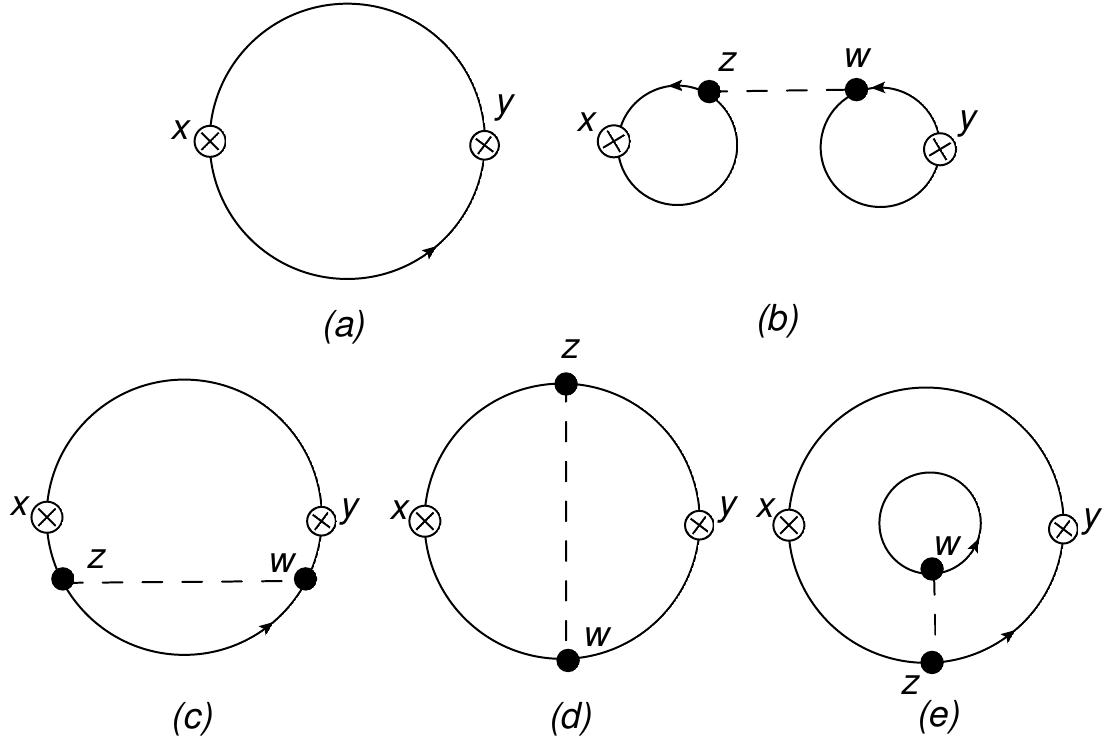} 
\par\end{centering}

\caption{\label{fig:LO-NLO-FeynDiag}The leading order and the next leading
order contribution to $\left\langle \bar{\psi}\psi_{x}\bar{\psi}\psi_{y}\right\rangle $
, the cross-dot vertices corresponds to $\bar{\psi}\psi$ insertion.
By comparing the large $r$ ($r\equiv|\vec{x}-\vec{y}|$) behavior
of the equal-time correlator of the fermion and the scalar field,
we want to show that diagram (b) dominates in the limit $r\rightarrow\infty$. }
\end{figure}

In this section, we evaluate the two-point function
of the renormalized isocurvature operator $\delta_{S}$, given by
Eq.~(\ref{eq:def_iso}). The average number density was computed
in \cite{Chung:2011ck}, the result is summarized in Appendix \ref{sec:prod_review}.
We only need to evaluate $\langle(\bar{\psi}\psi)_{x,r}(\bar{\psi}\psi)_{y,r}\rangle_{c}$.
Since we want to use the quantum computation to set the initial condition
for the subsequent classical fluid evolution, we will choose the time
of the evaluation such that both the quantum and the classical fluid
descriptions apply. We take $x^{0}=y^{0}=\eta_{f}$ at time after
the particle production ends, since the fluid description cannot describe
the particle production process. We will take the separation $|\vec{x}-\vec{y}|$
to be large enough such that the intersection of their past light-cone
$I^{-}(x)\cap I^{-}(y)$ lives deep within the inflationary era. This
ensures that the contributions from late-time short distance physics
(e.g. reheating, phase transition) are minimized. The relevant diagrams
for $\langle(\bar{\psi}\psi)_{x,r}(\bar{\psi}\psi)_{y,r}\rangle_{c}$
are given in Fig. (\ref{fig:LO-NLO-FeynDiag}). The crossed dot represent
$(\bar{\psi}\psi)_{x,r}$ insertion, the solid dot represent the Yukawa
interaction vertex, the dashed line represent the scalar $\sigma$
propagator, and the solid line represent the fermion propagator.

\subsection{Leading order result}

\label{sub:LO} We first consider the leading order diagram (a) in
Fig (\ref{fig:LO-NLO-FeynDiag}). The diagram is explicitly written
as 
\begin{eqnarray}
\langle\bar{\psi}\psi_{x}\bar{\psi}\psi_{y}\rangle_{(a)} & = & -\mbox{Tr}\left[\langle\psi_{x}\bar{\psi}_{y}\rangle\langle\psi_{y}\bar{\psi}_{x}\rangle\right]=\sum_{i,j}\bar{V}_{i,x}U_{j,x}\bar{U}_{j,y}V_{i,y}
\end{eqnarray}
 Using a contour integration technique, we can evaluate the mode-sum
analytically. The details are in given Appendix \ref{sec:large_r_fermion_correlator}.
The result%
\footnote{Note that we do not consider the the heavy mass case, $m_{\psi}\gg H_{inf}$
where $H_{inf}$ is the expansion rate during inflation, for the isocurvature
because the estimation of the particle production depends on how the
inflation ends as described in Section \ref{sec:Fermion-Isocurvature-Model}.
However, we provide the leading order of the two-point function to
develop better intuition for the behavior of super horizon modes of
$\psi$.%
} is 
\begin{eqnarray}
\langle\bar{\psi}\psi_{x}\bar{\psi}\psi_{y}\rangle_{LO} & = & \begin{cases}
\frac{1}{\pi^{4}a_{x}^{6}|\vec{x}-\vec{y}|^{6}}\left(1+O\left[\left(\frac{m_{\psi}}{H_{inf}}\right)^{2}\right]\right) & (m_{\psi}\ll H_{inf})\\
\frac{1}{\pi^{4}a_{x}^{6}|\vec{x}-\vec{y}|^{6}}(4\pi)\left(\frac{m_{\psi}}{H_{inf}}\right)^{3}\exp(-2\pi\frac{m_{\psi}}{H_{inf}}) & (m_{\psi}\gg H_{inf})
\end{cases}\label{eq:nxny_LO}
\end{eqnarray}
where $H_{inf}$ is the expansion rate during inflation. We can understand
this result by backtracking the two points $x,y$ to the time when
they were deep inside the horizon, and see what happened as they grow
apart.

In the heavy mass case $(m_{\psi}\gg H_{inf})$, the Compton radius
$m_{\psi}^{-1}$ is smaller than the horizon radius $H_{inf}^{-1}$.
The physical separation $r_{\mbox{phys}}$ will first grow to the
Compton wavelength, and trigger the exponential suppression factor
$\exp(-2m_{\psi}r_{phys})$ in the correlator. 
\begin{eqnarray}
\langle\bar{\psi}\psi_{x}\bar{\psi}\psi_{y}\rangle_{flat,m_{\psi}r_{phys}>1} & \sim & \frac{m_{\psi}^{3}}{4\pi^{3}r_{phys}^{3}}\exp(-2m_{\psi}r_{phys})
\end{eqnarray}
 As the physical separation $r_{\mbox{phys}}$ grows further to exceed
the horizon radius $H_{inf}^{-1}$, the correlator would freeze and
start decreasing as $(a_{r}/a_{\eta})^{6}$, where $a_{r}=1/(H_{inf}r)$
denote the scale factor at the horizon crossing. Substituting $a_{r}=\frac{1}{H_{inf}r}$
and $r_{phys}=H_{inf}^{-1}$, we recover the heavy mass formula: 
\begin{equation}
\left(\frac{a_{r}}{a_{\eta}}\right)^{6}\frac{m_{\psi}^{3}}{4\pi^{3}r_{phys}^{3}}\exp(-2m_{\psi}r_{phys})\sim\frac{1}{a_{x}^{6}r^{6}}\left(\frac{m_{\psi}}{H_{inf}}\right)^{3}\exp(-2\frac{m_{\psi}}{H_{inf}}).
\end{equation}

In the light mass case $(m_{\psi}\ll H_{inf})$, the physical distance
will cross the horizon radius first, without the exponential suppression
of $\exp(-2m_{\psi}r_{phys})$. From the flat space UV limit result
$\frac{1}{r_{phys}^{6}}$, 
\begin{equation}
\langle\bar{\psi}\psi_{x}\bar{\psi}\psi_{y}\rangle_{flat,mr_{phys}<1}\sim\frac{1}{r_{phys}^{6}}
\end{equation}
 we use $a_{r}=\frac{1}{H_{inf}r}$ and $r_{phys}=H_{inf}^{-1}$ to
obtain 
\begin{equation}
\left(\frac{a_{r}}{a_{\eta}}\right)^{6}\frac{1}{r_{phys}^{6}}\sim\frac{1}{a_{x}^{6}r^{6}}
\end{equation}
 Thus we recover the light mass result.

Unfortunately, the fractional relic density fluctuation at CMB scale%
\footnote{Since $\langle\delta\rho\delta\rho\rangle$ is frozen as long as the
two points are outside of horizon, we can extrapolate this large spatial
separation result obtained at the end of inflation to the recombination
time. %
} is too small 
\begin{eqnarray}
\frac{\langle\delta\rho_{x}\delta\rho_{y}\rangle}{\langle\bar{\rho}_{\psi}\rangle^{2}} & \sim & \frac{m_{\psi}^{2}/(\pi^{4}a^{6}r_{CMB}^{6})}{m_{\psi}^{2}m_{\psi}^{6}(a_{*}^{6}/a^{6})}\sim\left(\frac{1}{a_{*}m_{\psi}r_{CMB}}\right)^{6}.
\end{eqnarray}
 where $r_{CMB}$ is the comoving distance for typical CMB observation
scale and the subscript $*$ denotes the time when fermion production
ends. Let $a_{CMB}$ denotes the scale factor when CMB scale exits
the horizon then we have 
\begin{equation}
r_{CMB}^{-1}\sim a_{CMB}H_{inf}
\end{equation}
 Assuming the fermion production ends during reheating when $m_{\psi}=H(t_{*})$,
and $H\propto a^{-\alpha}$ during reheating, then we have 
\begin{equation}
\frac{a_{e}H_{inf}}{a_{*}m_{\psi}}\sim\frac{a_{e}H_{e}}{a_{*}H_{*}}\sim\left(\frac{a_{e}}{a_{*}}\right)^{1-\alpha}\sim\left(\frac{H_{e}}{H_{*}}\right)^{1-\frac{1}{\alpha}}
\end{equation}
 Assuming that inflation ends $60$ efolds after the CMB scale exits
horizon and a MD-like reheating, i.e., $\alpha=3/2$, then we have
\begin{equation}
\frac{\langle\delta\rho_{x}\delta\rho_{y}\rangle}{\langle\bar{\rho}_{\psi}\rangle^{2}}\sim\left(\frac{a_{CMB}H_{inf}}{a_{*}m_{\psi}}\right)^{6}\sim\left(\frac{a_{CMB}}{a_{e}}\frac{a_{e}H_{inf}}{a_{*}m_{\psi}}\right)^{6}\sim e^{-300}\left(\frac{H_{e}}{m_{\psi}}\right)^{2}
\end{equation}
 Using the fermion relic abundance formula (for $T_{RH}=10^{9}\mbox{GeV}$
and $g_{*}=100$ case) $\omega_{\psi}\sim(m_{\psi}/10^{10}\mbox{GeV})^{2}$,
we obtain 
\begin{equation}
\frac{\langle\delta\rho_{x}\delta\rho_{y}\rangle}{\rho_{tot}^{2}}\sim\omega_{\psi}^{2}\frac{\langle\delta\rho_{x}\delta\rho_{y}\rangle}{\langle\bar{\rho}_{\psi}\rangle^{2}}\sim e^{-300}\left(\frac{H_{e}}{10^{10}GeV}\right)^{2}
\end{equation}
 We thus find that generically the pure fermion isocurvature is very
small on scales relevant for the CMB.

\subsection{Next leading order result}

\label{sub:NLO}We consider the diagrams (b)-(e) in Fig.~\ref{fig:LO-NLO-FeynDiag},
which contain the effects of the Yukawa interaction to the fermion
production. We can perturbatively compute the diagrams using the ``in-in''
formalism (e.g.~see Refs. \cite{Weinberg:2005vy,CalzettaHu:2008}
and references therein).

Firstly, we estimate which diagram gives the largest contribution
when $x$ and $y$ have large spatial separations. From the fact that
equal-time correlator $\langle\sigma_{x}\sigma_{y}\rangle$ scales
as $r^{2\nu-3}$ where $\nu^{2}=9/4-m_{\sigma}^{2}/H^{2}$ from Eq.~(\ref{eq:light-scalar})
and $\langle\psi_{x}\bar{\psi}_{y}\rangle$ scales as $r^{-3}$, we
expect that diagrams that have fewer fermion lines stretched between
$x$ and $y$ decreases slower as $r\rightarrow\infty$. Thus, we
conclude diagram (b) gives the dominant contribution to the two-point
function.

For diagram (b), we expand it using commutators 
\begin{eqnarray}
I_{b}(x,y) & = & \langle(\bar{\psi}\psi)_{x,r}(\bar{\psi}\psi)_{y,r}\rangle_{c,diag(b)}\\
 & = & 4(i\lambda)^{2}\int^{x}(dz)\int^{y}(dw)\langle\bar{\psi}\psi_{[x}\bar{\psi}\psi_{z]}\rangle\langle\bar{\psi}\psi_{[y}\bar{\psi}\psi_{w]}\rangle\langle\sigma_{\{z}\sigma_{w\}}\rangle\nonumber \\
 &  & +4(i\lambda)^{2}\int^{x}(dz)\int^{y}(dw)\langle\bar{\psi}\psi_{\{x}\bar{\psi}\psi_{z\}}\rangle\langle\bar{\psi}\psi_{[y}\bar{\psi}\psi_{w]}\rangle\langle\sigma_{[w}\sigma_{z]}\rangle\Theta(w^{0}-z^{0})\nonumber \\
 &  & +4(i\lambda)^{2}\int^{x}(dz)\int^{y}(dw)\langle\bar{\psi}\psi_{[x}\bar{\psi}\psi_{z]}\rangle\langle\bar{\psi}\psi_{\{y}\bar{\psi}\psi_{w\}}\rangle\langle\sigma_{[z}\sigma_{w]}\rangle\Theta(z^{0}-w^{0})\\
 & \approx & (i\lambda)^{2}\int^{x}(dz)\int^{y}(dw)\langle[\bar{\psi}\psi_{x},\bar{\psi}\psi_{z}]\rangle\langle[\bar{\psi}\psi_{y},\bar{\psi}\psi_{w}]\rangle\langle\sigma_{\{z}\sigma_{w\}}\rangle
\end{eqnarray}
where $(dz)=\sqrt{-\mbox{det}\left(g_{\mu\nu}\right)}d^{4}z$, $[\cdots]$
means anti-symmetrization and $\{\cdots\}$ means symmetrization,
and we have implicitly assumed the PV regulator. From the scalar and
fermion mode functions in de Sitter spacetime, we know $\langle[\sigma_{x_{1}},\sigma_{x_{2}}]\rangle$
is suppressed by $a^{-2\nu}$ relative to $\langle\{\sigma_{x_{1}},\sigma_{x_{2}}\}\rangle$,
whereas $\langle[\bar{\psi}\psi_{x_{1}},\bar{\psi}\psi_{x_{2}}]\rangle$
is suppressed by $a^{-1}$ relative to $\langle\{\bar{\psi}\psi_{x_{1}},\bar{\psi}\psi_{x_{2}}\}\rangle$.
The last line is obtained by keeping only the dominant contribution.

Since the fermion particle production ends at $t_{*}$ and the previously
produced particles have been diluted away, we expect the $z$ and
$w$ integrals to peak around the time $t_{*}$. For late time and
large spatial separations, the scalar correlator $\langle\sigma_{\{z}\sigma_{w\}}\rangle$
is slowly varying with respect to changes in $z$ and $w$. Thus we
may approximately take $\langle\sigma_{\{z}\sigma_{w\}}\rangle=\langle\sigma_{\{z_{0}}\sigma_{w_{0}\}}\rangle$,
where $z_{0}=(t_{*},\vec{x})$ and $w_{0}=(t_{*},\vec{y})$, and factor
it outside of the $z,w$ integral: 
\begin{equation}
I_{b}(x,y)\approx(i\lambda)^{2}\langle\sigma_{\{z_{0}}\sigma_{w_{0}\}}\rangle[\int^{x}(dz)\langle[\bar{\psi}\psi_{x},\bar{\psi}\psi_{z}]\rangle][\int^{y}(dw)\langle[\bar{\psi}\psi_{y},\bar{\psi}\psi_{w}]\rangle]\label{eq:Ib}
\end{equation}

The remaining fermion integral $\int^{x}(dz)\langle[\bar{\psi}\psi_{x},\bar{\psi}\psi_{z}]\rangle$
is quadratically divergent. The counter-terms $\delta Z_{4}\sigma+\delta Z_{8}R\sigma$
in $(\bar{\psi}\psi)_{r}$ is in place to cancel such divergences.
Furthermore, our choice of the renormalization conditions given in
Section \ref{sub:Regularization-and-Renormalizati} ensures that a
constant shift in $\sigma$ is equivalent to a shift of the fermion
mass (see Eq.~(\ref{eq:ren-cond-Z8})). An explicit computation of
the fermion loop integral using the adiabatic subtraction is given
in Appendix \ref{sec:AppD:Relative-suppression}. Thus we have 
\begin{equation}
\langle(\delta_{S})_{r,x}(\delta_{S})_{r,y}\rangle_{NLO}\approx\omega_{\psi}^{2}\lambda^{2}[\partial_{m}\ln n_{\psi}|_{x}][\partial_{m}\ln n_{\psi}|_{y}]\langle\sigma_{\{(\vec{x},t_{*})}\sigma_{(\vec{y},t_{*})\}}\rangle\label{eq:2pt-NLO}
\end{equation}
where $t_{*}$ is the time when fermion production ends (i.e.~$m_{\psi}\sim H(t_{*})$)
and $\partial_{m}$ denotes the derivative with respect to $m_{\psi}$.
Note that $\langle(\delta_{S})_{r,x}(\delta_{S})_{r,y}\rangle_{NLO}$
freezes for $t>t_{*}$ since $\partial_{m}n_{\psi}$ and $n_{\psi}$
behaves as $a^{-3}$ after the fermion production ends. We will discuss
the numerical implications of this result below.

To summarize, we computed the isocurvature correlation function to
the next leading order, as in Eq.~(\ref{eq:2pt-NLO}). Intuitively,
the light scalar's quantum fluctuation modulate the fermion's mass,
which affect the fermion relic abundance. In the same line of thought,
we may extrapolate this result to estimate higher order corrections
\begin{equation}
\langle(\delta_{S})_{r,x}(\delta_{S})_{r,y}\rangle_{full}\approx\omega_{\psi}^{2}\frac{\langle n_{\psi}\left(m_{\psi}+\lambda\sigma(\vec{x},t_{*})\right)n_{\psi}\left(m_{\psi}+\lambda\sigma(\vec{y},t_{*})\right)\rangle_{\sigma}}{n_{\psi}^{2}}\label{eq:2pt-Full}
\end{equation}
where we have treated $n_{\psi}$ to be a function of its mass and
the expectation value is taken with respect of the $\sigma$ field.

\subsection{Isocurvature Power Spectrum}

\label{sub:Powerspectrum}

In the long wavelength limit, which corresponds to the low multipoles
in the angular CMB anisotropy, the temperature fluctuations dominantly
come from the Sach-Wolfe term \cite{Langlois:1999dw}, which is expressed
as

\begin{equation}
\frac{\Delta T}{T}=-\frac{1}{5}\zeta-\frac{2}{5}\delta_{S}.\label{eq:temp_fluc}
\end{equation}
 Then the power spectrum of the temperature fluctuations 
\begin{eqnarray}
\Delta_{\frac{\Delta T}{T}}^{2}(k) & \equiv & \frac{k^{3}}{2\pi^{2}}\int d^{3}x\left\langle \frac{\Delta T}{T}(t,\vec{x})\frac{\Delta T}{T}(t,\vec{0})\right\rangle e^{-i\vec{k}\cdot\vec{x}}=\frac{1}{25}\Delta_{\zeta}^{2}(k)+\frac{4}{25}\Delta_{\delta_{S}}^{2}(k),\\
\Delta_{\zeta}^{2}(k) & \equiv & \frac{k^{3}}{2\pi^{2}}\int d^{3}x\left\langle \zeta(t,\vec{x})\zeta(t,\vec{0})\right\rangle e^{-i\vec{k}\cdot\vec{x}},\\
\Delta_{\delta_{S}}^{2}(k) & \equiv & \frac{k^{3}}{2\pi^{2}}\int d^{3}x\left\langle \delta_{S}(t,\vec{x})\delta_{S}(t,\vec{0})\right\rangle e^{-i\vec{k}\cdot\vec{x}},
\end{eqnarray}
where the cross-correlation contribution $\left\langle \zeta\delta_{S}\right\rangle $
has been neglected because of the reason explained in Section \ref{sub:gauge-int}.
When the leading term approximation \eqref{eq:2pt-NLO} is valid,
Eq.~(\ref{eq:2pt-NLO}) yields the isocurvature power spectrum 
\begin{equation}
\Delta_{\delta_{S}}^{2}(t,k)=\omega_{\psi}^{2}(t)\lambda^{2}\left(\frac{\partial_{m}n_{\psi}(m_{\psi})}{n_{\psi}}\right)^{2}\Delta_{\sigma}^{2}(t_{*},k)+O(\lambda^{4}),\label{eq:Ps}
\end{equation}
which includes the extra factor $\omega_{\psi}^{2}$ due to the thermal
relics. Furthermore, when the mass of scalar field $\sigma$ is sufficiently
light such that $\sigma$ does not start its coherent oscillation
until the fermion particle production ends, i.e., $m_{\sigma}<H(t_{*})<H_{inf}$,
the power spectrum for $\sigma$ is

\begin{equation}
\Delta_{\sigma}^{2}(t_{*},k)\approx\frac{H^{2}(t_{k})}{4\pi^{2}}
\end{equation}
where $t_{k}$ is the time when the scale $k$ exits horizon. Note
that we have already shown that the correction of $m_{\sigma}$ due
to the fermion loop is negligible in Section \ref{sub:debyemass}.
Therefore, the isocurvature power spectrum becomes 
\begin{eqnarray}
\Delta_{\delta_{S}}^{2}(k) & \approx & \omega_{\psi}^{2}\lambda^{2}\left(\frac{\partial_{m}n_{\psi}(m_{\psi})}{n_{\psi}}\right)^{2}\frac{H^{2}(t_{k})}{4\pi^{2}}.\label{eq:iso_pow}
\end{eqnarray}
The currently known parametric bounds for this isocurvature power
spectrum is presented in Section \ref{sub:Parameter-bounds}.

\section{Result and Discussion}

\label{sec:result}

\subsection{Parameter bounds\label{sub:Parameter-bounds}}

\label{sub:bounds} In this subsection, we present the allowed parameter
region in the fermion isocurvature model from the observational constraints
using the dark matter relic abundance and the CDM isocurvature power-spectrum.
In this scenario, there are 5 independent parameters: $m_{\psi}$,
$H_{inf}$, $\lambda$, $T_{RH}$ and $m_{\sigma}$, where $H_{inf}$
is the Hubble scale during inflation and $T_{RH}$ is the reheating
temperature. We assume $H_{inf}$ and $T_{RH}$ are free parameters
governed entirely by the inflaton and the reheating sector. As discussed
in Section \ref{sec:Fermion-Isocurvature-Model}, as long as $m_{\sigma}\ll m_{\psi}$,
the exact value of the scalar mass $m_{\sigma}$ is numerically unimportant
in this model. Therefore, we are basically left with two parameter,
namely $\lambda$ and $m_{\psi}$.%
\footnote{Note that we implicitly assume that if $m_{\psi}$ and $T_{RH}$ are
such that the dark matter relic abundance is not saturated by the
$\psi$ energy density, the other CDM sector in Eq.~(\ref{eq:full_Lag})
is adjusted to provide the rest of the dark matter. Note that when
the $\psi$ dark matter abundance is small, no large tuning is needed
to make this occur since the well known WIMP miracle can saturate
the dark matter abundance.%
}

For the light fermion, $m_{\psi}<H_{inf}$, the fermion particle number
freezes when $H(t_{*})\sim m_{\psi}$ as reviewed in Appendix \ref{sec:prod_review}.
In particular, the Yukawa coupling works effectively as a mass shift
in our scenario $m_{eff}=\left|m_{\psi}+\lambda\sigma(t_{*})\right|$.
The fermion relic abundance \eqref{eq:relic-abundance} becomes 
\begin{eqnarray}
\Omega_{\psi}h^{2} & \sim & 3r\left(\frac{m_{eff}}{10^{11}\mbox{GeV}}\right)^{2}\left(\frac{T_{RH}}{10^{9}\mbox{GeV}}\right),\label{eq:Omega_psi}
\end{eqnarray}
where the extra factor $r$ comes from the difference in the effective
masses at $t_{*}$ and later time, at which the energy density of
$\psi$ is not negligible, such as the MD era. For example, if $\sigma$
is treated as a Gaussian random variable with $\sqrt{\left\langle \sigma^{2}\right\rangle }\sim H_{inf}/2\pi$,
we can approximate $r\approx m_{\psi}/\left\langle m_{eff}\right\rangle $
and write
\begin{equation}
\Omega_{\psi}h^{2}\sim\begin{cases}
\left(\frac{m_{\psi}}{10^{11}\mbox{GeV}}\right)^{2}\left(\frac{T_{RH}}{10^{9}\mbox{GeV}}\right) & \;\mbox{if }\, m_{\psi}>\lambda H_{\inf}/2\pi\\
\frac{2\pi m_{\psi}}{\lambda H_{inf}}\left(\frac{\lambda H_{inf}}{10^{11}\mbox{GeV}}\right)^{2}\left(\frac{T_{RH}}{10^{9}\mbox{GeV}}\right) & \;\mbox{if }\, m_{\psi}<\lambda H_{\inf}/2\pi
\end{cases},\label{eq:Omega_Psi}
\end{equation}
where $O(1)$ factors are neglected.

\begin{figure}
\begin{centering}
\includegraphics[width=0.6\linewidth]{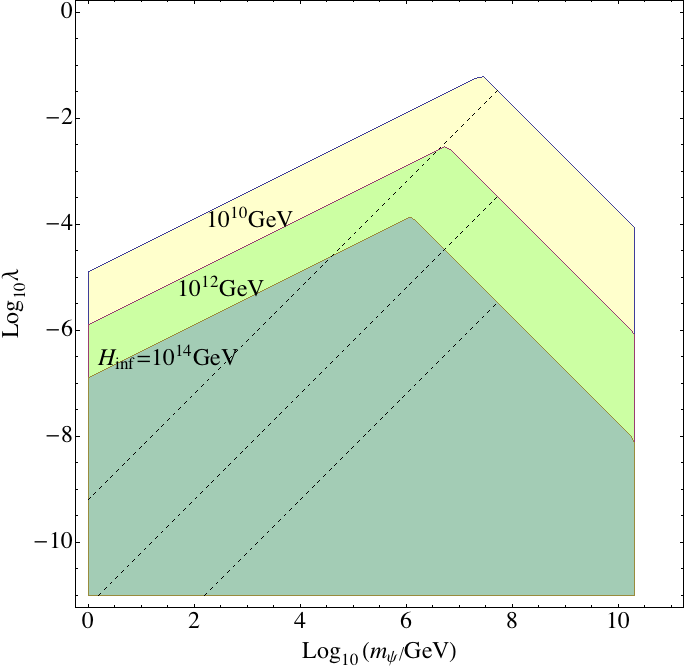} 
\par\end{centering}

\caption{\label{fig:bound}Bounds on the fermion mass and Yukawa coupling for
various inflationary Hubble scales. The vertical bound corresponds
to the total dark matter relic density constraint, the right diagonal
and the left diagonal bounds correspond to the constraints from the
CDM isocurvature and the scalar annihilation using Eq.~(\ref{eq:scalar_constraint2}),
respectively. The splitting dashed lines in each region separates
the small mass and large mass correction regime. In this plot, we
set $T_{RH}=10^{9}GeV.$ }
\end{figure}

Furthermore, from the result (\ref{eq:iso_pow}) in Sec. \ref{sub:Powerspectrum},
the fractional isocurvature amplitude \cite{Bean:2006io} becomes
\begin{equation}
\alpha_{S}\equiv\frac{\Delta_{\delta_{S}}^{2}}{\Delta_{\zeta}^{2}+\Delta_{\delta_{S}}^{2}}\sim\frac{\lambda^{2}}{2}\left(\frac{m_{\psi}}{10^{4}\mbox{GeV}}\right)^{2}\left(\frac{H}{10^{13}\mbox{GeV}}\right)^{2}\left(\frac{T_{RH}}{10^{9}\mbox{GeV}}\right)^{2},\label{eq:Alpha_psi}
\end{equation}
where we have used 
\begin{equation}
\frac{\partial_{m}n_{\psi}}{n_{\psi}}\sim\begin{cases}
m_{\psi}^{-1} & \mbox{for }m_{\psi}>\lambda H_{inf}/2\pi\\
2\pi\lambda^{-1}H_{inf}^{-1} & \mbox{for }m_{\psi}<\lambda H_{inf}/2\pi
\end{cases},
\end{equation}
because the number density $n_{\psi}$ at the time $t_{*}$ is determined
by only one dimensionful scale $m_{eff}\sim H(t_{*})$. The current
observational bound \cite{Ade:2013lta,Ade:2013rta,Komatsu:2008ex,Komatsu:2010in,Larson:2010gs,Bean:2006io}
of the isocurvature for the uncorrelated case, i.e. $\left\langle \zeta\delta_{S}\right\rangle =0$,
is $\alpha_{S}<0.016\,(95\%\mbox{ CL})$ from the Planck+WP9 combined
data, which yields the constraints on the parameters $\lambda$ and
$m_{\psi}$. Combining the above consideration, we have the parameter
plot shown in Fig. \ref{fig:bound}.

The case that $m_{\psi}<\lambda H_{inf}/(2\pi)$ (which we will refer
to as large mass correction regime) is potentially the most interesting
case because the fermion number density $n_{\psi}$ depends on $|m_{\psi}+\lambda\sigma|$,
not $m_{\psi}+\lambda\sigma$ as the sign of the fermion mass is irrelevant
for particle production%
\footnote{The sign of the fermion mass changes under a chiral transformation.%
}. This may lead to interesting features such as large non-Gaussianities
when the effective mass varies from negative to positive depending
on the local Hubble patches at $t_{*}$. However, this parametric
region has couple of problems: 1) the perturbative calculation of
$n_{\psi}$ may be unsuitable since we are not resuming the large
mass corrections; 2) Eq.~(\ref{eq:gencond}) may not be satisfied.
Hence, for the rest of this section, we primarily focus on the case
that $m_{\psi}>\lambda H_{inf}/(2\pi)$, which we will refer to as
the small mass correction regime.

\subsection{Non-Gaussianities}

\label{sub:NG} 
\begin{figure}
\centering{}\includegraphics[width=0.5\linewidth]{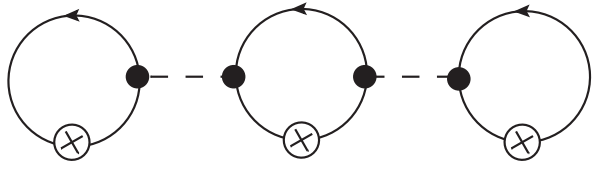}\caption{\label{fig:3pt_function}The two leading order diagrams to 3-point
function $\langle\delta_{S}\delta_{S}\delta_{S}\rangle$ . The cross-dot
vertices corresponds to $\bar{\psi}\psi/n_{\psi}$ insertion.}
\end{figure}

In this subsection, we compute the bi-spectrum $B_{S}(\vec{p}_{1},\vec{p}_{2},\vec{p}_{3})$
defined by 
\begin{equation}
\left(2\pi\right)^{3}\delta^{(3)}(\sum_{i}\vec{p}_{i})B_{S}(\vec{p}_{1},\vec{p}_{2},\vec{p}_{3})=\int d^{3}x_{1}d^{3}x_{2}d^{3}x_{3}e^{-i\sum_{i}\vec{p}_{i}\cdot\vec{x_{i}}}\left\langle \delta_{S}(\vec{x}_{1})\delta_{S}(\vec{x}_{2})\delta_{S}(\vec{x}_{3})\right\rangle .
\end{equation}
 The fermion density fluctuation is intrinsically non-Gaussian since
$n_{\psi}$ is the non-linear function of $\sigma$, which is treated
as a Gaussian random variable. When the effective mass fluctuation
due to $\lambda\sigma$ is small, we can Taylor-expand the number
density with respect to $\lambda\sigma$,

\begin{equation}
n_{\psi}\left(m_{\psi}+\lambda\sigma\right)=n_{\psi}\left(m_{\psi}\right)+\lambda\left(\partial_{m_{\psi}}n_{\psi}(m_{\psi})\right)\sigma+\frac{1}{2}\lambda^{2}\left(\partial_{m_{\psi}}^{2}n_{\psi}(m_{\psi})\right)\sigma^{2}+O(\lambda^{3}).
\end{equation}
 Then the bispectrum is written as 
\begin{eqnarray}
B_{S}(\vec{p}_{1},\vec{p}_{2},\vec{p}_{3}) & = & \lambda^{4}\omega_{\psi}^{3}\frac{\left(\partial_{m}n_{\psi}\right)^{2}\left(\partial_{m}^{2}n_{\psi}\right)}{n_{\psi}^{3}}\left[\Delta_{\sigma}^{2}(p_{1})\Delta_{\sigma}^{2}(p_{2})+\mbox{2 perms}\right]+O(\lambda^{6}),\label{eq:Bs}
\end{eqnarray}
which is shown diagrammatically in Fig. \ref{fig:3pt_function}. Now
we compare this with the observational non-Gaussianities using the
conventional non-Gaussian parameter $f_{NL}$ defined by 
\begin{equation}
B_{\zeta}(\vec{p}_{1},\vec{p}_{2},\vec{p}_{3})\equiv\frac{6}{5}f_{NL}\left[\Delta_{\zeta}^{2}(p_{1})\Delta_{\zeta}^{2}(p_{2})+\mbox{2 perms}\right].\label{eq:Bzeta}
\end{equation}

Identifying $B_{\zeta}$ as the bispectrum of the temperature fluctuation
using Eq.(\ref{eq:temp_fluc}) and comparing it with $B_{S}$, we
find in the squeezed triangle limit 
\begin{equation}
f_{NL}^{S}=\frac{8\, B_{S}}{B_{\zeta}|_{f_{NL}=1}}=8\frac{5}{6}\lambda^{4}\omega_{\psi}^{3}\frac{\left(\partial_{m}n_{\psi}\right)^{2}\left(\partial_{m}^{2}n_{\psi}\right)}{n_{\psi}^{3}}\frac{\Delta_{\sigma}^{2}(p_{1})\Delta_{\sigma}^{2}(p_{2})+2\text{ perms.}}{\Delta_{\zeta}^{2}(p_{1})\Delta_{\zeta}^{2}(p_{2})+2\text{ perms.}}.\label{eq:fnl}
\end{equation}
The factor $8$ arises because the radiation transfer function for
isocurvature is twice larger than that for adiabatic perturbation
for the low multipoles of the CMB anisotropy as shown in Eq.~(\ref{eq:temp_fluc}).
Although the isocurvature non-Gaussianities parameter $f_{NL}^{S}$
should not be compared directly with $f_{NL}$ defined by the curvature
perturbation \cite{Komatsu:2001rj}, this can be done with the extra
$O(1)$ correction factor \cite{Hikage:2008sk,Kawasaki:2008sn,Takahashi:2009cx,Chung:2011xd,Hikage:2012be,Hikage:2012tf}.
The reason why $\partial_{m}^{2}n_{\psi}$ appears instead of a first
derivative is because of the squeezed triangle limit allows the short
distance propagator to become important.

In order to obtain the functional structure of $n_{\psi}(m,H;t)$,
which relies on the background behavior, we specialize to the case
of the inflaton coherent oscillation reheating scenarios, in which
the total fermion number freezes during the reheating. During the
early stage of the reheating when the inflaton field oscillates coherently,
the equation of state of the inflaton is zero and the background behaves
like the matter dominated (MD) era. After approximating the early
stage of the reheating to the MD-like era (i.e. inflaton coherent
oscillations period), we get (see Eq.~(\ref{eq:energy-den})) 
\begin{equation}
n_{\psi}(t)\sim\frac{m_{\psi}^{3}}{3\pi^{2}}\left(\frac{a(t_{m})}{a_{t}}\right)^{3}\sim m_{\psi}H_{e}^{2}\left(\frac{a_{e}}{a_{t}}\right)^{3}
\end{equation}
 However, this leading order result gives $\partial_{m_{\psi}}^{2}n_{\psi}=0$
which renders $f_{NL}^{S}=0$ via Eq.(\ref{eq:fnl}).

To find the non-zero result of $f_{NL}^{S}$, we need to study the
mass dependence of $n_{\psi}$ in more detail, which in turn requires
the knowledge of $|\beta_{k}(t;m)|^{2}$. To this point, we have approximated
our spectrum by $|\beta_{k}(t;m)|^{2}\sim1/2\Theta(k_{*}-k)$, where
$k_{*}=a(t_{*})m$ and $t_{*}$is the time when $m=H$. However, in
general the spectrum should contain more than one characteristic scale,
such as $k_{e}=a(t_{e})H_{e}$ where $t_{e}$ marks the end of inflation.
Thus, in general, the number density should contain a fudge factor
$f(\frac{m}{H_{e}})$ i.e. 
\begin{equation}
n_{\psi}\sim m_{\psi}H_{e}^{2}\left(\frac{a_{e}}{a_{t}}\right)^{3}f(\frac{m_{\psi}}{H_{e}})
\end{equation}
 and $f(0)=1$. This higher order correction to $n_{\psi}$ would
render $\partial_{m}^{2}n_{\psi}\neq0$ for the MD-like reheating
scenario.

For simplicity, if we assume that $f(x)=1+a_{1}x$, then in the limit
where $\Delta_{\sigma}^{2},\,\Delta_{\zeta}^{2},\,\mbox{and }\Delta_{\delta_{S}}^{2}$
are scale invariant, we find 
\begin{eqnarray}
f_{NL}^{S} & \sim & a_{1}\left(\frac{\alpha_{S}(\lambda,m_{\psi},H_{e},T_{RH})}{0.02}\right)^{2}\left(\frac{\Omega_{\psi}h^{2}(m_{\psi},T_{RH})}{10^{-7}}\right)^{-1}\left(\frac{m_{\psi}/H_{e}}{10^{-1}}\right).\label{eq:fnleq}
\end{eqnarray}
Although we would naively guess $a_{1}\sim O(1)$, the justification
of the Taylor expansion for $f(x)$ and the estimation of the coefficient
$a_{1}$ will be left for future work since the main thrust of this
work is the computation of isocurvature perturbations and not the
non-Gaussianities. The maximum $f_{NL}$ for the $m_{\psi}\gtrsim\lambda H_{inf}/(2\pi)$
case (consistent with small mass correction case) is achieved when
this inequality is saturated and $\alpha_{S}$ is at its phenomenological
maximum. We find this maximum to be at 
\begin{equation}
f_{NL,\mbox{max}}^{S}\sim O(100)a_{1}\frac{m_{\psi}}{H_{inf}/(2\pi)}.
\end{equation}
Recall that our scenario assumes that $2\pi m_{\psi}/H_{inf}<1$.
Hence, although $f_{NL}^{S}$ cannot be made arbitrarily large, there
may exist a parametric regime in which $f_{NL}^{S}$ is observable
depending on $a_{1}$. Note that this extremum value corresponds to
making the inhomogeneities $O(1)$ while staying consistent with phenomenology
through the $\omega_{\psi}$ dilution factor: i.e. at this parametric
point, the fermion abundance is $\Omega_{\psi}h^{2}\approx10^{-6}$
while most of the CDM is made up of assumed dark matter different
from $\psi$.

\section{Natural Suppression of Gravitational Coupling to the Inflaton}

\label{sub:gauge-int} 

As briefly discussed in \ref{sec:Fermion-Isocurvature-Model}, the
gravity induced coupling of the fermion to the inflaton give a suppressed
contribution to the isocurvature correlation function. We would like
to consider this in more detail in this section. In addition, the
argument below also shows that $\langle\bar{\psi}\psi\zeta\rangle$
cross-correlation is negligible, justifying the classification of
this fermionic isocurvature perturbations as uncorrelated.

First, consider the $\zeta\psi\psi$ interaction given by Eq.~(\ref{eq:Interaction_ZPsiPsi})
following the argument given in Ref. \cite{Chung:2013sla}. In this
case, the most important coupling term is $a^{2}\zeta\delta_{ij}T_{\psi}^{ij}\in\mathcal{H}_{int}$
because the other interactions are derivatively suppressed, and decays
as $O(1/a^{2})$ or faster. Since $\zeta$ also freezes outside the
horizon, using the similar argument given surrounding Eq.~(\ref{eq:Ib})
we can factor the $\zeta$ correlation function out of the dominantly
contributing integral, which corresponds to the diagram (b). Then
we have 
\begin{eqnarray}
I_{\zeta\psi\psi}(x,y) & \approx & (i)^{2}\langle\zeta_{\{z_{0}}\zeta_{w_{0}\}}\rangle[\int_{t_{r}}^{t}dt_{z}\int d^{3}z\, a^{3}(t_{z})\langle[\bar{\psi}\psi_{x},T_{\psi\, i}^{\, i}(z)]\rangle]\nonumber \\
 &  & \times[\int_{t_{r}}^{t}dt_{w}\int d^{3}w\, a^{3}(t_{w})\langle[\bar{\psi}\psi_{y},T_{\psi\, i}^{\, i}(w)]\rangle]+O\left(\frac{a^{2}(t_{r})}{a^{2}(t)}\right)\label{eq:Ib_zPsiPsi}
\end{eqnarray}
where $z_{0}=\left(t_{*},\vec{x}\right)$, $w_{0}=(t_{*},\vec{y})$,
$t=x^{0}=y^{0}$, and $t_{r}$ denotes the time that the comoving
distance $r=\left|\vec{x}-\vec{y}\right|$ crosses the horizon during
inflation. In the integral, we have assumed the PV regulator. Note
that $\lambda\int(dz)T_{\psi\, i}^{\, i}$ is a generator of the spatial
dilatation, $x^{i}\to(1+\lambda)x^{i}$ which is an element of diffeomorphism.
Thus, we have
\begin{equation}
\int_{-\infty}^{t}dt_{z}\int d^{3}z\, a^{3}(t_{z})\langle[\bar{\psi}\psi_{x},T_{\psi\, i}^{\, i}(z)]\rangle=0
\end{equation}
because $\bar{\psi}\psi$ is a diffeomorphism invariant scalar. Indeed,
this is a Ward identity similar to that of Ref. \cite{Chung:2013sla}.
Although the integral in Eq.~(\ref{eq:Ib_zPsiPsi}) does not completely
vanish (because of the time integral limit being $t_{r}$ and not
$-\infty$), the mode function of $\psi$ decays as $1/a^{3}$ (as
shown Appendix \ref{sec:large_r_fermion_correlator}) because of the
classical conformal symmetry characterizing the massless fermionic
sector%
\footnote{Thus, the result is different for a scalar case, which is minimally
coupled to gravity. In particular, the cross-correlation for the light
scalar case is computed in Ref. \cite{Chung:2013sla} and is
\[
\left\langle \zeta\left(t,\vec{x}\right)\sigma^{2}\left(t,\vec{y}\right)\right\rangle \sim O\left(\left(\frac{a(t_{r})}{a(t)}\right)^{3-2\nu},\left(\frac{a(t_{r})}{a(t)}\right)^{2}\right),
\]
where $\nu\equiv\sqrt{\frac{9}{4}-\frac{m_{\sigma}^{2}}{H^{2}}}$.%
}, we have 
\begin{equation}
\int_{t_{r}}^{t}dt_{z}\int d^{3}z\, a^{3}(t_{z})\langle[\bar{\psi}\psi_{x},T_{\psi\, i}^{\, i}(z)]\rangle\sim O\left(\frac{a^{3}(t_{r})}{a^{3}(t)}\right).
\end{equation}
In a similar manner, we can have

\begin{equation}
\left\langle \zeta_{x}\left(\bar{\psi}\psi\right)_{y}\right\rangle \sim O\left(\frac{a^{2}(t_{r})}{a^{2}(t)}\right).
\end{equation}
Therefore, we can conclude that large scale density perturbations
of $\psi$ particles generated by $\zeta$ interaction and the curvature
and isocurvature cross-correlation via the $\zeta\mbox{\ensuremath{\bar{\psi}}\ensuremath{\ensuremath{\psi}}}$
are negligible.

\section{Summary and Conclusion}

\label{sec:conclusion}

In this work, we have presented a fermionic isocurvature scenario
which contains fermionic field fluctuation information during inflation.
To our knowledge, this is the first work that describes isocurvature
inhomogeneities of fermionic fields during inflation. Because massless
free fermions have a tree-level conformal symmetry, such isocurvature
models must couple to a conformal symmetry breaking sector. Because
the $\zeta$ sector coupling to fermion $\psi$ is suppressed due
to the dilatation symmetry, an additional scalar sector $\sigma$
is coupled to $\psi$ (with mass $m_{\psi})$ through a Yukawa coupling
with strength $\lambda$. Composite operator renormalization in curved
spacetime plays an important role in determining the isocurvature
perturbations. We have computed the fermion isocurvature two point
correlation function which has its dominant contribution in the long
wavelength limit coming at one loop 1PI level. We have also estimated
the local non-Gaussianity and found a value that is promising for
observability for a particular corner of the parameter space.

As far as the existence proof inspired ``minimal'' model of this
paper is concerned, a large phenomenologically viable parameter region
spanned by $\{\lambda,m_{\psi}\}$ exists for various inflationary
models controlled by $\{H_{inf},T_{RH}\}$. The large $\lambda$ parameter
region is bounded either by current CMB constraints on isocurvature
perturbations or the constraint of $\sigma$ not decaying to $\psi$.
The large $m_{\psi}$ region is constrained by the relic abundance
non-overclosure. The small $m_{\psi}$ region is constrained by requiring
that $\sigma$ not decay to $\psi$ (for a fixed $\lambda$ and $H_{inf}$).
The large non-Gaussianity parametric region is associated with largest
$\lambda$ consistent with isocurvature bounds and the simplifying
assumption $m_{\psi}\gtrsim\lambda H_{inf}/(2\pi)$. This intuitively
corresponds to a large fermion inhomogeneity (i.e.~$\delta\rho_{\psi}/\bar{\rho}_{\psi}\sim O(1)$)
with a tiny $\bar{\rho}_{\psi}/(\bar{\rho}_{\psi}+\bar{\rho}_{m})$
where $\bar{\rho}_{m}$ corresponds to an adiabatic cold dark matter
component that helps saturate the phenomenologically measured cold
dark matter abundance.

Our results regarding the gravitational fermion production give good
dynamical intuition on many models with dynamical fermions existing
during inflation. One shortcoming of the explicit model used in the
current work is the tuning of the $\sigma$ sector imposed to keep
it light and to prevent any $\sigma$ decay into $\psi$. In addition
to model building issues, it would be interesting to consider in the
future non-Gaussianities from such models more completely and carefully
beyond the estimation presented in this work. It may also be interesting
to see what UV model fermionic sector built independently of cosmological
motivation can be constrained using the analysis presented in this
paper. 
\begin{acknowledgments}
This work was supported in part by the DOE through grant DE-FG02-95ER40896,
Wisconsin Alumni Research Foundation, and National Science Foundation
under Grant No. NSF PHY11-25915. We thank the hospitality and support
of KIAS where part of this work was accomplished.
\end{acknowledgments}
\appendix

\section{Scalar and Spinor fields in Curved spacetime}

First we list the relevant results about scalar field. Consider the
following action 
\begin{equation}
S=\int d^{4}x\sqrt{\left|g\right|}\,\left\{ -\frac{1}{2}g^{\alpha\beta}\partial_{\alpha}\phi\partial_{\beta}\phi-\frac{1}{2}m^{2}\phi^{2}-\frac{1}{2}\xi R\phi^{2}\right\} ,
\end{equation}
This gives rises to equation of motion 
\begin{equation}
\frac{1}{\sqrt{\left|g\right|}}\partial_{\mu}(g^{\mu\nu}\sqrt{\left|g\right|}\partial_{\nu}\phi)-(m^{2}+\xi R)\phi=0
\end{equation}
 Scalar product between two solutions are defined as 
\begin{equation}
(\phi_{1},\phi_{2})=-i\int_{\Sigma}[\phi_{1}\partial_{\mu}\phi_{2}^{*}-\phi_{2}\partial_{\mu}\phi_{1}^{*}]\sqrt{\left|g_{\Sigma}\right|}d\Sigma^{\mu}
\end{equation}
 where $\Sigma$ is a spacelike hypersurface.

For FRW metric, we can use mode decomposition 
\begin{equation}
\phi(x)=\int d^{3}k(c_{\vec{k}}u_{\vec{k}}(x)+c_{\vec{k}}^{\dagger}u_{\vec{k}}^{*}(x))\label{eq:scalar-mode-decomp}
\end{equation}
 with the normalization condition 
\begin{eqnarray}
[c_{\vec{k}},c_{\vec{p}}^{\dagger}] & = & \delta^{3}(\vec{k}-\vec{p})\\
(u_{\vec{k}},u_{\vec{p}}) & = & \delta^{3}(\vec{k}-\vec{p})
\end{eqnarray}
 The mode functions can be written explicitly as 
\begin{eqnarray}
u_{\vec{k}}(x) & = & \frac{e^{i\vec{k}\cdot\vec{x}}}{(2\pi)^{3/2}a(\eta)}f_{k}(\eta)\\
f_{k}\partial_{\eta}f_{k}^{*}-f_{k}^{*}\partial_{\eta}f_{k} & = & i
\end{eqnarray}
 The time-part of the mode function obeys the differential equation
\begin{equation}
\frac{d^{2}}{d\eta^{2}}f_{k,\eta}+\{k^{2}+a_{\eta}^{2}[m^{2}+(\xi-\frac{1}{6}R(\eta))]\}f_{k,\eta}=0
\end{equation}
 where $R(\eta)=6a^{-1}\partial_{\eta}^{2}a$, and $\eta$ is the
conformal time. For de Sitter spacetime, the mode solution for a minimally
coupled scalar $\left(\xi=0\right)$ is 
\begin{equation}
f_{k}(\eta)=\frac{1}{\sqrt{2k}}\sqrt{\frac{\pi}{2}\left(\frac{k}{aH}\right)}e^{i\frac{\pi}{2}(\nu+\frac{1}{2})}H_{\nu}^{(1)}(\frac{k}{aH})
\end{equation}
 where $\nu^{2}=\frac{9}{4}-\frac{m^{2}}{H^{2}}$.

The following relations of first kind of Hankel functions are useful
\begin{eqnarray}
H_{\nu}^{(1)}(z) & \rightarrow & -i\frac{\Gamma(\nu)}{\pi}\left(\frac{2}{z}\right)^{\nu}\quad(z\rightarrow0)\\
H_{\nu}^{(1)}(z) & \rightarrow & \sqrt{\frac{2}{\pi z}}e^{-i\frac{\pi}{2}(\nu+\frac{1}{2})}e^{iz}\quad(z\rightarrow\infty)
\end{eqnarray}

From the mode expansion, we may construct the equal-time correlator
in dS spacetime. In particular, we are interested in the large separation
limit. For light scalar, when $\nu$ is real, we have 
\begin{equation}
\langle\sigma_{x}\sigma_{y}\rangle\approx\frac{H^{2}}{8\pi}\frac{\Gamma(\frac{3}{2}-\nu)}{\Gamma(\frac{3}{2})\Gamma(1-\nu)\sin(\nu\pi)}(aHr)^{2\nu-3}\label{eq:light-scalar}
\end{equation}
 For heavy scalar, when $\nu=i\alpha$ and if $\alpha\sim\frac{m}{H}\gg1$,
then 
\begin{equation}
\langle\sigma_{x}\sigma_{y}\rangle\approx\frac{H^{3/2}m^{1/2}}{\pi^{3/2}}e^{-\frac{m}{H}\pi}\sin[2\frac{m}{H}\ln(aHr)-\frac{1}{4}\pi](aHr)^{-3}
\end{equation}

Next, we give the result for spinor field. Consider the free Dirac
field $\psi$ action 
\begin{equation}
S=\int(dx)\left(i\bar{\psi}\gamma^{\mu}\nabla_{\mu}\psi-m\bar{\psi}\psi\right).\label{eq:DiracLagrangian}
\end{equation}
 where $(dx)=d^{4}x\sqrt{|g_{x}|}$ and $\gamma^{\mu}\equiv\gamma^{a}e_{a}^{\,\mu}$
with vierbein $e_{a}^{\,\mu}$. The covariant derivatives for $\psi$
is defined by

\begin{equation}
\nabla_{\mu}\psi=\partial_{\mu}\psi+\frac{1}{2}\omega_{\mu}^{ab}\Sigma_{ab}\psi
\end{equation}
 and the spin-connection is defined by 
\begin{equation}
\omega_{\mu}^{ab}=e_{\,\nu}^{a}\nabla_{\mu}e^{b\nu}
\end{equation}
 and the Lorentz generator on the spinor field is given by 
\begin{equation}
\Sigma_{ab}=-\frac{1}{4}[\gamma_{a},\gamma_{b}],
\end{equation}
where the $\gamma$ matrices satisfy the $\left\{ \gamma_{a},\gamma_{b}\right\} =-2\eta_{ab}$
with $\eta\equiv\mbox{diag}(-1,1,1,1)$. Note that the sign convention
is chosen such that $[\Sigma^{12},\Sigma^{23}]=\Sigma^{13}$.

Extremizing the action with respect to $\delta\bar{\psi}$ and $\delta\psi$
yields the equations of motion: 
\begin{eqnarray}
(i\gamma^{\mu}\nabla_{\mu}-m)\psi=0,\qquad\nabla_{\mu}\bar{\psi}(-i\gamma^{\mu})-\bar{\psi}m=0.\label{eq:fermion-EOM}
\end{eqnarray}
 The solution space can be endowed with a scalar product as 
\begin{equation}
(\psi_{1},\psi_{2})_{\Sigma}=\int d\Sigma n_{\mu}\bar{\psi}_{1}\gamma^{\mu}\psi_{2}
\end{equation}
 in which $\Sigma$ is an arbitrary space-like hypersurface, $d\Sigma$
is the volume 3-form on this hypersurface computed with the induced
metric, and $n_{\mu}$ is the future-pointing time-like unit vector
normal to $\Sigma$. The current conservation condition 
\begin{equation}
\nabla_{\mu}(\bar{\psi}_{1}\gamma^{\mu}\psi_{2})=0
\end{equation}
 implies the integral in the scalar product is independent of the
choice of $\Sigma$.

If we adopt the Dirac basis for the $\gamma$ matrices, i.e. 
\begin{equation}
\gamma^{0}=\left(\begin{array}{cc}
I & 0\\
0 & -I
\end{array}\right),\quad\gamma^{i}=\left(\begin{array}{cc}
0 & \sigma^{i}\\
-\sigma^{i} & 0
\end{array}\right)
\end{equation}
 the mode functions can be written as 
\begin{eqnarray}
U_{\vec{k},r}(x) & = & \frac{1}{a_{x}^{3/2}}\frac{e^{i\vec{k}\cdot\vec{x}}}{(2\pi)^{3/2}}\left(\begin{array}{c}
u_{A,k,x^{0}}\\
r\, u_{B,k,x^{0}}
\end{array}\right)\otimes h_{\hat{k},r}\label{eq:U-mode}\\
V_{\vec{k},r}(x) & = & -i\gamma^{2}U_{\vec{k},r}^{*}(x)=\frac{1}{a_{x}^{3/2}}\frac{e^{-i\vec{k}\cdot\vec{x}}}{(2\pi)^{3/2}}\left(\begin{array}{c}
r\, u_{B,k,x^{0}}^{*}\\
-u_{A,k,x^{0}}^{*}
\end{array}\right)\otimes(-i\sigma_{2})h_{\hat{k},r}^{*}\label{eq:V-mode}
\end{eqnarray}
 where $h_{\hat{k},r}$ is eigenvector of $\hat{k}\cdot\vec{\sigma}$.
The normalization conditions requires 
\begin{eqnarray}
h_{\hat{k},r}^{\dagger}h_{\hat{k},s} & = & \delta_{rs}\\
|u_{A,k,\eta}|^{2}+|u_{B,k,\eta}|^{2} & = & 1.\label{eq:uAuB-norm}
\end{eqnarray}
 The time dependent parts of the mode functions obey the following
equation 
\begin{equation}
i\frac{d}{d\eta}\left(\begin{array}{c}
u_{A}\\
u_{B}
\end{array}\right)=\left(\begin{array}{cc}
am & k\\
k & -am
\end{array}\right)\left(\begin{array}{c}
u_{A}\\
u_{B}
\end{array}\right).\label{eq:Dirac Equation}
\end{equation}
 In the special case of the de Sitter background with Bunch-Davies
boundary condition, we have 
\begin{eqnarray}
\left(\begin{array}{c}
u_{A}\\
u_{B}
\end{array}\right)_{k,\eta}^{in} & = & \left(\begin{array}{c}
\sqrt{\frac{\pi}{4}(\frac{k}{aH_{e}})}e^{i\frac{\pi}{2}(1-i\frac{m}{H_{e}})})H_{\frac{1}{2}-i\frac{m}{H_{e}}}^{(1)}(\frac{k}{aH})\\
\sqrt{\frac{\pi}{4}(\frac{k}{aH_{e}})}e^{i\frac{\pi}{2}(1+i\frac{m}{H})})H_{\frac{1}{2}+i\frac{m}{H_{e}}}^{(1)}(\frac{k}{aH})
\end{array}\right)\label{eq:spinor-dS}\\
\mbox{if}\,|kx^{0}|\ll1 & \longrightarrow & \left(\begin{array}{c}
\frac{1}{\sqrt{2\pi}}e^{\frac{\pi}{2}\frac{m}{H}}e^{-im(t-t_{e})+i\frac{m}{H}\ln(2k/a_{e}H)}\Gamma(\frac{1}{2}-i\frac{m}{H})\\
\frac{1}{\sqrt{2\pi}}e^{-\frac{\pi}{2}\frac{m}{H}}e^{+im(t-t_{e})-i\frac{m}{H}\ln(2k/a_{e}H)}\Gamma(\frac{1}{2}+i\frac{m}{H})
\end{array}\right)\label{eq:dS-mode-end-1}
\end{eqnarray}

Since the interaction picture operator $\psi(x)$ obeys the same classical
equations, Eq.~(\ref{eq:fermion-EOM}), we can expand the operator
using $\{U_{i},V_{i}\}$ as the basis: 
\begin{equation}
\psi(x)=\sum_{i}a_{i}U_{i}(x)+b_{i}^{\dagger}V_{i}(x)\label{eq:fermion-mode-decomp}
\end{equation}
 and the normalization conditions on $U_{i},V_{i}$ gives the usual
canonical anti-commutation relations of the creation and annihilation
operators.

The first order WKB approximation is defined as 
\begin{equation}
\left(\begin{array}{c}
u_{A}\\
u_{B}
\end{array}\right)_{k,\eta}^{WKB}=\left(\begin{array}{c}
\sqrt{\frac{\omega+am}{2\omega}}\\
\sqrt{\frac{\omega-am}{2\omega}}
\end{array}\right)e^{-i\int^{\eta}\omega d\eta'}\label{eq:WKB-mode}
\end{equation}
 In the following, when we talk about fermion particle, we are implicitly
referring to the WKB-mode.

Thus one can introduce the time-dependent Bogoliubov coefficients
$\{\alpha_{k,\eta},\beta_{k,\eta}\}$ between the in-modes and WKB-modes:
\begin{equation}
\left(\begin{array}{c}
u_{A}\\
u_{B}
\end{array}\right)_{k,\eta}^{in}=\alpha_{k,\eta}\left(\begin{array}{c}
u_{A}\\
u_{B}
\end{array}\right)_{k,\eta}^{WKB}+\beta_{k,\eta}\left(\begin{array}{c}
u_{B}^{*}\\
-u_{A}^{*}
\end{array}\right)^{WKB}.\label{eq:bogo-trans}
\end{equation}
 Clearly, $(\alpha,\beta)\rightarrow(1,0)$ as $\eta\rightarrow-\infty$
. We may also note that the Bogoliubov coefficients obey normalization
condition as 
\begin{equation}
|\alpha_{k,\eta}|^{2}+|\beta_{k,\eta}|^{2}=1.\label{eq:bogo-norm}
\end{equation}
 in agreement with fermion statistics.

Using Eq.~(\ref{eq:bogo-trans}), (\ref{eq:WKB-mode}) and (\ref{eq:Dirac Equation}),
we can derive the evolution equation for the Bogoliubov coefficients,
as shown in Eq.~(\ref{eq:Bogo-EOM}).

\section{Review of fermion particle production}

\label{sec:prod_review}In this section, we give a brief review of
the main result about fermion production during inflation \cite{Chung:2011ck}.
The fermion number density can be obtained by solving this equations
of Bogoliubov coefficients 
\begin{equation}
\partial_{\eta}\left(\begin{array}{c}
\alpha_{k,\eta}\\
\beta_{k,\eta}
\end{array}\right)=\frac{a^{2}m\, k\, H}{2\omega^{2}}\left(\begin{array}{cc}
0 & e^{2i\int^{\eta}\omega d\eta'}\\
-e^{-i\int^{\eta}\omega d\eta'} & 0
\end{array}\right)\left(\begin{array}{c}
\alpha_{k,\eta}\\
\beta_{k,\eta}
\end{array}\right)\label{eq:Bogo-EOM}
\end{equation}
 We define the non-adiabaticity for a mode $k$ as $\epsilon_{k,\eta}=\frac{m\, k_{p}\, H}{\omega_{p}^{3}}$,
where subscript $p$ stand for ``physical'', $\omega_{p}=\omega/a$
etc. As the system evolves from an initial vacuum condition of $\left(\begin{array}{c}
\alpha_{k,\eta}\\
\beta_{k,\eta}
\end{array}\right)=\left(\begin{array}{c}
1\\
0
\end{array}\right)$, $\beta_{k,\eta}$ will only increase significantly when $\epsilon_{k,\eta}\sim O(1)$.
This implies the following results, 
\begin{enumerate}
\item In the heavy mass limit ($m_{\psi}\gg H_{inf}$), $\epsilon_{k,\eta}$
is always suppressed by $\frac{H}{m_{\psi}}$, we get $|\beta_{k,\eta}|^{2}\sim\exp[-C\frac{m_{\psi}}{H(\eta_{k})}]\ll1$,
where $C$ is some order one constant and $H(\eta_{k})$ is the Hubble
rate at the most non-adiabatic moment for mode $k$. 
\item In the light mass limit ($m_{\psi}\ll H_{inf}$), $\epsilon_{k,\eta}$
is largest when $k_{p}\sim m_{\psi}$, we call this time $\eta_{k}$.
If $m_{\psi}<H(\eta_{k})$, we have $|\beta_{k}|^{2}\sim\frac{1}{2}$,
otherwise it is suppressed by $\exp[-C\frac{m_{\psi}}{H(\eta_{k})}]$
as well. 
\end{enumerate}
Since the heavy fermion production is always exponentially suppressed
by $m_{\psi}/H$ ratio, we focus on the light fermion case. The energy
density at time $t$ is given by 
\begin{equation}
\rho(t)\sim\frac{m_{\psi}^{4}}{3\pi^{2}}\left(\frac{a(t_{*})}{a(t)}\right)^{3},\label{eq:energy-den}
\end{equation}
 where $t_{*}$ is the time when $H(t)=m_{\psi}$. If $t_{*}$ occurs
during reheating, one get the relic abundance today time as 
\begin{equation}
\Omega_{\psi}h^{2}\sim3\left(\frac{m_{\psi}}{10^{11}\mbox{GeV}}\right)^{2}\left(\frac{T_{RH}}{10^{9}\mbox{GeV}}\right).\label{eq:relic-abundance}
\end{equation}

\section{Asymptotic behavior of $\langle\psi_{x}\bar{\psi}_{y}\rangle$ at
large $r$}

\label{sec:large_r_fermion_correlator}In this section we derive the
result about leading order contribution to $\langle n_{\psi,x}n_{\psi,y}\rangle$,
i.e. Eq.~(\ref{eq:nxny_LO}). By Wick contraction, this reduces to
computing the field correlator $\langle\psi_{x}\bar{\psi}_{y}\rangle$.
The standard way to compute the correlator is to plug in the mode
decomposition Eq.~(\ref{eq:fermion-mode-decomp}) and compute the
mode functions $\{U_{i},V_{i}\}$. The difficulties lie in how to
obtain the mode functions on a curved spacetime. For inflationary
background spacetime, one can use the de Sitter spacetime as an approximation
and obtain exact analytic solutions. However, it is unclear how do
these mode solutions evolve after inflation ends. Such postinflationary
solutions are relevant for our computation because the particle production
freezes out after the end of inflation. Here we give an approach that
answers this question.

First, we plug in the mode decomposition to the equal-time correlator:
\begin{eqnarray}
 &  & \langle\psi_{x}\bar{\psi}_{y}\rangle\nonumber \\
 & = & \int d^{3}k\frac{1}{a_{x}^{3}}\frac{e^{i\vec{k}\cdot\vec{r}}}{(2\pi)^{3}}\left(\begin{array}{cc}
|u_{A,k,x^{0}}|^{2}\otimes I_{2} & -\, u_{A,k,x^{0}}u_{B,k,x^{0}}^{*}\otimes(\hat{k}\cdot\vec{\sigma})\\
u_{B,k,x^{0}}u_{A,k,x^{0}}^{*}\otimes(\hat{k}\cdot\vec{\sigma}) & -|u_{B,k,x^{0}}|^{2}\otimes I_{2}
\end{array}\right)
\end{eqnarray}
 where we have performed the spin-sum in the last step. Since 
\begin{eqnarray}
\int d^{3}k\frac{e^{i\vec{k}\cdot\vec{r}}}{(2\pi)^{3}}|u_{A,k,x^{0}}|^{2} & = & \int d^{3}k\frac{e^{i\vec{k}\cdot\vec{r}}}{(2\pi)^{3}}(1-|u_{B,k,x^{0}}|^{2})\label{eq:dummy3}\\
 & = & \delta^{3}(\vec{r})-\int d^{3}k\frac{e^{i\vec{k}\cdot\vec{r}}}{(2\pi)^{3}}|u_{B,k,x^{0}}|^{2}
\end{eqnarray}
 and $\vec{r}\neq0$, we see the diagonal elements are the same. Then
we perform the angular integral $d^{2}\hat{k}$. Recall that 
\begin{eqnarray}
\int d^{3}k\, e^{i\vec{k}\cdot\vec{r}}f(k) & = & \int4\pi k^{2}dk\,\frac{\sin(kr)}{kr}f(k)\\
\int d^{3}k\, e^{i\vec{k}\cdot\vec{r}}\hat{k}_{i}f(k) & = & (-i\hat{r}_{i}\partial_{r})\int4\pi k^{2}dk\,\frac{\sin(kr)}{kr}\frac{f(k)}{k}
\end{eqnarray}
 After the angular integral, we have 
\begin{eqnarray}
\langle\psi_{x}\bar{\psi}_{y}\rangle & = & \int\frac{4\pi k^{2}dk}{(2\pi)^{3}}\left(\begin{array}{cc}
A & B\\
B^{*} & C
\end{array}\right)\label{eq:equal-time-2pt-ferm}\\
A & = & |u_{A,k,\eta}|^{2}\cdot\frac{\sin(kr)}{kr}\\
B & = & (i\hat{r}\cdot\vec{\sigma})u_{A,k,\eta}u_{B,k,\eta}^{*}\cdot\partial_{r}\frac{\sin(kr)}{kr}\frac{1}{k}\\
C & = & -|u_{B,k,\eta}|^{2}\cdot\frac{\sin(kr)}{kr}
\end{eqnarray}
 It is sufficient to study these two integrals for the diagonal and
off-diagonal elements. 
\begin{eqnarray}
I_{11} & = & I_{22}=\int_{0}^{\infty}\frac{4\pi k^{2}dk}{(2\pi)^{3}}\,|u_{A,k,\eta}|^{2}\cdot\frac{\sin(kr)}{kr}\label{eq:I1}\\
I_{12} & = & I_{21}^{*}=\partial_{r}\int_{0}^{\infty}\frac{4\pi k^{2}dk}{(2\pi)^{3}}\, u_{A,k,\eta}u_{B,k,\eta}^{*}\frac{\sin(kr)}{kr}\frac{1}{k}\label{eq:I2}
\end{eqnarray}
 Now, we only need to find the mode function $u_{A},u_{B}$, and perform
the mode sum.

Let's consider the mode functions first. Since we are interested in
evaluating the fermion field correlator at a time when the fermion
production has ended, i.e. when $m\gg H(x^{0})$ and in the limit
$r\rightarrow\infty$, we can make the following approximations about
the mode functions $\{u_{A,k,x^{0}},u_{B,k,x^{0}}\}$. First, since
the particle production has stopped, the non-adiabatic parameter is
suppressed by $\frac{H(t)}{m}$, thus we can approximately replace
the Bogoliubov coefficients by their late time asymptotic values,
i.e. 
\begin{equation}
\alpha_{k,x^{0}}\approx\alpha_{k},\,\,\,\,\,\,\beta_{k,x^{0}}\approx\beta_{k}.
\end{equation}
 Second, since we want to capture the particle production effect on
the correlator and the produced particles are non-relativistic at
the time of production, by the time $x^{0}$ which is sufficiently
long after the production has ended, we may approximate the produced
modes all have $k\ll a(x^{0})m$. Thus, the WKB modes can be approximated
by 
\begin{equation}
\left(\begin{array}{c}
u_{A}\\
u_{B}
\end{array}\right)_{k,\eta,IR}^{WKB}=\left(\begin{array}{c}
\sqrt{\frac{\omega+am}{2\omega}}\\
\sqrt{\frac{\omega-am}{2\omega}}
\end{array}\right)e^{-i\int^{\eta}\omega d\eta'}\rightarrow\left(\begin{array}{c}
\frac{1}{\sqrt{2}}\\
0
\end{array}\right)e^{-i\int^{\eta}\omega d\eta'}.
\end{equation}
 Combining these two approximations, we have

\begin{equation}
\left(\begin{array}{c}
u_{A}\\
u_{B}
\end{array}\right)_{k,\eta,IR}^{in}\approx\left(\begin{array}{c}
\alpha_{k}\frac{1}{\sqrt{2}}e^{-i\int^{\eta}\omega d\eta'}\\
-\beta_{k}\frac{1}{\sqrt{2}}e^{i\int^{\eta}\omega d\eta'}
\end{array}\right)
\end{equation}
 Thus we can easily evaluate $I_{11},I_{12}$: 
\begin{eqnarray}
2\pi^{2}I_{11,IR} & = & \frac{1}{r}\mbox{Im}\int_{0}^{\infty}kdk\,\frac{1}{2}[1-n(k)]\cdot e^{ikr}
\end{eqnarray}
 We note that for the contribution from $1$ vanishes 
\begin{equation}
\frac{1}{r}\mbox{Im}\int_{0}^{\infty}kdk\,[1]\cdot e^{ikr}=\frac{1}{r}\mbox{Im}\int_{0}^{\infty}(is)ids\,[1]\cdot e^{-sr}=0
\end{equation}
 For the contribution from $n(k)$ , we may assume it to be a real
analytic function on $\mathbb{R}^{+}$and can be analytically continuated
to upper-right quadrant of the complex $k$ plane. The location of
singularity of $n(k)$ determines contour of $k$. For example, we
may consider the $n(k)$ for heavy fermion case ($m>H_{inf}$): 
\begin{equation}
n(k)_{heavy}=\exp\left[-\frac{4(k/a_{nad})^{2}}{mH}-\frac{4m}{H}\right]
\end{equation}
 where $a_{nad}$ is at the non-adiabatic time point. In this case,
the non-adiabatic time is the transition from de Sitter era to the
reheating era, i.e. $a_{nad}=a_{e}$. One can apply steepest descent
to find that 
\begin{eqnarray}
 &  & 2\pi^{2}I_{11,heavy,IR}\nonumber \\
 & \approx & -\frac{1}{r}\exp[-\frac{4m}{H}-\frac{1}{16}mHr^{2}](a_{e}^{2}mH)\mbox{Im}[-i\frac{1}{4}\sqrt{mH}a_{e}r\frac{1}{2}\sqrt{\pi}]\label{eq:intermedstep1}\\
 & = & \frac{1}{8}\sqrt{\pi}a_{e}^{3}(mH)^{\frac{3}{2}}\exp[-\frac{4m}{H}-\frac{1}{16}a_{e}^{2}mHr^{2}]
\end{eqnarray}
 For light fermion, we may approximate the number density spectrum
as 
\begin{equation}
n(k)_{light}=\frac{1}{1+\exp(\frac{k^{2}}{(a_{nad}m)^{2}})}
\end{equation}
 where the non-adiabatic point occurs when $H$ drops below $m$,
i.e. $a_{nad}=a(\eta_{*})=a_{*}$. This ansatz is only used to mimic
the cut-off of the spectrum at $k\sim a_{nad}m$. The singularity
lies at 
\begin{equation}
\frac{k^{2}}{a_{*}^{2}m^{2}}=(2n+1)\pi i,\qquad n=0,1,2\cdots
\end{equation}
 or $k_{*,n}=a_{*}m\sqrt{(2n+1)\pi}e^{\frac{\pi}{4}i}$. Again, one
can perform the steepest descent around the $n=0$ singularity $k_{*}=a_{*}m\sqrt{\pi}e^{\frac{\pi}{4}i}$.
Let $\delta=(k-k_{*})/a_{*}m$, we have 
\begin{eqnarray}
2\pi^{2}I_{11,light,IR} & = & \pi a_{*}^{3}\frac{m^{2}}{a_{*}r}\exp[-\sqrt{\frac{\pi}{2}}a_{*}mr]\cos(\sqrt{\frac{\pi}{2}}a_{*}mr)\label{eq:I1-light-IR}
\end{eqnarray}
 For both the heavy and light fermion case, $I_{11}\propto\exp(-a_{*}Mr)$,
where $a_{*}M$ is the scale that $n(k)$ cuts off. We should also
remind ourself that the UV vacuum contributions also exist, which
scales as 
\begin{equation}
I_{11,UV}\propto\exp[-a_{\eta}mr]
\end{equation}
 due to the singularity at $k=a_{\eta}m$ in the mode functions $u_{A}^{WKB},u_{B}^{WKB}$.
Thus we have shown that the diagonal element of Eq.~(\ref{eq:equal-time-2pt-ferm})
is always exponentially suppressed.

Next, we turn to look at the off diagonal element $I_{12}$. Unlike
the $I_{11}$ case, whose integrand $|u_{A}|^{2}$ has constant asymptotic
value in the IR region, the $I_{12}$'s IR contribution 
\begin{equation}
u_{A,k,\eta}u_{B,k,\eta}^{*}=\alpha_{k}\beta_{k}^{*}e^{-2i\int^{\eta}\omega d\eta'}
\end{equation}
 contains $e^{-2imt}$ time dependence. Physically, if we decompose
the in-state into WKB vacuum and excitation state 
\begin{equation}
|\mbox{in,vac}\rangle=\sim|\mbox{WKB,vac}\rangle+\sim|\mbox{WKB,2-particles}\rangle+\sim|\mbox{WKB,4-particles}\rangle
\end{equation}
 then this term comes from the interference term 
\begin{equation}
\langle\mbox{WKB},\mbox{vac}|\psi_{x}\bar{\psi}_{y}|\mbox{WKB},\mbox{2-particles}\rangle\in\langle\mbox{in},\mbox{vac}|\psi_{x}\bar{\psi}_{y}|\mbox{in},\mbox{vac}\rangle.
\end{equation}
If we care about $r$ large enough, for example corresponding to the
CMB observation scale at recombination, we may assume the relevant
$k$ scale exit horizon and become non-relativistic during inflation.
Thus we may safely use the dS mode function to evaluate $I_{12,IR,CMB}$.

Recall that during dS era, we have Eq.~(\ref{eq:spinor-dS}), where
we choose the end of inflation time $t_{e}$ as the reference point.
Thus 
\begin{equation}
u_{A,k,\eta}u_{B,k,\eta}^{*}=\frac{1}{2\pi}e^{-2im(t-t_{e})+2i\frac{m}{H}\ln(2k/a_{e}H)}\Gamma^{2}(\frac{1}{2}-i\frac{m}{H})
\end{equation}
 Performing the integral using steepest descent, we find the leading
contribution comes from $k\sim0$ singularity in $u_{A,k,\eta}u_{B,k,\eta}^{*}$.
We note that the $k$ dependent phase factor $e^{2i\frac{m}{H}\ln(2k/H)}$
cannot be absorbed by a redefinition of the mode functions $u_{A,k,\eta},u_{B,k,\eta}$,
since this phase factor depends on the relative phase of $u_{A,k,\eta},u_{B,k,\eta}$
which is fixed by the Bunch-Davies initial condition.

Plugging in the Eq.~(\ref{eq:I2}), we have 
\begin{eqnarray}
 &  & 2\pi^{2}I_{12,IR}\nonumber \\
 & = & -e^{-2im(t-t(r))+i\phi(\frac{m}{H})}r^{-3}\sqrt{\frac{2\pi\frac{m}{H}}{\sinh(2\pi\frac{m}{H})}\left(1+\left(\frac{m}{H}\right)^{2}\right)}\label{eq:I2-IR}
\end{eqnarray}
 where $\phi(\frac{m}{H})=\mbox{Arg}(\Gamma(2+ix)\Gamma(\frac{1}{2}-ix))$
and $t(r)$ is the time when $a(t_{r})Hr=4$. We may consider the
light mass limit 
\begin{equation}
2\pi^{2}I_{12,IR,light}\approx-e^{-2im(t-t(r))}r^{-3}
\end{equation}
 and the heavy mass limit 
\begin{equation}
2\pi^{2}I_{12,IR,heavy}\approx-(4\pi)^{\frac{1}{2}}\left(\frac{m}{H}\right)^{\frac{3}{2}}\exp(-\pi\frac{m}{H})e^{-2im(t-t(r))}r^{-3}
\end{equation}

We may also consider the effect of having an IR cut-off $k_{IR}$,
which is the scale that exit horizon at the beginning of inflation.
Such an IR cut-off will introduce a $\exp(-k_{IR}r)$ type of exponential
suppression factor. However, for observable universe with comoving
radius $R_{obs}$, as long as $k_{IR}R_{obs}\ll1$, we may ignore
this suppression factor.

After evaluating the matrix element for the fermion correlators, we
find that 
\begin{enumerate}
\item For the light fermion case, i.e. $m\ll H_{inf}$, in the limit $r\rightarrow\infty$
\begin{equation}
\langle\psi_{x}\bar{\psi}_{y}\rangle\approx\frac{1}{a_{x}^{3}}\frac{1}{2\pi^{2}}\left(\begin{array}{cc}
A & B\\
B^{*} & A
\end{array}\right)\label{eq:ABBA}
\end{equation}
 where 
\begin{eqnarray}
A & = & \frac{1}{2}\pi a_{*}^{3}\frac{m^{2}}{a_{*}r}\exp[-\sqrt{\frac{\pi}{2}}a_{*}mr]\cos(\sqrt{\frac{\pi}{2}}a_{*}mr)\\
B & = & -i\hat{r}\cdot\vec{\sigma}e^{-2im(t-t_{r})}r^{-3}
\end{eqnarray}

where $a_{*}$ in evaluated at $\eta_{*}$. 

\item For the heavy fermion case, i.e. $m\gg H_{inf}$, in the limit $r\rightarrow\infty$,
we find in Eq.~(\ref{eq:ABBA}) 
\begin{eqnarray}
A & = & \frac{1}{16}\sqrt{\pi}a_{e}^{3}(mH_{e})^{\frac{3}{2}}\exp[-\frac{4m}{H_{e}}-\frac{1}{16}a_{e}^{2}mH_{e}r^{2}]\\
B & = & -i\hat{r}\cdot\vec{\sigma}(4\pi)^{\frac{1}{2}}\left(\frac{m}{H_{e}}\right)^{\frac{3}{2}}\exp(-\pi\frac{m}{H_{e}})e^{-2im(t-t(r))}r^{-3}
\end{eqnarray}
 and $a_{e}$ is evaluated at the end of inflation. 
\end{enumerate}
Finally, we plug in the field correlator to $\langle n_{\psi,x}n_{\psi,y}\rangle$,
and drop the term that are exponentially suppressed when $r\rightarrow\infty$,
to get Eq.~(\ref{eq:nxny_LO}).

\section{Relative suppression of Commutators\label{sec:AppD:Relative-suppression}}

In this subsection, we want compare the dependence on the scale factor
$a(t)$ between $\langle in|[O_{x},O_{y}]|in\rangle$ and $\langle in|\{O_{x},O_{y}\}|in\rangle$,
where $O_{x}$ is a bosonic hermitian operator and $x,y$ are spacetime
points located near the end of inflation. For simplicity, we take
$H$ as a constant. In particular, we are interested in the cases
where $O=\sigma,\bar{\psi}\psi,\zeta$. We want to show that the commutator
of $O$ suffers from additional suppression factor compared to the
anti-commutator.

In general, the diagonal matrix elements of products of hermitian
operator obeys 
\begin{equation}
(\langle in|O_{x}O_{y}|in\rangle)^{*}=\langle in|O_{y}O_{x}|in\rangle
\end{equation}
 therefore 
\begin{eqnarray}
\langle in|[O_{x},O_{y}]|in\rangle & = & 2i\mbox{Im}\langle in|O_{x}O_{y}|in\rangle\\
\langle in|\{O_{x},O_{y}\}|in\rangle & = & 2\mbox{Re}\langle in|O_{x}O_{y}|in\rangle
\end{eqnarray}
 We can just study $\langle in|O_{x}O_{y}|in\rangle$. We may use
the mode expansion of the field operator to evaluate such an expression,
and focus on modes that are outside of horizon at both times $\eta_{x},\eta_{y}$.

We shall first take $O=\sigma$, and we assume that the scalar is
light, i.e. $m_{\sigma}<\frac{3}{2}H$, such that $\nu$ is real:
\begin{eqnarray}
\langle in|\sigma_{x}\sigma_{y}|in\rangle & = & \int4\pi k^{2}dk\frac{[\int d^{2}\hat{k}e^{i\vec{k}\cdot(\vec{x}-\vec{y})}]}{(2\pi)^{3}a_{x}^{3/2}a_{y}^{3/2}}\frac{1}{H}\frac{\pi}{4}[J_{x}J_{y}+Y_{x}Y_{y}+i(Y_{x}J_{y}-J_{x}Y_{y})]
\end{eqnarray}
 where $J_{x}=J_{\nu}(\frac{k}{a_{x}H}),Y_{x}=Y_{\nu}(\frac{k}{a_{x}H})$
are the first and second kinds of Bessel functions with real values.
The $d^{2}\hat{k}$ is the angular integral with normalization $\int d^{2}\hat{k}=1$,
and $\int d^{2}\hat{k}e^{i\vec{k}\cdot(\vec{x}-\vec{y})}=\sin(kr)/kr$
is real. If we focus on the $k$ modes that are outside of horizon,
i.e. $k/aH\ll1$, we may use the small argument expansion of the Bessel
function, i.e. when ($0<z<\sqrt{1+\nu}$) 
\begin{eqnarray}
J_{\nu}(z) & \approx & \frac{1}{\Gamma(\alpha+1)}\left(\frac{z}{2}\right)^{\nu}\\
Y_{\nu}(z) & \approx & -\frac{\Gamma(\alpha)}{\pi}\left(\frac{2}{z}\right)^{\nu}.
\end{eqnarray}
 Then, under the common scaling of $a_{x}\rightarrow\lambda a_{x},a_{y}\rightarrow\lambda a_{y}$,
with $\lambda$ increasing, we see the various term in the correlator
scales as 
\begin{eqnarray}
a_{x}^{-3/2}a_{y}^{-3/2}J_{x}J_{y} & \propto & \lambda^{-2\nu-3}\\
a_{x}^{-3/2}a_{y}^{-3/2}Y_{x}Y_{y} & \propto & \lambda^{2\nu-3}\\
a_{x}^{-3/2}a_{y}^{-3/2}(Y_{x}J_{y}-J_{x}Y_{y}) & \propto & \lambda^{-3}
\end{eqnarray}
 Thus, we see under this common scaling, the IR contribution to the
two point functions are 
\begin{eqnarray}
\langle in|\{\sigma_{x},\sigma_{y}\}|in\rangle_{IR} & = & 2\int_{IR}4\pi k^{2}dk\frac{[\int d^{2}\hat{k}e^{i\vec{k}\cdot(\vec{x}-\vec{y})}]}{(2\pi)^{3}a_{x}^{3/2}a_{y}^{3/2}}\frac{1}{H}\frac{\pi}{4}(J_{x}J_{y}+Y_{x}Y_{y})\propto\lambda^{2\nu-3}\\
\langle in|[\sigma_{x},\sigma_{y}]|in\rangle_{IR} & = & 2i\int_{IR}4\pi k^{2}dk\frac{[\int d^{2}\hat{k}e^{i\vec{k}\cdot(\vec{x}-\vec{y})}]}{(2\pi)^{3}a_{x}^{3/2}a_{y}^{3/2}}\frac{1}{H}\frac{\pi}{4}(Y_{x}J_{y}-J_{x}Y_{y})\propto\lambda^{-3}
\end{eqnarray}
 Thus, we have shown under the scaling $a\rightarrow\lambda a$, the
commutator of $\sigma$ is suppressed by $\lambda^{-2\nu}$ factor
relative to its anti-commutator. For small mass scalar, $\lambda^{-2\nu}\approx\lambda^{-3+\frac{2m^{2}}{3H^{2}}}$.

For the case of $O=\zeta$, we have similar statements as the scalar
case with $\nu=\frac{3}{2}$, i.e. $\langle[\zeta_{x},\zeta_{y}]\rangle_{IR}$
is suppressed by $\lambda^{-3}$ relative to $\langle\{\zeta_{x},\zeta_{y}\}\rangle_{IR}$
under the scaling of $a\rightarrow\lambda a$.

Next, we consider the case of $O=\bar{\psi}\psi$. Using the mode
decomposition Eq.(\ref{eq:fermion-mode-decomp}) and mode functions
Eq.~(\ref{eq:U-mode},\ref{eq:V-mode}), we have 
\begin{eqnarray}
\langle\bar{\psi}\psi_{x}\bar{\psi}\psi_{y}\rangle & = & \sum_{i,j}\frac{1}{a_{x}^{3}a_{y}^{3}}\frac{e^{i(\vec{k}_{i}+\vec{k}_{j})\cdot(\vec{x}-\vec{y})}}{(2\pi)^{6}}[h_{i}^{T}(i\sigma_{2})h_{j}][h_{j}^{\dagger}(-i\sigma_{2})h_{i}^{*}]F_{ij,x}F_{ij,y}^{*}\label{eq:psipsipsipsi}
\end{eqnarray}
 where 
\begin{eqnarray}
F_{ij,x} & = & r_{i}u_{B,i,x}u_{A,j,x}+(i\leftrightarrow j)\\
F_{ij,x}F_{ij,y}^{*} & = & 2[r_{i}u_{B,i,x}u_{A,j,x}+(i\leftrightarrow j)](r_{i}u_{B,i,y}^{*}u_{A,j,y}^{*})\\
 & = & 2[u_{B,i,x}u_{A,j,x}u_{B,i,y}^{*}u_{A,j,y}^{*}+r_{i}r_{j}u_{B,i,x}u_{A,j,x}u_{B,j,y}^{*}u_{A,i,y}^{*}].\label{eq:FF}
\end{eqnarray}
 We note that in Eq.~(\ref{eq:psipsipsipsi}), the factor $e^{i(\vec{k}_{i}+\vec{k}_{j})\cdot(\vec{x}-\vec{y})}$
after angular average is real, and the factor $[h_{i}^{T}(i\sigma_{2})h_{j}][h_{j}^{\dagger}(-i\sigma_{2})h_{i}^{*}]=|[h_{i}^{T}(i\sigma_{2})h_{j}]|^{2}$
is also real, thus the imaginary and real part of $F_{ij,x}F_{ij,y}^{*}$
correspond to the commutator and anti-commutator respectively.

Next, we consider the two terms in Eq.~(\ref{eq:FF}) one by one,
using explicit expression of Eq.~(\ref{eq:dS-mode-end-1}) to get
\begin{eqnarray}
u_{B,i,x}u_{A,j,x}u_{B,i,y}^{*}u_{A,j,y}^{*} & = & \sqrt{\frac{\pi}{4}\frac{k_{i}}{a_{x}H}}\sqrt{\frac{\pi}{4}\frac{k_{j}}{a_{x}H}}\sqrt{\frac{\pi}{4}\frac{k_{i}}{a_{y}H}}\sqrt{\frac{\pi}{4}\frac{k_{j}}{a_{y}H}}\nonumber \\
 &  & (J_{+,i,x}+iY_{+,i,x})(J_{-,j,x}+iY_{-,j,x})(J_{-,i,y}-iY_{-,i,y})(J_{+,j,y}-iY_{+,j,y})
\end{eqnarray}
 where 
\begin{equation}
J_{\pm,i,x}=J_{\frac{1}{2}\pm i\frac{m}{H}}(\frac{k_{i}}{a_{x}H}),\quad Y_{\pm,i,x}=Y_{\frac{1}{2}\pm i\frac{m}{H}}(\frac{k_{i}}{a_{x}H}).
\end{equation}
 Using the small $z$ expansion of Bessel function again, where $\mbox{Re }(\nu)=\frac{1}{2}$
in all the cases, we can extract its scaling behavior under $a\rightarrow\lambda a$,
\begin{eqnarray}
 &  & (J_{+,i,x}+iY_{+,i,x})(J_{-,j,x}+iY_{-,j,x})(J_{-,i,y}-iY_{-,i,y})(J_{+,j,y}-iY_{+,j,y})\nonumber \\
 & = & Y_{+,i,x}Y_{-,j,x}Y_{-,i,y}Y_{+,j,y}\cdots\cdots\propto\lambda^{2},\mbox{real}\nonumber \\
 &  & -iJ_{+,i,x}Y_{-,j,x}Y_{-,i,y}Y_{+,j,y}-iY_{+,i,x}J_{-,j,x}Y_{-,i,y}Y_{+,j,y}\cdots\cdots\propto\lambda^{1},\mbox{imaginary}\nonumber \\
 &  & +iY_{+,i,x}Y_{-,j,x}J_{-,i,y}Y_{+,j,y}+iY_{+,i,x}Y_{-,j,x}Y_{-,i,y}J_{+,j,y}\cdots\cdots\propto\lambda^{1},\mbox{imaginary}\nonumber \\
 &  & +\mbox{terms subdominant in \ensuremath{\lambda}expansion.}
\end{eqnarray}
 Thus the imaginary part is suppressed by $\lambda^{-1}$ relative
to the real part. We can do similar analysis to the second part $r_{i}r_{j}u_{B,i,x}u_{A,j,x}u_{B,j,y}^{*}u_{A,i,y}^{*}$
in Eq.~(\ref{eq:FF}) and found the same behavior. Thus, for $\bar{\psi}\psi$
operator, we have the following scaling law 
\begin{eqnarray}
\langle\{\bar{\psi}\psi_{x},\bar{\psi}\psi_{y}\}\rangle_{IR} & \propto & \lambda^{-6}\\
\langle[\bar{\psi}\psi_{x},\bar{\psi}\psi_{y}]\rangle_{IR} & \propto & \lambda^{-7}.
\end{eqnarray}

Thus, we see the commutator for $\bar{\psi}\psi$ gives additional
suppression of $a^{-1}$ factor compared with the anti-commutator,
whereas the commutator for $\sigma$ and $\zeta$ gives additional
suppression of $a^{-3}$ factor.

\section{Explicit check of the mass insertion formula}

In this section, we show that the particle production part of the
following equation holds using the adiabatic subtraction. 
\begin{equation}
-i\int^{y}(dw)\langle[\bar{\psi}\psi_{x},\bar{\psi}\psi_{z}]\rangle=\partial_{m}\langle\bar{\psi}\psi_{x}\rangle=\partial_{m}n_{\Psi}(x)\label{eq:mass-insertion}
\end{equation}

Expressing both side of Eq.(\ref{eq:mass-insertion}) using the mode
sum, we see the left hand side is 
\begin{equation}
-i\int^{y}(dw)\langle[\bar{\psi}\psi_{x},\bar{\psi}\psi_{w}]\rangle=\frac{16}{a_{x}^{3}}\int^{y^{0}}dw^{0}\, a_{w}\int\frac{d^{3}k}{(2\pi)^{3}}\mbox{Im}[(u_{A,k}u_{B,k})_{x}(u_{A,k}u_{B,k})_{w}^{*}]\label{eq:LHS}
\end{equation}
 and the right hand side is 
\begin{equation}
\partial_{m}\langle\bar{\psi}\psi_{x}\rangle=\frac{2}{a_{x}^{3}}\int\frac{d^{3}k}{(2\pi)^{3}}\partial_{m}(|u_{B}|^{2}-|u_{A}|^{2})\label{eq:RHS}
\end{equation}
 Thus, we only need to check for each given $k$, the following equation
is right 
\begin{equation}
\partial_{m}(|u_{B}|^{2}-|u_{A}|^{2})=8\int^{y^{0}}dw^{0}\, a_{w}\mbox{Im}[(u_{A,k}u_{B,k})_{x}(u_{A,k}u_{B,k})_{w}^{*}]
\end{equation}
 From the left hand side, we have 
\begin{equation}
\partial_{m}(|u_{B}|^{2}-|u_{A}|^{2})=-2\mbox{Re}\left[\left(\begin{array}{cc}
u_{A}^{*} & u_{B}^{*}\end{array}\right)\sigma_{3}\frac{\partial}{\partial m}\left(\begin{array}{c}
u_{A}\\
u_{B}
\end{array}\right)_{k,x}\right]
\end{equation}
 and upon expressing mode function at time $x^{0}$ in term of evolution
operator acting on the initial value, we have 
\begin{eqnarray}
\frac{\partial}{\partial m}\left(\begin{array}{c}
u_{A}\\
u_{B}
\end{array}\right)_{k,x} & = & -i\int_{\eta_{i}}^{x^{0}}dz^{0}\, U(x^{0}\leftarrow z^{0})\,\frac{\partial}{\partial m}\left(\begin{array}{cc}
am & k\\
k & -am
\end{array}\right)\, U(z^{0}\leftarrow\eta_{i})\left(\begin{array}{c}
u_{A}\\
u_{B}
\end{array}\right)_{k,i}
\end{eqnarray}
 Combining these two expression, we can obtain the desired result
after some algebra.

However, the remaining $d^{3}k$ integrals in Eq.~(\ref{eq:LHS})
and Eq.~(\ref{eq:RHS}) are UV divergent. To make them finite, we
express both side in terms of Bogoliubov coefficients and dropped
the pure vacuum contribution to get 
\begin{eqnarray}
-i\int^{x^{0}}(dw)\langle[\bar{\psi}\psi_{x},\bar{\psi}\psi_{w}]\rangle & \approx & 16\int\frac{d^{3}k}{(2\pi a_{x})^{3}}(\frac{am}{\omega_{k}})_{x}\int^{x}d\eta_{w}\, a_{w}(\frac{am}{\omega})_{w}\mbox{Im}[(\alpha\beta)_{x}(\alpha\beta)_{w}^{*}]\\
\partial_{m}\langle\bar{\psi}\psi_{x}\rangle & \approx & \frac{2}{a_{x}^{3}}\int\frac{d^{3}k}{(2\pi)^{3}}\partial_{m}[2|\beta_{k,x}|^{2}\frac{a_{x}m}{\omega_{k,x}}]\approx\frac{4}{a_{x}^{3}}\int\frac{d^{3}k}{(2\pi)^{3}}(\frac{a_{x}m}{\omega_{k,x}})\partial_{m}|\beta_{k,x}|^{2}
\end{eqnarray}
 Now, we only need to check 
\begin{equation}
\partial_{m}|\beta_{k,x}|^{2}=4\int^{x}d\eta_{w}\, a_{w}(\frac{am}{\omega})_{w}\mbox{Im}[(\alpha\beta)_{x}(\alpha\beta)_{w}^{*}]
\end{equation}
 Suppose, $x^{0}$ is late enough such that $\beta_{k,x}$ is constant
and equals to its value at asymptotic future $\beta_{k}$, then we
get 
\begin{eqnarray}
\partial_{m}|\beta_{k}|^{2} & = & 4\int_{\eta_{i}}^{x^{0}}dz^{0}a_{z}\frac{am}{\omega}\mbox{Im}(\alpha_{k}\beta_{k})_{x}(\alpha\beta)_{z}^{*}
\end{eqnarray}
 Thus, Eq.~(\ref{eq:mass-insertion}) is compatible with the Bogoliubov
projection.

\section{Gravitational Interaction}

\label{sub:model}

Here we derive the gravitational interaction. Consider the action
\begin{eqnarray}
S & = & S_{EH}+S_{\phi}+S_{\sigma}+S_{\psi}\\
 & = & \int(dx)\left\{ \frac{1}{2}M_{p}^{2}R+[-\frac{1}{2}g^{\mu\nu}\partial_{\mu}\phi\partial_{\nu}\phi-V(\phi)]+[-\frac{1}{2}g^{\mu\nu}\partial_{\mu}\sigma\partial_{\nu}\sigma-\frac{1}{2}m_{\sigma}^{2}\sigma^{2}]\right.\nonumber \\
 &  & \left.+\bar{\psi}(i\gamma^{\mu}\nabla_{\mu}-m_{\psi})\psi-\lambda\sigma\bar{\psi}\psi\right\} ,\label{eq:full_Lag-2}
\end{eqnarray}
where $M_{p}^{2}=\frac{1}{8\pi G}=1$. The metric is given in ADM
formalism%
\footnote{We use $(-+++)$ sign convention for the metric, and physical time
$t$ .%
} \cite{Arnowitt:1962} by 
\begin{equation}
g_{\mu\nu}=\left(\begin{array}{cc}
-N^{2}+h_{ij}N^{i}N^{j} & h_{ij}N^{j}\\
h_{ij}N^{j} & h_{ij}
\end{array}\right),\quad g^{\mu\nu}=\left(\begin{array}{cc}
-N^{-2} & N^{i}N^{-2}\\
N^{i}N^{-2} & h^{ij}-N^{i}N^{j}N^{-2}
\end{array}\right),
\end{equation}
where $h_{ij}$ is the metric tensor on the constant time hypersurface,
and $h^{ij}$ is the inverse metric. We use Latin indices $i,j\cdots$
for objects on the 3-dimensional constant time hypersurface, and we
use $h_{ij}$ and $h^{ij}$ to raise and lower the indices. Then we
use the Hamiltonian and the momentum constraints to determine the
lapse function $N$ and the shift vector $N^{i}$: 
\begin{eqnarray}
0 & = & \frac{1}{N}[R^{(3)}-\frac{1}{N^{2}}(E_{ij}E^{ij}-E^{2})]-2NT^{00}\label{eq:H_constraint}\\
0 & = & \frac{2}{N}\nabla_{i}^{(3)}[\frac{1}{N}(E^{ij}-Eh^{ij})]+2N^{j}T^{00}+2T^{0j},\label{eq:P_constraint}
\end{eqnarray}
where $T^{\mu\nu}$ is the total matter stress tensor, $R^{(3)}$
is the Ricci scalar calculated with the three-metric $h_{ij}$, and
\begin{eqnarray}
E_{ij} & = & \frac{1}{2}(\dot{h}_{ij}-\nabla_{i}^{(3)}N_{j}-\nabla_{j}^{(3)}N_{i}).\\
E & = & E_{ij}h^{ij}.
\end{eqnarray}

In order to consider the perturbation around the background configuration
\begin{equation}
\phi^{(0)}=\bar{\phi}(t),\quad\sigma^{(0)}=0,\quad g_{\mu\nu}^{(0)}=\left(\begin{array}{cc}
-1 & 0\\
0 & a^{2}(t)\delta_{ij}
\end{array}\right)
\end{equation}
where the background fields satisfy the background equations of motion
\begin{eqnarray}
3H^{2} & = & \frac{1}{2}\dot{\bar{\phi}}^{2}+V(\bar{\phi})\\
\dot{H} & = & -\frac{1}{2}\dot{\bar{\phi}}^{2}\\
\ddot{\bar{\phi}}+3H\dot{\bar{\phi}}+V'(\bar{\phi}) & = & 0,
\end{eqnarray}
we choose the comoving gauge, defined by %
\footnote{In this section, Latin indices $i,j$ are raised and lowered by $\delta_{ij}$,
and repeated indices are contracted.%
} 
\begin{equation}
\delta\phi=0,\quad\gamma_{ii}=0,\quad\partial_{i}\gamma_{ij}=0
\end{equation}
where 
\begin{equation}
h_{ij}=a^{2}(t)[e^{\Gamma}]_{ij},\quad\Gamma_{ij}=2\zeta\delta_{ij}+\gamma_{ij}.\label{eq:h_ij}
\end{equation}
Then we solve the constraint equations (\ref{eq:H_constraint}) and
(\ref{eq:P_constraint}) perturbatively using $\zeta$ and $\gamma$,
and putting their solutions for $N$ and $N^{i}$ back into the action,
we can get the perturbed action: 
\begin{equation}
S^{(C)}=S_{\zeta\zeta}^{(C)}+S_{\sigma\sigma}^{(C)}+S_{\psi\psi}^{(C)}+S_{\gamma\gamma}^{(C)}+S_{\zeta\zeta\zeta}^{(C)}+S_{\zeta\sigma\sigma}^{(C)}+S_{\zeta\psi\psi}^{(C)}+S_{\zeta\sigma\sigma}^{(C)}\cdots.
\end{equation}
For the interaction terms $S_{\zeta\sigma\sigma}^{(C)}$ and $S_{\zeta\psi\psi}^{(C)}$,
we need the solutions of $N$ and $N^{i}$ up to linear order in $\zeta$
\begin{equation}
N^{(1,C)}=1+\frac{\dot{\zeta}}{H},\quad N_{i}^{(1,C)}=\partial_{i}[-\frac{\zeta}{H}+\epsilon\frac{a^{2}}{\nabla^{2}}\dot{\zeta}],
\end{equation}
where $\epsilon\equiv\dot{H}/H^{2}.$ Hence, the metric perturbations
becomes 
\begin{equation}
\delta g_{\mu\nu}^{(C)}=\left(\begin{array}{cc}
-2\frac{\dot{\zeta}}{H} & (-\frac{\zeta}{H}+\epsilon\frac{a^{2}}{\nabla^{2}}\dot{\zeta})_{,i}\\
(-\frac{\zeta}{H}+\epsilon\frac{a^{2}}{\nabla^{2}}\dot{\zeta})_{,i} & a^{2}\left(\delta_{ij}2\zeta+\gamma_{ij}\right)
\end{array}\right),\label{eq:del_g_C-1}
\end{equation}
and we have the $\zeta$-matter cubic interaction action 
\begin{equation}
S_{\zeta\sigma\sigma}^{(C)}+S_{\zeta\psi\psi}^{(C)}=\frac{1}{2}\int d^{4}x\sqrt{-g}\left(T_{\sigma}^{\mu\nu}+T_{\psi}^{\mu\nu}\right)\delta g_{\mu\nu}^{(C)},\label{eq:Interaction_ZPsiPsi}
\end{equation}
where $T_{\sigma}^{\mu\nu}$ and $T_{\psi}^{\mu\nu}$ is the stress
energy tensors for $\sigma$ and $\psi$, respectively, which are
written as 
\begin{eqnarray}
T_{\sigma}^{\mu\nu} & = & g^{\mu\alpha}g^{\nu\beta}\partial_{\alpha}\sigma\partial_{\beta}\sigma+g^{\mu\nu}\mathcal{L}_{\sigma},\label{eq:scal-stress}\\
T_{\psi}^{\mu\nu} & = & -\frac{i}{2}[\bar{\psi}\gamma^{(\mu}\nabla^{\nu)}\psi-\nabla^{(\mu}(\bar{\psi})\gamma^{\nu)}\psi]+g^{\mu\nu}\mbox{Re}\left(\mathcal{L}_{\psi}\right).
\end{eqnarray}
 Particularly, up to the cubic interaction, $\mathcal{L}_{int}=-\mathcal{H}_{int}$.
Thus $S_{\zeta\sigma\sigma}^{(C)}+S_{\zeta\psi\psi}^{(C)}=-\int dt\, H_{\zeta\sigma\sigma}(t)+H_{\zeta\psi\psi}(t).$

\bibliographystyle{JHEP}
\bibliography{ref,misc,ConsistencyRelation,Curvaton,Non-Gaussianity,Inflation_general,wimpzilla,deltaN,gravitino_isocurvature}

\providecommand{\href}[2]{#2}\begingroup\raggedright\begin{thebibliography}{100}

\bibitem{Ade:2013xsa}
{\bf Planck Collaboration} Collaboration, P.~Ade {\em et.~al.}, {\it {Planck
  2013 results. I. Overview of products and scientific results}},
  \href{http://xxx.lanl.gov/abs/1303.5062}{{\tt arXiv:1303.5062}}.

\bibitem{Ade:2013lta}
{\bf Planck Collaboration} Collaboration, P.~Ade {\em et.~al.}, {\it {Planck
  2013 results. XVI. Cosmological parameters}},
  \href{http://xxx.lanl.gov/abs/1303.5076}{{\tt arXiv:1303.5076}}.

\bibitem{Ade:2013rta}
{\bf Planck Collaboration} Collaboration, P.~Ade {\em et.~al.}, {\it {Planck
  2013 results. XXII. Constraints on inflation}},
  \href{http://xxx.lanl.gov/abs/1303.5082}{{\tt arXiv:1303.5082}}.

\bibitem{Ade:2013sta}
{\bf Planck Collaboration} Collaboration, P.~Ade {\em et.~al.}, {\it {Planck
  2013 results. XXIII. Isotropy and Statistics of the CMB}},
  \href{http://xxx.lanl.gov/abs/1303.5083}{{\tt arXiv:1303.5083}}.

\bibitem{Ade:2013ydc}
{\bf Planck Collaboration} Collaboration, P.~Ade {\em et.~al.}, {\it {Planck
  2013 Results. XXIV. Constraints on primordial non-Gaussianity}},
  \href{http://xxx.lanl.gov/abs/1303.5084}{{\tt arXiv:1303.5084}}.

\bibitem{Hinshaw:2012fq}
G.~Hinshaw, D.~Larson, E.~Komatsu, D.~Spergel, C.~Bennett, {\em et.~al.}, {\it
  {Nine-Year Wilkinson Microwave Anisotropy Probe (WMAP) Observations:
  Cosmological Parameter Results}},
  \href{http://xxx.lanl.gov/abs/1212.5226}{{\tt arXiv:1212.5226}}.

\bibitem{Komatsu:2010fb}
{\bf WMAP Collaboration} Collaboration, E.~Komatsu {\em et.~al.}, {\it
  {Seven-Year Wilkinson Microwave Anisotropy Probe (WMAP) Observations:
  Cosmological Interpretation}},  {\em Astrophys.J.Suppl.} {\bf 192} (2011) 18,
  [\href{http://xxx.lanl.gov/abs/1001.4538}{{\tt arXiv:1001.4538}}].

\bibitem{Brown:2009uy}
{\bf QUaD collaboration} Collaboration, .~M. Brown {\em et.~al.}, {\it
  {Improved measurements of the temperature and polarization of the CMB from
  QUaD}},  {\em Astrophys.J.} {\bf 705} (2009) 978--999,
  [\href{http://xxx.lanl.gov/abs/0906.1003}{{\tt arXiv:0906.1003}}].

\bibitem{Reichardt:2008ay}
C.~Reichardt, P.~Ade, J.~Bock, J.~R. Bond, J.~Brevik, {\em et.~al.}, {\it {High
  resolution CMB power spectrum from the complete ACBAR data set}},  {\em
  Astrophys.J.} {\bf 694} (2009) 1200--1219,
  [\href{http://xxx.lanl.gov/abs/0801.1491}{{\tt arXiv:0801.1491}}].

\bibitem{Fowler:2010cy}
{\bf ACT Collaboration} Collaboration, J.~Fowler {\em et.~al.}, {\it {The
  Atacama Cosmology Telescope: A Measurement of the 600 ell 8000 Cosmic
  Microwave Background Power Spectrum at 148 GHz}},  {\em Astrophys.J.} {\bf
  722} (2010) 1148--1161, [\href{http://xxx.lanl.gov/abs/1001.2934}{{\tt
  arXiv:1001.2934}}].

\bibitem{Lueker:2009rx}
M.~Lueker, C.~Reichardt, K.~Schaffer, O.~Zahn, P.~Ade, {\em et.~al.}, {\it
  {Measurements of Secondary Cosmic Microwave Background Anisotropies with the
  South Pole Telescope}},  {\em Astrophys.J.} {\bf 719} (2010) 1045--1066,
  [\href{http://xxx.lanl.gov/abs/0912.4317}{{\tt arXiv:0912.4317}}].

\bibitem{Percival:2007yw}
W.~J. Percival, S.~Cole, D.~J. Eisenstein, R.~C. Nichol, J.~A. Peacock, {\em
  et.~al.}, {\it {Measuring the Baryon Acoustic Oscillation scale using the
  SDSS and 2dFGRS}},  {\em Mon.Not.Roy.Astron.Soc.} {\bf 381} (2007)
  1053--1066, [\href{http://xxx.lanl.gov/abs/0705.3323}{{\tt
  arXiv:0705.3323}}].

\bibitem{Eisenstein:2005su}
{\bf SDSS Collaboration} Collaboration, D.~J. Eisenstein {\em et.~al.}, {\it
  {Detection of the baryon acoustic peak in the large-scale correlation
  function of SDSS luminous red galaxies}},  {\em Astrophys.J.} {\bf 633}
  (2005) 560--574, [\href{http://xxx.lanl.gov/abs/astro-ph/0501171}{{\tt
  astro-ph/0501171}}].

\bibitem{Starobinsky:1980te}
A.~A. Starobinsky, {\it {A New Type of Isotropic Cosmological Models Without
  Singularity}},  {\em Phys.Lett.} {\bf B91} (1980) 99--102.

\bibitem{Sato:1980yn}
K.~Sato, {\it {First Order Phase Transition of a Vacuum and Expansion of the
  Universe}},  {\em Mon.Not.Roy.Astron.Soc.} {\bf 195} (1981) 467--479.

\bibitem{Linde:1981mu}
A.~D. Linde, {\it {A New Inflationary Universe Scenario: A Possible Solution of
  the Horizon, Flatness, Homogeneity, Isotropy and Primordial Monopole
  Problems}},  {\em Phys.Lett.} {\bf B108} (1982) 389--393.

\bibitem{Mukhanov:1981xt}
V.~F. Mukhanov and G.~Chibisov, {\it {Quantum Fluctuation and Nonsingular
  Universe. (In Russian)}},  {\em JETP Lett.} {\bf 33} (1981) 532--535.

\bibitem{Albrecht:1982wi}
A.~Albrecht and P.~J. Steinhardt, {\it {Cosmology for Grand Unified Theories
  with Radiatively Induced Symmetry Breaking}},  {\em Phys.Rev.Lett.} {\bf 48}
  (1982) 1220--1223.

\bibitem{Hawking:1982my}
S.~Hawking and I.~Moss, {\it {FLUCTUATIONS IN THE INFLATIONARY UNIVERSE}},
  {\em Nucl.Phys.} {\bf B224} (1983) 180.

\bibitem{Guth1982}
A.~H. Guth and S.~Pi, {\it {Fluctuations in the New Inflationary Universe}},
  {\em Phys.Rev.Lett.} {\bf 49} (1982) 1110--1113.

\bibitem{Starobinsky:1982ee}
A.~A. Starobinsky, {\it {Dynamics of Phase Transition in the New Inflationary
  Universe Scenario and Generation of Perturbations}},  {\em Phys.Lett.} {\bf
  B117} (1982) 175--178.

\bibitem{Bardeen:1983qw}
J.~M. Bardeen, P.~J. Steinhardt, and M.~S. Turner, {\it {Spontaneous Creation
  of Almost Scale - Free Density Perturbations in an Inflationary Universe}},
  {\em Phys.Rev.} {\bf D28} (1983) 679.

\bibitem{Silk:1986vc}
J.~Silk and M.~S. Turner, {\it {Double Inflation}},  {\em Phys.Rev.} {\bf D35}
  (1987) 419.

\bibitem{Polarski:1994rz}
D.~Polarski and A.~A. Starobinsky, {\it {Isocurvature perturbations in multiple
  inflationary models}},  {\em Phys.Rev.} {\bf D50} (1994) 6123--6129,
  [\href{http://xxx.lanl.gov/abs/astro-ph/9404061}{{\tt astro-ph/9404061}}].

\bibitem{Langlois:1999dw}
D.~Langlois, {\it {Correlated adiabatic and isocurvature perturbations from
  double inflation}},  {\em Phys.Rev.} {\bf D59} (1999) 123512,
  [\href{http://xxx.lanl.gov/abs/astro-ph/9906080}{{\tt astro-ph/9906080}}].

\bibitem{Yamaguchi:2001zh}
M.~Yamaguchi, {\it {Density fluctuations and primordial black holes formation
  in natural double inflation in supergravity}},  {\em Phys.Rev.} {\bf D64}
  (2001) 063503, [\href{http://xxx.lanl.gov/abs/hep-ph/0105001}{{\tt
  hep-ph/0105001}}].

\bibitem{Lyth:2001nq}
D.~H. Lyth and D.~Wands, {\it {Generating the curvature perturbation without an
  inflaton}},  {\em Phys.Lett.} {\bf B524} (2002) 5--14,
  [\href{http://xxx.lanl.gov/abs/hep-ph/0110002}{{\tt hep-ph/0110002}}].

\bibitem{Enqvist:2001zp}
K.~Enqvist and M.~S. Sloth, {\it {Adiabatic CMB perturbations in pre - big bang
  string cosmology}},  {\em Nucl.Phys.} {\bf B626} (2002) 395--409,
  [\href{http://xxx.lanl.gov/abs/hep-ph/0109214}{{\tt hep-ph/0109214}}].

\bibitem{Moroi:2001ct}
T.~Moroi and T.~Takahashi, {\it {Effects of cosmological moduli fields on
  cosmic microwave background}},  {\em Phys.Lett.} {\bf B522} (2001) 215--221,
  [\href{http://xxx.lanl.gov/abs/hep-ph/0110096}{{\tt hep-ph/0110096}}].

\bibitem{Lyth:2002my}
D.~H. Lyth, C.~Ungarelli, and D.~Wands, {\it {The primordial density
  perturbation in the curvaton scenario}},  {\em Phys. Rev.} {\bf D67} (2003)
  023503, [\href{http://xxx.lanl.gov/abs/astro-ph/0208055}{{\tt
  astro-ph/0208055}}].

\bibitem{Langlois:2013dh}
D.~Langlois and T.~Takahashi, {\it {Density Perturbations from Modulated Decay
  of the Curvaton}},  \href{http://xxx.lanl.gov/abs/1301.3319}{{\tt
  arXiv:1301.3319}}.

\bibitem{Enqvist:2012tc}
K.~Enqvist, D.~G. Figueroa, and R.~N. Lerner, {\it {Curvaton Decay by Resonant
  Production of the Standard Model Higgs}},  {\em JCAP} {\bf 1301} (2013) 040,
  [\href{http://xxx.lanl.gov/abs/1211.5028}{{\tt arXiv:1211.5028}}].

\bibitem{Harigaya:2012up}
K.~Harigaya, M.~Ibe, M.~Kawasaki, and T.~T. Yanagida, {\it {Non-Gaussianity
  from Attractor Curvaton}},  \href{http://xxx.lanl.gov/abs/1211.3535}{{\tt
  arXiv:1211.3535}}.

\bibitem{Enomoto:2012uy}
S.~Enomoto, K.~Kohri, and T.~Matsuda, {\it {Non-Gaussianity in the unified
  curvaton mechanism : The generalized curvaton mechanism that comprehends
  modulation at the transition}},
  \href{http://xxx.lanl.gov/abs/1210.7118}{{\tt arXiv:1210.7118}}.

\bibitem{Enqvist:2012xn}
K.~Enqvist, R.~N. Lerner, O.~Taanila, and A.~Tranberg, {\it {Spectator field
  dynamics in de Sitter and curvaton initial conditions}},  {\em JCAP} {\bf
  1210} (2012) 052, [\href{http://xxx.lanl.gov/abs/1205.5446}{{\tt
  arXiv:1205.5446}}].

\bibitem{Dimopoulos:2011gb}
K.~Dimopoulos, K.~Kohri, D.~H. Lyth, and T.~Matsuda, {\it {The inflating
  curvaton}},  {\em JCAP} {\bf 1203} (2012) 022,
  [\href{http://xxx.lanl.gov/abs/1110.2951}{{\tt arXiv:1110.2951}}].

\bibitem{Alabidi:2010ba}
L.~Alabidi, K.~Malik, C.~T. Byrnes, and K.-Y. Choi, {\it {How the curvaton
  scenario, modulated reheating and an inhomogeneous end of inflation are
  related}},  {\em JCAP} {\bf 1011} (2010) 037,
  [\href{http://xxx.lanl.gov/abs/1002.1700}{{\tt arXiv:1002.1700}}].

\bibitem{Lin:2010ua}
C.~Lin and Y.~Wang, {\it {Quadra-Spectrum and Quint-Spectrum from Inflation and
  Curvaton Models}},  {\em JCAP} {\bf 1007} (2010) 011,
  [\href{http://xxx.lanl.gov/abs/1004.0461}{{\tt arXiv:1004.0461}}].

\bibitem{Assadullahi:2007uw}
H.~Assadullahi, J.~Valiviita, and D.~Wands, {\it {Primordial non-Gaussianity
  from two curvaton decays}},  {\em Phys.Rev.} {\bf D76} (2007) 103003,
  [\href{http://xxx.lanl.gov/abs/0708.0223}{{\tt arXiv:0708.0223}}].

\bibitem{Moroi:2002rd}
T.~Moroi and T.~Takahashi, {\it {Cosmic density perturbations from late
  decaying scalar condensations}},  {\em Phys.Rev.} {\bf D66} (2002) 063501,
  [\href{http://xxx.lanl.gov/abs/hep-ph/0206026}{{\tt hep-ph/0206026}}].

\bibitem{Bartolo:2002vf}
N.~Bartolo and A.~R. Liddle, {\it {The Simplest curvaton model}},  {\em
  Phys.Rev.} {\bf D65} (2002) 121301,
  [\href{http://xxx.lanl.gov/abs/astro-ph/0203076}{{\tt astro-ph/0203076}}].

\bibitem{Seckel:1985tj}
D.~Seckel and M.~S. Turner, {\it {Isothermal Density Perturbations in an Axion
  Dominated Inflationary Universe}},  {\em Phys.Rev.} {\bf D32} (1985) 3178.

\bibitem{Preskill:1982cy}
J.~Preskill, M.~B. Wise, and F.~Wilczek, {\it {Cosmology of the Invisible
  Axion}},  {\em Phys.Lett.} {\bf B120} (1983) 127--132.

\bibitem{Abbott:1982af}
L.~Abbott and P.~Sikivie, {\it {A Cosmological Bound on the Invisible Axion}},
  {\em Phys.Lett.} {\bf B120} (1983) 133--136.

\bibitem{Dine:1982ah}
M.~Dine and W.~Fischler, {\it {The Not So Harmless Axion}},  {\em Phys.Lett.}
  {\bf B120} (1983) 137--141.

\bibitem{Steinhardt:1983ia}
P.~J. Steinhardt and M.~S. Turner, {\it {Saving the Invisible Axion}},  {\em
  Phys.Lett.} {\bf B129} (1983) 51.

\bibitem{Turner:1985si}
M.~S. Turner, {\it {Cosmic and Local Mass Density of Invisible Axions}},  {\em
  Phys.Rev.} {\bf D33} (1986) 889--896.

\bibitem{Kolb:1990vq}
E.~W. Kolb and M.~S. Turner, {\it {The Early universe}},  {\em Front.Phys.}
  {\bf 69} (1990) 1--547.

\bibitem{Fox:2004kb}
P.~Fox, A.~Pierce, and S.~D. Thomas, {\it {Probing a QCD string axion with
  precision cosmological measurements}},
  \href{http://xxx.lanl.gov/abs/hep-th/0409059}{{\tt hep-th/0409059}}.

\bibitem{Beltran:2006sq}
M.~Beltran, J.~Garcia-Bellido, and J.~Lesgourgues, {\it {Isocurvature bounds on
  axions revisited}},  {\em Phys.Rev.} {\bf D75} (2007) 103507,
  [\href{http://xxx.lanl.gov/abs/hep-ph/0606107}{{\tt hep-ph/0606107}}].

\bibitem{Hertzberg:2008wr}
M.~P. Hertzberg, M.~Tegmark, and F.~Wilczek, {\it {Axion Cosmology and the
  Energy Scale of Inflation}},  {\em Phys.Rev.} {\bf D78} (2008) 083507,
  [\href{http://xxx.lanl.gov/abs/0807.1726}{{\tt arXiv:0807.1726}}].

\bibitem{Linde:1996gt}
A.~D. Linde and V.~F. Mukhanov, {\it {Nongaussian isocurvature perturbations
  from inflation}},  {\em Phys. Rev.} {\bf D56} (1997) 535--539,
  [\href{http://xxx.lanl.gov/abs/astro-ph/9610219}{{\tt astro-ph/9610219}}].

\bibitem{Chung:1998zb}
D.~J. Chung, E.~W. Kolb, and A.~Riotto, {\it {Superheavy dark matter}},  {\em
  Phys.Rev.} {\bf D59} (1999) 023501,
  [\href{http://xxx.lanl.gov/abs/hep-ph/9802238}{{\tt hep-ph/9802238}}]. In
  *Venice 1999, Neutrino telescopes, vol. 2* 217-237.

\bibitem{Chung:2004nh}
D.~J. Chung, E.~W. Kolb, A.~Riotto, and L.~Senatore, {\it {Isocurvature
  constraints on gravitationally produced superheavy dark matter}},  {\em
  Phys.Rev.} {\bf D72} (2005) 023511,
  [\href{http://xxx.lanl.gov/abs/astro-ph/0411468}{{\tt astro-ph/0411468}}].

\bibitem{Chung:2011xd}
D.~J. Chung and H.~Yoo, {\it {Isocurvature Perturbations and Non-Gaussianity of
  Gravitationally Produced Nonthermal Dark Matter}},
  \href{http://xxx.lanl.gov/abs/1110.5931}{{\tt arXiv:1110.5931}}.

\bibitem{Bartolo:2001cw}
N.~Bartolo, S.~Matarrese, and A.~Riotto, {\it {Nongaussianity from inflation}},
   {\em Phys.Rev.} {\bf D65} (2002) 103505,
  [\href{http://xxx.lanl.gov/abs/hep-ph/0112261}{{\tt hep-ph/0112261}}].

\bibitem{Kawasaki:2008sn}
M.~Kawasaki, K.~Nakayama, T.~Sekiguchi, T.~Suyama, and F.~Takahashi, {\it
  {Non-Gaussianity from isocurvature perturbations}},  {\em JCAP} {\bf 0811}
  (2008) 019, [\href{http://xxx.lanl.gov/abs/0808.0009}{{\tt
  arXiv:0808.0009}}].

\bibitem{Langlois:2008vk}
D.~Langlois, F.~Vernizzi, and D.~Wands, {\it {Non-linear isocurvature
  perturbations and non-Gaussianities}},  {\em JCAP} {\bf 0812} (2008) 004,
  [\href{http://xxx.lanl.gov/abs/0809.4646}{{\tt arXiv:0809.4646}}].

\bibitem{Kofman:1989ed}
L.~Kofman, G.~R. Blumenthal, H.~Hodges, and J.~R. Primack, {\it {GENERATION OF
  NONFLAT AND NONGAUSSIAN PERTURBATIONS FROM INFLATION}},  {\em ASP Conf.Ser.}
  {\bf 15} (1991) 339--351.

\bibitem{Geyer:2004bx}
B.~Geyer, D.~Robaschik, and J.~Eilers, {\it {Target mass corrections for
  virtual Compton scattering at twist-2 and generalized, non-forward
  Wandzura-Wilczek and Callan-Gross relations}},  {\em Nucl.Phys.} {\bf B704}
  (2005) 279--331, [\href{http://xxx.lanl.gov/abs/hep-ph/0407300}{{\tt
  hep-ph/0407300}}].

\bibitem{Ferrer:2004nv}
F.~Ferrer, S.~Rasanen, and J.~Valiviita, {\it {Correlated isocurvature
  perturbations from mixed inflaton-curvaton decay}},  {\em JCAP} {\bf 0410}
  (2004) 010, [\href{http://xxx.lanl.gov/abs/astro-ph/0407300}{{\tt
  astro-ph/0407300}}].

\bibitem{Boubekeur:2005fj}
L.~Boubekeur and D.~Lyth, {\it {Detecting a small perturbation through its
  non-Gaussianity}},  {\em Phys.Rev.} {\bf D73} (2006) 021301,
  [\href{http://xxx.lanl.gov/abs/astro-ph/0504046}{{\tt astro-ph/0504046}}].

\bibitem{Barbon:2006us}
J.~Barbon and C.~Hoyos-Badajoz, {\it {Dynamical Higgs potentials with a
  landscape}},  {\em Phys.Rev.} {\bf D73} (2006) 126002,
  [\href{http://xxx.lanl.gov/abs/hep-th/0602285}{{\tt hep-th/0602285}}].

\bibitem{Lyth:2006gd}
D.~H. Lyth, {\it {Non-gaussianity and cosmic uncertainty in curvaton-type
  models}},  {\em JCAP} {\bf 0606} (2006) 015,
  [\href{http://xxx.lanl.gov/abs/astro-ph/0602285}{{\tt astro-ph/0602285}}].

\bibitem{Koyama:2007if}
K.~Koyama, S.~Mizuno, F.~Vernizzi, and D.~Wands, {\it {Non-Gaussianities from
  ekpyrotic collapse with multiple fields}},  {\em JCAP} {\bf 0711} (2007) 024,
  [\href{http://xxx.lanl.gov/abs/0708.4321}{{\tt arXiv:0708.4321}}].

\bibitem{Lalak:2007vi}
Z.~Lalak, D.~Langlois, S.~Pokorski, and K.~Turzynski, {\it {Curvature and
  isocurvature perturbations in two-field inflation}},  {\em JCAP} {\bf 0707}
  (2007) 014, [\href{http://xxx.lanl.gov/abs/0704.0212}{{\tt
  arXiv:0704.0212}}].

\bibitem{Huang:2007hh}
M.-x. Huang, G.~Shiu, and B.~Underwood, {\it {Multifield DBI Inflation and
  Non-Gaussianities}},  {\em Phys.Rev.} {\bf D77} (2008) 023511,
  [\href{http://xxx.lanl.gov/abs/0709.3299}{{\tt arXiv:0709.3299}}].

\bibitem{Lehners:2008vx}
J.-L. Lehners, {\it {Ekpyrotic and Cyclic Cosmology}},  {\em Phys.Rept.} {\bf
  465} (2008) 223--263, [\href{http://xxx.lanl.gov/abs/0806.1245}{{\tt
  arXiv:0806.1245}}].

\bibitem{Beltran:2008tc}
M.~Beltr{\'a}n, {\it {Isocurvature, non-Gaussianity, and the curvaton model}},
  {\em Physical Review D} {\bf 78} (2008), no.~2 023530.

\bibitem{Kawasaki:2008jy}
M.~Kawasaki, K.~Nakayama, and F.~Takahashi, {\it {Non-Gaussianity from Baryon
  Asymmetry}},  {\em JCAP} {\bf 0901} (2009) 002,
  [\href{http://xxx.lanl.gov/abs/0809.2242}{{\tt arXiv:0809.2242}}].

\bibitem{Langlois:2009jp}
D.~Langlois and L.~Sorbo, {\it {Primordial perturbations and non-Gaussianities
  from modulated trapping}},  {\em JCAP} {\bf 0908} (2009) 014,
  [\href{http://xxx.lanl.gov/abs/0906.1813}{{\tt arXiv:0906.1813}}].

\bibitem{Chen:2010xka}
X.~Chen, {\it {Primordial Non-Gaussianities from Inflation Models}},  {\em
  Adv.Astron.} {\bf 2010} (2010) 638979,
  [\href{http://xxx.lanl.gov/abs/1002.1416}{{\tt arXiv:1002.1416}}].

\bibitem{Langlois:2010fe}
D.~Langlois and T.~Takahashi, {\it {Primordial Trispectrum from Isocurvature
  Fluctuations}},  {\em JCAP} {\bf 1102} (2011) 020,
  [\href{http://xxx.lanl.gov/abs/1012.4885}{{\tt arXiv:1012.4885}}].

\bibitem{Langlois:2011zz}
D.~Langlois and A.~Lepidi, {\it {General treatment of isocurvature
  perturbations and non-Gaussianities}},  {\em JCAP} {\bf 1101} (2011) 008,
  [\href{http://xxx.lanl.gov/abs/1007.5498}{{\tt arXiv:1007.5498}}].

\bibitem{Mulryne:2011ni}
D.~Mulryne, S.~Orani, and A.~Rajantie, {\it {Non-Gaussianity from the hybrid
  potential}},  {\em Phys.Rev.} {\bf D84} (2011) 123527,
  [\href{http://xxx.lanl.gov/abs/1107.4739}{{\tt arXiv:1107.4739}}].

\bibitem{Gong:2011cd}
J.-O. Gong and H.~M. Lee, {\it {Large non-Gaussianity in non-minimal
  inflation}},  {\em JCAP} {\bf 1111} (2011) 040,
  [\href{http://xxx.lanl.gov/abs/1105.0073}{{\tt arXiv:1105.0073}}].

\bibitem{DeSimone:2012gq}
A.~De~Simone, H.~Perrier, and A.~Riotto, {\it {Non-Gaussianities from the
  Standard Model Higgs}},  {\em JCAP} {\bf 1301} (2013) 037,
  [\href{http://xxx.lanl.gov/abs/1210.6618}{{\tt arXiv:1210.6618}}].

\bibitem{Enqvist:2012vx}
K.~Enqvist and S.~Rusak, {\it {Modulated preheating and isocurvature
  perturbations}},  {\em JCAP} {\bf 1303} (2013) 017,
  [\href{http://xxx.lanl.gov/abs/1210.2192}{{\tt arXiv:1210.2192}}].

\bibitem{Kawasaki:2011pd}
M.~Kawasaki, T.~Kobayashi, and F.~Takahashi, {\it {Non-Gaussianity from
  Curvatons Revisited}},  {\em Phys.Rev.} {\bf D84} (2011) 123506,
  [\href{http://xxx.lanl.gov/abs/1107.6011}{{\tt arXiv:1107.6011}}].

\bibitem{Kawasaki:2013ae}
M.~Kawasaki and K.~Nakayama, {\it {Axions : Theory and Cosmological Role}},
  \href{http://xxx.lanl.gov/abs/1301.1123}{{\tt arXiv:1301.1123}}.

\bibitem{Nurmi:2013xv}
S.~Nurmi, C.~T. Byrnes, and G.~Tasinato, {\it {A non-Gaussian landscape}},
  \href{http://xxx.lanl.gov/abs/1301.3128}{{\tt arXiv:1301.3128}}.

\bibitem{Kawasaki:2006gs}
M.~Kawasaki, F.~Takahashi, and T.~Yanagida, {\it {Gravitino overproduction in
  inflaton decay}},  {\em Phys.Lett.} {\bf B638} (2006) 8--12,
  [\href{http://xxx.lanl.gov/abs/hep-ph/0603265}{{\tt hep-ph/0603265}}].

\bibitem{Kawasaki:2006hm}
M.~Kawasaki, F.~Takahashi, and T.~Yanagida, {\it {The Gravitino-overproduction
  problem in inflationary universe}},  {\em Phys.Rev.} {\bf D74} (2006) 043519,
  [\href{http://xxx.lanl.gov/abs/hep-ph/0605297}{{\tt hep-ph/0605297}}].

\bibitem{Pradler:2006hh}
J.~Pradler and F.~D. Steffen, {\it {Constraints on the Reheating Temperature in
  Gravitino Dark Matter Scenarios}},  {\em Phys.Lett.} {\bf B648} (2007)
  224--235, [\href{http://xxx.lanl.gov/abs/hep-ph/0612291}{{\tt
  hep-ph/0612291}}].

\bibitem{Pradler:2006qh}
J.~Pradler and F.~D. Steffen, {\it {Thermal gravitino production and collider
  tests of leptogenesis}},  {\em Phys.Rev.} {\bf D75} (2007) 023509,
  [\href{http://xxx.lanl.gov/abs/hep-ph/0608344}{{\tt hep-ph/0608344}}].

\bibitem{Endo:2007ih}
M.~Endo, F.~Takahashi, and T.~Yanagida, {\it {Anomaly-induced inflaton decay
  and gravitino-overproduction problem}},  {\em Phys.Lett.} {\bf B658} (2008)
  236--240, [\href{http://xxx.lanl.gov/abs/hep-ph/0701042}{{\tt
  hep-ph/0701042}}].

\bibitem{Endo:2007sz}
M.~Endo, F.~Takahashi, and T.~Yanagida, {\it {Inflaton Decay in Supergravity}},
   {\em Phys.Rev.} {\bf D76} (2007) 083509,
  [\href{http://xxx.lanl.gov/abs/0706.0986}{{\tt arXiv:0706.0986}}].

\bibitem{Takahashi:2009dr}
T.~Takahashi, M.~Yamaguchi, J.~Yokoyama, and S.~Yokoyama, {\it {Gravitino Dark
  Matter and Non-Gaussianity}},  {\em Phys.Lett.} {\bf B678} (2009) 15--19,
  [\href{http://xxx.lanl.gov/abs/0905.0240}{{\tt arXiv:0905.0240}}]. * Brief
  entry *.

\bibitem{Takahashi:2009cx}
T.~Takahashi, M.~Yamaguchi, and S.~Yokoyama, {\it {Primordial Non-Gaussianity
  in Models with Dark Matter Isocurvature Fluctuations}},  {\em Phys.Rev.} {\bf
  D80} (2009) 063524, [\href{http://xxx.lanl.gov/abs/0907.3052}{{\tt
  arXiv:0907.3052}}].

\bibitem{Starobinsky:1986fxa}
A.~A. Starobinsky, {\it {Multicomponent de Sitter (Inflationary) Stages and the
  Generation of Perturbations}},  {\em JETP Lett.} {\bf 42} (1985) 152--155.

\bibitem{Sasaki:1995aw}
M.~Sasaki and E.~D. Stewart, {\it {A General analytic formula for the spectral
  index of the density perturbations produced during inflation}},  {\em
  Prog.Theor.Phys.} {\bf 95} (1996) 71--78,
  [\href{http://xxx.lanl.gov/abs/astro-ph/9507001}{{\tt astro-ph/9507001}}].

\bibitem{Sasaki:1998ug}
M.~Sasaki and T.~Tanaka, {\it {Superhorizon scale dynamics of multiscalar
  inflation}},  {\em Prog.Theor.Phys.} {\bf 99} (1998) 763--782,
  [\href{http://xxx.lanl.gov/abs/gr-qc/9801017}{{\tt gr-qc/9801017}}].

\bibitem{Chung:2011ck}
D.~J. Chung, L.~L. Everett, H.~Yoo, and P.~Zhou, {\it {Gravitational Fermion
  Production in Inflationary Cosmology}},
  \href{http://xxx.lanl.gov/abs/1109.2524}{{\tt arXiv:1109.2524}}.

\bibitem{Kuzmin:1998kk}
V.~Kuzmin and I.~Tkachev, {\it {Matter creation via vacuum fluctuations in the
  early universe and observed ultrahigh-energy cosmic ray events}},  {\em
  Phys.Rev.} {\bf D59} (1999) 123006,
  [\href{http://xxx.lanl.gov/abs/hep-ph/9809547}{{\tt hep-ph/9809547}}].

\bibitem{Kuzmin:1999zk}
V.~A. Kuzmin and I.~I. Tkachev, {\it {Ultrahigh-energy cosmic rays and
  inflation relics}},  {\em Phys.Rept.} {\bf 320} (1999) 199--221,
  [\href{http://xxx.lanl.gov/abs/hep-ph/9903542}{{\tt hep-ph/9903542}}].

\bibitem{Chung:2013sla}
D.~J.~H. Chung, H.~Yoo, and P.~Zhou, {\it {Quadratic Isocurvature
  Cross-Correlation, Ward Identity, and Dark Matter}},
  \href{http://xxx.lanl.gov/abs/1303.6024}{{\tt arXiv:1303.6024}}.

\bibitem{birrell1982ix}
N.~Birrell and P.~Davies, {\em Quantum Fields in Curved Space}.
\newblock Cambridge Monographs on Mathematical Physics. Cambridge University
  Press, 1984.

\bibitem{DeWitt1975}
B.~S. DeWitt, {\it {Quantum Field Theory in Curved Space-Time}},  {\em
  Phys.Rept.} {\bf 19} (1975) 295--357.

\bibitem{peebles1980large}
P.~Peebles, {\em The Large-scale Structure of the Universe}.
\newblock Princeton series in physics. PRINCETON University Press, 1980.

\bibitem{Efstathiou:1986}
G.~{Efstathiou} and J.~R. {Bond}, {\it {Isocurvature cold dark matter
  fluctuations}},  {\em Monthly Notices of the Royal Astronomical Society} {\bf
  218} (Jan., 1986) 103--121.

\bibitem{Liddle:2000cg}
A.~R. Liddle and D.~Lyth, {\it {Cosmological inflation and large scale
  structure}}, .

\bibitem{Green:2013rd}
D.~Green, M.~Lewandowski, L.~Senatore, E.~Silverstein, and M.~Zaldarriaga, {\it
  {Anomalous Dimensions and Non-Gaussianity}},
  \href{http://xxx.lanl.gov/abs/1301.2630}{{\tt arXiv:1301.2630}}.

\bibitem{Ellis:1990iu}
J.~R. Ellis, J.~L. Lopez, and D.~V. Nanopoulos, {\it {Confinement of fractional
  charges yields integer charged relics in string models}},  {\em Phys.Lett.}
  {\bf B247} (1990) 257.

\bibitem{Benakli:1998ut}
K.~Benakli, J.~R. Ellis, and D.~V. Nanopoulos, {\it {Natural candidates for
  superheavy dark matter in string and M theory}},  {\em Phys. Rev.} {\bf D59}
  (1999) 047301, [\href{http://xxx.lanl.gov/abs/hep-ph/9803333}{{\tt
  hep-ph/9803333}}].

\bibitem{Kusenko:1997si}
A.~Kusenko and M.~E. Shaposhnikov, {\it {Supersymmetric Q balls as dark
  matter}},  {\em Phys.Lett.} {\bf B418} (1998) 46--54,
  [\href{http://xxx.lanl.gov/abs/hep-ph/9709492}{{\tt hep-ph/9709492}}].

\bibitem{Han:1998pa}
T.~Han, T.~Yanagida, and R.-J. Zhang, {\it {Adjoint messengers and perturbative
  unification at the string scale}},  {\em Phys.Rev.} {\bf D58} (1998) 095011,
  [\href{http://xxx.lanl.gov/abs/hep-ph/9804228}{{\tt hep-ph/9804228}}].

\bibitem{Dvali:1999tq}
G.~Dvali, {\it {Infrared hierarchy, thermal brane inflation and superstrings as
  superheavy dark matter}},  {\em Phys.Lett.} {\bf B459} (1999) 489--496,
  [\href{http://xxx.lanl.gov/abs/hep-ph/9905204}{{\tt hep-ph/9905204}}].

\bibitem{Hamaguchi:1998nj}
K.~Hamaguchi, Y.~Nomura, and T.~Yanagida, {\it {Longlived superheavy dark
  matter with discrete gauge symmetries}},  {\em Phys.Rev.} {\bf D59} (1999)
  063507, [\href{http://xxx.lanl.gov/abs/hep-ph/9809426}{{\tt
  hep-ph/9809426}}].

\bibitem{Hamaguchi:1999cv}
K.~Hamaguchi, K.~Izawa, Y.~Nomura, and T.~Yanagida, {\it {Longlived superheavy
  particles in dynamical supersymmetry breaking models in supergravity}},  {\em
  Phys.Rev.} {\bf D60} (1999) 125009,
  [\href{http://xxx.lanl.gov/abs/hep-ph/9903207}{{\tt hep-ph/9903207}}].

\bibitem{Coriano:2001mg}
C.~Coriano, A.~E. Faraggi, and M.~Plumacher, {\it {Stable superstring relics
  and ultrahigh-energy cosmic rays}},  {\em Nucl.Phys.} {\bf B614} (2001)
  233--253, [\href{http://xxx.lanl.gov/abs/hep-ph/0107053}{{\tt
  hep-ph/0107053}}].

\bibitem{Cheng:2002iz}
H.-C. Cheng, K.~T. Matchev, and M.~Schmaltz, {\it {Radiative corrections to
  Kaluza-Klein masses}},  {\em Phys.Rev.} {\bf D66} (2002) 036005,
  [\href{http://xxx.lanl.gov/abs/hep-ph/0204342}{{\tt hep-ph/0204342}}].

\bibitem{Shiu:2003ta}
G.~Shiu and L.-T. Wang, {\it {D matter}},  {\em Phys.Rev.} {\bf D69} (2004)
  126007, [\href{http://xxx.lanl.gov/abs/hep-ph/0311228}{{\tt
  hep-ph/0311228}}].

\bibitem{Berezinsky:2008bg}
V.~Berezinsky, M.~Kachelriess, and M.~Solberg, {\it {Supersymmetric superheavy
  dark matter}},  {\em Phys.Rev.} {\bf D78} (2008) 123535,
  [\href{http://xxx.lanl.gov/abs/0810.3012}{{\tt arXiv:0810.3012}}].

\bibitem{Kephart:2001ix}
T.~W. Kephart and Q.~Shafi, {\it {Family unification, exotic states and
  magnetic monopoles}},  {\em Phys.Lett.} {\bf B520} (2001) 313--316,
  [\href{http://xxx.lanl.gov/abs/hep-ph/0105237}{{\tt hep-ph/0105237}}].

\bibitem{Kephart:2006zd}
T.~W. Kephart, C.-A. Lee, and Q.~Shafi, {\it {Family unification, exotic states
  and light magnetic monopoles}},  {\em JHEP} {\bf 0701} (2007) 088,
  [\href{http://xxx.lanl.gov/abs/hep-ph/0602055}{{\tt hep-ph/0602055}}].

\bibitem{Bartolo:2004if}
N.~Bartolo, E.~Komatsu, S.~Matarrese, and A.~Riotto, {\it {Non-Gaussianity from
  inflation: Theory and observations}},  {\em Phys.Rept.} {\bf 402} (2004)
  103--266, [\href{http://xxx.lanl.gov/abs/astro-ph/0406398}{{\tt
  astro-ph/0406398}}].

\bibitem{Kawasaki:2008pa}
M.~Kawasaki, K.~Nakayama, T.~Sekiguchi, T.~Suyama, and F.~Takahashi, {\it {A
  General Analysis of Non-Gaussianity from Isocurvature Perturbations}},  {\em
  JCAP} {\bf 0901} (2009) 042, [\href{http://xxx.lanl.gov/abs/0810.0208}{{\tt
  arXiv:0810.0208}}].

\bibitem{Weinberg:2008ug}
S.~Weinberg, {\em {Cosmology}}.
\newblock Oxford University Press, USA, Apr., 2008.

\bibitem{zinn}
J.~Zinn-Justin, {\em Quantum Field Theory and Critical Phenomena}.
\newblock International Series of Monographs on Physics. Clarendon Press, 2002.

\bibitem{Weinberg:2010wq}
S.~Weinberg, {\it {Ultraviolet Divergences in Cosmological Correlations}},
  {\em Phys.Rev.} {\bf D83} (2011) 063508,
  [\href{http://xxx.lanl.gov/abs/1011.1630}{{\tt arXiv:1011.1630}}].

\bibitem{Chung:1998bt}
D.~J. Chung, {\it {Classical inflation field induced creation of superheavy
  dark matter}},  {\em Phys.Rev.} {\bf D67} (2003) 083514,
  [\href{http://xxx.lanl.gov/abs/hep-ph/9809489}{{\tt hep-ph/9809489}}].

\bibitem{Chung:2001cb}
D.~J. Chung, P.~Crotty, E.~W. Kolb, and A.~Riotto, {\it {On the gravitational
  production of superheavy dark matter}},  {\em Phys.Rev.} {\bf D64} (2001)
  043503, [\href{http://xxx.lanl.gov/abs/hep-ph/0104100}{{\tt
  hep-ph/0104100}}].

\bibitem{Weinberg:2005ww}
S.~Weinberg, {\it {Quantum contributions to cosmological correlations}},  {\em
  Physical Review D} {\bf 72} (2005), no.~4 043514.

\bibitem{Calzetta:2008tw}
E.~A. Calzetta and B.-L. Hu, {\em {Nonequilibrium quantum field theory}}.
\newblock Cambridge Univ Pr, 2008.

\bibitem{Sikivie:2006ni}
P.~Sikivie, {\it {Axion Cosmology}},  {\em Lect.Notes Phys.} {\bf 741} (2008)
  19--50, [\href{http://xxx.lanl.gov/abs/astro-ph/0610440}{{\tt
  astro-ph/0610440}}].

\bibitem{Weinberg:2005vy}
S.~Weinberg, {\it {Quantum contributions to cosmological correlations}},  {\em
  Phys.Rev.} {\bf D72} (2005) 043514,
  [\href{http://xxx.lanl.gov/abs/hep-th/0506236}{{\tt hep-th/0506236}}].

\bibitem{CalzettaHu:2008}
E.~Calzetta and B.~Hu, {\em Nonequilibrium Quantum Field Theory}.
\newblock Cambridge Monographs on Mathematical Physics. Cambridge University
  Press, 2008.

\bibitem{Bean:2006io}
R.~Bean, J.~Dunkley, and E.~Pierpaoli, {\it {Constraining isocurvature initial
  conditions with WMAP 3-year data}},  {\em Physical Review D} {\bf 74} (2006),
  no.~6 063503.

\bibitem{Komatsu:2008ex}
E.~Komatsu, J.~Dunkley, M.~R. Nolta, C.~L. Bennett, B.~Gold, G.~Hinshaw,
  N.~Jarosik, D.~Larson, M.~Limon, L.~Page, D.~N. Spergel, M.~Halpern, R.~S.
  Hill, A.~Kogut, S.~S. Meyer, G.~S. Tucker, J.~L. Weiland, E.~Wollack, and
  E.~L. Wright, {\it {Five-Year Wilkinson Microwave Anisotropy Probe (WMAP)
  Observations: Cosmological Interpretation}},  {\em arXiv} {\bf astro-ph}
  (Mar., 2008).

\bibitem{Komatsu:2010in}
E.~Komatsu, K.~M. Smith, J.~Dunkley, C.~L. Bennett, B.~Gold, G.~Hinshaw,
  N.~Jarosik, D.~Larson, M.~R. Nolta, L.~Page, D.~N. Spergel, M.~Halpern, R.~S.
  Hill, A.~Kogut, M.~Limon, S.~S. Meyer, N.~Odegard, G.~S. Tucker, J.~L.
  Weiland, E.~Wollack, and E.~L. Wright, {\it {SEVEN-YEAR WILKINSON MICROWAVE
  ANISOTROPY PROBE (WMAP1) OBSERVATIONS: COSMOLOGICAL INTERPRETATION}},  {\em
  arXiv} {\bf astro-ph.CO} (Jan., 2010).

\bibitem{Larson:2010gs}
D.~Larson, J.~Dunkley, G.~Hinshaw, E.~Komatsu, M.~Nolta, C.~Bennett, B.~Gold,
  M.~Halpern, R.~Hill, and N.~Jarosik, {\it {Seven-year Wilkinson Microwave
  Anisotropy Probe (WMAP) observations: power spectra and WMAP-derived
  parameters}},  {\em The Astrophysical Journal Supplement Series} {\bf 192}
  (2011) 16.

\bibitem{Komatsu:2001rj}
E.~Komatsu and D.~N. Spergel, {\it {Acoustic signatures in the primary
  microwave background bispectrum}},  {\em Phys. Rev.} {\bf D63} (2001) 063002,
  [\href{http://xxx.lanl.gov/abs/astro-ph/0005036}{{\tt astro-ph/0005036}}].

\bibitem{Hikage:2008sk}
C.~Hikage, K.~Koyama, T.~Matsubara, T.~Takahashi, and M.~Yamaguchi, {\it
  {Limits on Isocurvature Perturbations from Non-Gaussianity in WMAP
  Temperature Anisotropy}},  {\em Mon.Not.Roy.Astron.Soc.} {\bf 398} (2009)
  2188--2198, [\href{http://xxx.lanl.gov/abs/0812.3500}{{\tt
  arXiv:0812.3500}}]. * Brief entry *.

\bibitem{Hikage:2012be}
C.~Hikage, M.~Kawasaki, T.~Sekiguchi, and T.~Takahashi, {\it {CMB constraint on
  non-Gaussianity in isocurvature perturbations}},
  \href{http://xxx.lanl.gov/abs/1211.1095}{{\tt arXiv:1211.1095}}.

\bibitem{Hikage:2012tf}
C.~Hikage, M.~Kawasaki, T.~Sekiguchi, and T.~Takahashi, {\it {Extended analysis
  of CMB constraints on non-Gaussianity in isocurvature perturbations}},  {\em
  JCAP} {\bf 1303} (2013) 020, [\href{http://xxx.lanl.gov/abs/1212.6001}{{\tt
  arXiv:1212.6001}}].

\bibitem{Arnowitt:1962}
R.~Arnowitt, S.~Deser, and C.~Misner, {\it {The dynamics of general
  relativity}},  in {\em Gravitation: An Introduction to Current Research}
  (L.~Witten, ed.), ch.~7, pp.~227--265.
\newblock Wiley, 1962.
\newblock \href{http://xxx.lanl.gov/abs/0405109}{{\tt 0405109}}.

\end{thebibliography}\endgroup
 
\end{document}